\documentclass{llncs}
\usepackage{url}

\usepackage{etex}
\usepackage{theorem}
\usepackage{mathtools,xspace}
\usepackage{caption}
\usepackage{amsmath}
\usepackage{amsfonts}
\usepackage{amssymb}
\usepackage{afterpage}
\usepackage{pdflscape}

\usepackage[all]{xy}
\SelectTips{cm}{}  
\xyoption{v2}
\xyoption{curve}
\xyoption{2cell}
\UseAllTwocells
\SilentMatrices

\usepackage{wrapfig}
\usepackage{floatflt}
\usepackage{tikz}
\usetikzlibrary{automata,positioning,arrows,shapes,snakes,backgrounds,fit}

\usepackage{cancel}

\usepackage{makeidx}  

\usepackage{stmaryrd}

\usepackage{rotating}

\newcounter{asmenumi}
\newcommand{\initarr}{\raisebox{.7em}{\rotatebox{180}{$!$}}}

\newcommand{\congrightarrow}{\mathrel{\stackrel{
           \raisebox{.5ex}{$\scriptstyle\cong\,$}}{
           \raisebox{0ex}[0ex][0ex]{$\rightarrow$}}}}

\newcommand{\iso}{\congrightarrow}

\newcommand{\pow}{\mathcal{P}}

\newcommand{\dist}{\mathcal{D}}
\newcommand{\lift}{\mathcal{L}}
\newcommand{\giry}{\mathcal{G}}
\newcommand{\place}{\underline{\phantom{n}}\,} 
\newcommand{\Sets}{\mathbf{Sets}}

\newcommand{\Meas}{\mathbf{Meas}}

\newcommand{\lang}{L}

\newcommand{\langB}{L^{\mathrm{B}}}
\newcommand{\langp}{L^{\mathsf{p}}}

\newcommand{\sigalg}{\mathfrak{F}}
\newcommand{\FSigma}{F_{\Sigma}}

\newcommand{\op}{\mathrm{op}}

\newcommand{\accstate}{\raisebox{.65ex}{\xymatrix{*+<.95em>[o][F=]{}}}}
\newcommand{\xyaccstate}{*+<.95em>[o][F=]{}}
\newcommand{\nonaccstate}{\raisebox{.65ex}{\xymatrix{*+<.95em>[o][F-]{}}}}
\newcommand{\xynonaccstate}{*+<.95em>[o][F-]{}}

\newcommand{\ProbRun}{\mathrm{ProbRun}}

\newcommand{\obj}[1]{|#1|}

\newlength{\WIDTHXRIGHTARROW}
\def\KAR#1{\settowidth{\WIDTHXRIGHTARROW}{\ensuremath{#1}\,\,\,\,\,}%
           \hspace{0.5\WIDTHXRIGHTARROW}\raisebox{0.1ex}[0ex][0ex]{$\shortmid$}\hspace{-0.5\WIDTHXRIGHTARROW}#1}
\newcommand{\xkrightarrow}[1]{\KAR{\xrightarrow{#1}}}
\newcommand{\xkleftarrow}[1]{\KAR{\xleftarrow{#1}}}

\newcommand{\defarrow}{\stackrel{\text{def.}}{\Leftrightarrow}}

\newcommand{\kto}[0]{\mathrel{\mspace{-4mu} \to \mspace{-16mu} \raisebox{0.1ex}[0ex][0ex]{$\shortmid$} \mspace{6mu}}}

\newcommand{\Kl}[0]{\mathcal{K}\mspace{-1mu}\ell}

\newcommand{\id}[0]{\mathrm{id}}
\newcommand{\tr}[0]{{\sf tr}}
\newcommand{\trB}[0]{\tr^{\mathsf{B}}}
\newcommand{\trp}[0]{\tr^{\mathsf{p}}}
\newcommand{\trinf}[0]{{\sf tr^{\infty}}}

\newcommand{\dtr}[0]{{\sf dtr}}

\newcommand\Acc{\mathsf{Acc}}

\newcommand{\myalph}{\mathsf{A}}

\newcommand{\Treeinf}[0]{\mathsf{Tree}_{\infty}}
\newcommand{\Treefin}[0]{\mathsf{Tree}_{\mathsf{fin}}}
\newcommand{\Treeplus}{\Treefin^+}

\newcommand{\Fplus}{F^{+}}
\newcommand{\Fplusinf}{F^{\plusinf}}
\newcommand{\Fplusplusinf}{F^{+\plusinf}}

\newcommand{\empseq}[0]{\langle\rangle}

\newcommand{\Cyl}{\mathrm{Cyl}}

\newcommand{\DelSt}{\mathrm{DelSt}}
\newcommand{\DelStij}[2]{\DelSt^{(#1)}_{#2}}

\newcommand{\kar}{\ar|-*\dir{|}}

\newcommand{\sol}{\mathrm{sol}}

\newcommand{\multFplusinf}{\mu^{\Fplusinf}}

\newcommand{\plusinf}{{\oplus}}
\newcommand{\fincoalg}{\zeta}
\newcommand{\initalg}{\iota}

\newcommand{\plustoinf}{\tau}

\newcommand{\iji}[1]{(#1)}
\newcommand{\ijj}[1]{#1}
\newcommand{\Fj}[1]{F^\ddagger_{\ijj{#1}}}
\newcommand{\Fij}[2]{F^{\iji{#1}}_{\ijj{#2}}}
\newcommand{\Xij}[2]{X^{\iji{#1}}_{\ijj{#2}}}
\newcommand{\FSigmaplus}{\FSigma^+}
\newcommand{\FSigmaplusinf}{\FSigma^{\plusinf}}
\newcommand{\FSigmaplusplusinf}{\FSigma^{+\plusinf}}
\newcommand{\FSigmaj}[1]{(\FSigma)^\ddagger_{\ijj{#1}}}
\newcommand{\FSigmaij}[2]{(\FSigma)^{\iji{#1}}_{\ijj{#2}}}

\newcommand{\aij}[2]{{\alpha^{\iji{#1}}_{\ijj{#2}}}}
\newcommand{\bij}[2]{{\beta^{\iji{#1}}_{\ijj{#2}}}}
\newcommand{\bi}[1]{\beta_{\ijj{#1}}}

\newcommand{\myl}{\tilde{\ell}}
\newcommand{\mylij}[2]{\myl^{(#1)}_{#2}}
\newcommand{\lij}[2]{l^{(#1)}_{#2}}

\newcommand{\redeco}{p}

\newcommand{\redecoij}[2]{{\redeco^{(#1)}_{#2}}}
\newcommand{\pij}[2]{\redecoij{#1}{#2}}
\newcommand{\pj}[1]{{\redeco_{#1}}}

\newcommand{\decomp}{\mathsf{decomp}}
\newcommand{\decompij}[2]{\decomp^{(#1)}_{#2}}

\newcommand{\Run}{\mathrm{Run}}
\newcommand{\AccRun}{\mathrm{AccRun}}

\newcommand{\AccTree}{\mathrm{AccTree}}
\newcommand{\AccTreeij}[2]{\AccTree^{(#1)}_{#2}}
\newcommand{\AccTreeBj}[1]{\AccTree_{#1}}

\newcommand{\qedhere}{\ifmmode\tag*{\qed}\else\hfill\qed\fi}

\newif\ifignore 
\ignorefalse

\makeatletter
  \def\@thmcountersep{.}
\makeatother

\theorembodyfont{\itshape}
\newtheorem{mytheorem}{Theorem}[section]
\newtheorem{mylemma}[mytheorem]{Lemma}
\newtheorem{myproposition}[mytheorem]{Proposition}
\newtheorem{mysublemma}[mytheorem]{Sublemma}

\theorembodyfont{\rmfamily}

\newtheorem{myremark}[mytheorem]{Remark}
\newtheorem{myexample}[mytheorem]{Example}
\newtheorem{mydefinition}[mytheorem]{Definition}
\newtheorem{myassumption}[mytheorem]{Assumption}

\spnewtheorem*{myproof}{Proof}{\itshape}{\rmfamily}

\begin{document}

\title{Categorical B\"uchi and Parity Conditions via Alternating Fixed Points of Functors}         

\author{
Natsuki Urabe
\inst{1}
\and
Ichiro Hasuo
\inst{2}
}
%
%
%
\institute{University of Tokyo, Japan
\and
National Institute of Informatics, Tokyo, Japan
}

\maketitle

\begin{abstract}
Categorical studies of recursive data structures and their associated reasoning principles have mostly focused on two extremes: initial algebras and induction, and final coalgebras and coinduction. In this paper we study their in-betweens. We formalize notions of alternating fixed points of functors using constructions that are similar to that of free monads. We find their use in categorical modeling of accepting run trees under the B\"uchi and parity acceptance condition. This modeling abstracts away from states of an automaton; it can thus be thought of as the ``behaviors'' of systems with the B\"uchi or parity conditions, in a way that follows the tradition of coalgebraic modeling of system behaviors.
\end{abstract}

\section{Introduction}\label{sec:intro}
\begin{wrapfigure}[3]{r}[5pt]{1.7cm}
\small
\vspace*{-1.0cm}
\hspace{-5mm}
\begin{tikzpicture}[shorten >=1pt, node distance=.8cm, initial text=, bend angle=30, accepting/.style={double distance=2.0pt}, every loop/.style={min distance=5mm,looseness=5}]
\node [state, initial,   label={above:$x$}, minimum size=0pt] (x)                  {};
\node [state, accepting, label={above:$y$},  label={[label distance=1.0mm]60:$\mathcal{A}$}, minimum size=0pt] (y) [right = 4mm of x] {};
\path [->] (x) edge [bend left] node             [above]  {$b$}  (y)
               edge [in=310, out=230, loop] node [below]  {$a$}  ()
           (y) edge [bend left] node             [below]  {$a$}  (x)
               edge [in=310, out=230, loop] node [below]  {$b$}  ();
\end{tikzpicture}
\end{wrapfigure}
\subsubsection*{B\"uchi Automata }
The \emph{B\"uchi condition} is a common acceptance condition for automata for infinite words.
Let $x_{i}\in X$ be a state of an automaton $\mathcal{A}$ and $a_{i}\in\myalph$ be a character, for each $i\in\omega$. An infinite run $x_{0}\!\xrightarrow{a_{0}}\!x_{1}\!\xrightarrow{a_{1}}\!\cdots$ satisfies the B\"uchi condition if $x_{i}$ is an accepting state (usually denoted by $\accstate$) for infinitely many $i$. An example of a B\"uchi automaton is shown on the right. The word $(ba)^{\omega}$ is accepted, while $ba^{\omega}$ is not.
A function that assigns each $x\in X$ the set of accepted words from $x$ is called the \emph{trace semantics} of 
the B\"uchi automaton.

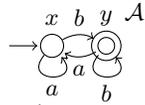
\begin{wrapfigure}[3]{r}[5pt]{1.5cm}
\small
\vspace*{-1.0cm}
\hspace{-4mm}
\xymatrix@R=1.2em@C=.8em{
 {F Y} \ar@{-->}[r]_{} \ar@{}[dr]|{}  & 
 {FZ}  \\
 {Y} \ar[u]^{d} \ar@{-->}[r]^{} & 
 {Z} \ar[u]^{\fincoalg}_{\cong} 
 }
\end{wrapfigure}
\subsubsection*{Categorical Modeling}
The main goal of this paper is to give a categorical characterization of such runs under the B\"uchi condition. This is in the line of the established field of categorical studies of finite and infinite datatypes: it is well-known that finite trees form an initial algebra, and infinite trees form a final coalgebra; and finite/infinite words constitute a special case. These categorical characterizations offer powerful reasoning principles of \emph{(co)induction} for both definition and proof. While the principles are  categorically simple ones corresponding to universality of initial/final objects, they have proved powerful and useful in many different branches of computer science, such as functional programming and process theory. See the diagram on the right above illustrating coinduction: given a functor $F$, its final coalgebra $\zeta\colon Z\iso FZ$ has a unique homomorphism to it from an arbitrary $F$-coalgebra $d\colon Y\to FY$. In many examples, a final coalgebra is described as a set of ``infinite $F$-trees.''

Extension of such (co)algebraic characterizations of data structures to the B\"uchi condition is not straightforward, however. A major reason is the non-local character of 
the B\"uchi condition: its satisfaction cannot be reduced to a local, one-step property of the run. For example, one possible attempt of capturing the B\"uchi condition is as a suitable subobject of the set $\Run(X)=(\myalph\times X)^{\omega}$ of all runs (including nonaccepting ones). The latter set admits clean categorical characterization as a final coalgebra
\begin{math}
 \Run(X)\iso F\bigl(\Run(X)\bigr)
\end{math}
for the functor $F=(\myalph\times X)\times\place$. Specifying its subset according to 
the B\"uchi condition seems hard if we insist on the coalgebraic language 
which is centered around the local notion of transition represented by a coalgebra structure morphism $c\colon X\to FX$.

There have been some research efforts in this direction, namely the categorical characterization of the B\"uchi condition. In~\cite{CianciaV12} the authors insisted on finality and 
%
characterize languages of \emph{Muller automata} (a generalization of B\"uchi automata)
by a final coalgebra in $\Sets^2$. 
Their characterization however relies on the lasso characterization of the B\"uchi condition that works only in the setting of finite state spaces. In~\cite{urabeSH16coalgebraictrace} we presented an alternative characterization that 
covers infinite state spaces and automata with probabilistic branching. The key idea was the departure from coinduction, that is, reasoning that relies on the universal property of greatest fixed points. Note that a final coalgebra $\zeta\colon Z\iso FZ$ is a ``categorical greatest fixed point'' for a functor $F$. 

Our framework in~\cite{urabeSH16coalgebraictrace} was built on top of the so-called \emph{Kleisli approach}  to trace semantics of coalgebras~\cite{powerT99coalgebraicfoundation,jacobs04tracesemantics,hasuoJ05cotextfree,hasuoJS07generictrace}. There a system is a coalgebra in a Kleisli category $\Kl(T)$, where $T$ represents the kind of branching the system exhibits (nondeterminism, probability, etc.). A crucial fact in this approach is that homsets of the category $\Kl(T)$ come with a natural order structure. Specifically, in~\cite{urabeSH16coalgebraictrace},  we characterized trace semantics under the B\"uchi condition 
as 
in the diagrams (\ref{eq:fig:nestedFP}) below%
\footnote{We write $f:X\kto Y$ for a Kleisli arrow $f\in\Kl(T)(X,Y)$ and $\overline{F}:\Kl(T)\to\Kl(T)$ for 
a lifting of the functor $F$ over $\Kl(T)$, for distinction.}, where i) 
$X_1$ (resp.\ $X_2$) is the set of nonaccepting (resp.\ accepting) states of the B\"uchi automaton (i.e.\ $X=X_1+X_2$),
and ii) the two diagrams form a \emph{hierarchical equation system} (HES), that is roughly a planar representation of nested and alternating fixed points. 
In the HES, we first calculate the least fixed point for the left diagram, and 
then calculate the greatest fixed point for the right diagram with $u_1$ replaced by the obtained least fixed point.
Note that the order of calculating fixed points matters.
%
\begin{equation}\label{eq:fig:nestedFP}
{\small
\!\!\!\!\!\xymatrix@R=1.4em@C=2.7em{
 {F X} \kar[r]^*!/^3pt/{\scriptsize\mbox{$\overline{F}[u_1,u_2]$}} \ar@{}[dr]|{\color{red}=_\mu}  & 
 {FZ}  \\
 {X_1} \kar[u]^{c_1} \kar[r]^{u_1} & 
 {Z} \kar[u]^{J\fincoalg}_{\cong} 
 }
 \qquad
  \xymatrix@R=1.4em@C=2.7em{
 {F X} \kar[r]^*!/^3pt/{\scriptsize\mbox{$\overline{F}[u_1,u_2]$}} \ar@{}[dr]|{\color{blue}=_\nu}  & 
 {FZ}  \\
 {X_2} \kar[u]^{c_2} \kar[r]^{u_2} & 
 {Z} \kar[u]^{J\fincoalg}_{\cong} 
 }}
\end{equation}

\subsubsection*{Contributions: Decorated Trace Semantics by Categorical Datatypes}
In this paper we introduce an alternative categorical characterization to the one in~\cite{urabeSH16coalgebraictrace} for the B\"uchi conditions, where we do not need alternating fixed points in homsets. This is made possible by suitably refining the value domain, from a final coalgebra to a novel categorical datatypes $\Fplusplusinf 0$ and $\Fplus(\Fplusplusinf 0)$ that have the B\"uchi condition built in them. Diagrammatically the characterization looks as in~(\ref{eq:fig:GFP}) below. Note that we ask for the greatest fixed point in both squares. 
\begin{equation}\label{eq:fig:GFP}
{\small
\xymatrix@R=1.2em@C=2.3em{
 {F X} \kar[r]^(.3)*!/^3pt/{\scriptsize\mbox{$\overline{F}(v_1+v_2)$}} \ar@{}[dr]|{\color{blue}=_\nu}  & 
 {F (\Fplus(\Fplusplusinf 0)+\Fplusplusinf 0)}  \\
 {X_1} \kar[u]^{c_1} \kar[r]^{v_1} & 
 {\Fplus(\Fplusplusinf 0)} \kar[u]^{J(\bi{1})_0}_{\cong} 
 }
 \qquad
 \xymatrix@R=1.2em@C=2.3em{
 {F X} \kar[r]^(.3)*!/^3pt/{\scriptsize\mbox{$\overline{F}(v_1+v_2)$}} \ar@{}[dr]|{\color{blue}=_\nu}  & 
 {F (\Fplus(\Fplusplusinf 0)+\Fplusplusinf 0)}  \\
 {X_2} \kar[u]^{c_2} \kar[r]^{v_2} & 
 {\Fplusplusinf 0} \kar[u]^{J(\bi{2})_0}_{\cong} 
}}
\end{equation}
%

The functors $\Fplus$ and $\Fplusplusinf $ used in the datatypes 
are obtained by applying two operations
$(\place)^+$ and $(\place)^\plusinf$ to a functor $F$.
For an endofunctor $G$ on a category $\mathbb{C}$ with enough initial algebras, 
$G^+ X$ is given by the carrier object of a (choice of) an initial $G(\place+X)$-algebra for each $X\in\mathbb{C}$. 
The universality of initial algebras allows one to define $G^+f:G^+X\to G^+Y$ for each $f:X\to Y$ and 
extend $G^+$ to a functor $G^+:\mathbb{C}\to\mathbb{C}$. 
This definition
is much similar to that of a \emph{free monad} $G^*$,
where $G^*X$ is the carrier object of an initial $G(\place)+X$-algebra for $X\in\mathbb{C}$.
The operation $(\place)^\plusinf$ is defined similarly: for $G:\mathbb{C}\to\mathbb{C}$ and $X\in\mathbb{C}$, 
$G^\plusinf X$ is given by the carrier object of a final $G(\place+X)$-coalgebra.
This construction resembles to that of \emph{free completely iterative algebras}~\cite{MILIUS20051}.

%
%

The constructions of $\Fplus(\Fplusplusinf 0)$ and $\Fplusplusinf 0$ has a clear intuitive meaning. 
For the specific example of $\myalph$-labeled nondeterministic B\"uchi automata, $T=\pow$, $F=\myalph\times(\place)$, $\Fplus(\Fplusplusinf 0)
\cong\Fplusplusinf 0\cong(\myalph^+)^\omega$. 
%
Hence an element in $\Fplus(\Fplusplusinf 0)$ or $\Fplusplusinf 0$ is identified with an infinite sequence of finite words. 
We understand it as an infinite word ``decorated'' with information about how accepting states are visited,
by considering that an accepting state is visited at each splitting between finite words.
For example, we regard $(a_0a_1)(a_2a_3a_4)(a_5a_6)(a_7)\ldots\in(\myalph^+)^\omega\cong\Fplusplusinf 0$ 
as an infinite word 
decorated as follows.
\vspace{-5mm}
\begin{equation}\label{eq:1801201738}
\vspace{-4mm}
\xymatrix@R=.6em@C=1.2em{
 {} & {} & {} & {} & {} & {} & {} & {} & {}\\
 \xyaccstate \ar[r]^{a_0} &
 \xynonaccstate \ar[r]^{a_1} & 
 \xyaccstate \ar[r]^{a_2} \ar@{..}[u] \ar@{..}[d] &
 \xynonaccstate \ar[r]^{a_3} & 
 \xynonaccstate \ar[r]^{a_4} & 
 \xyaccstate  \ar[r]^{a_5} \ar@{..}[u] \ar@{..}[d]&
 \xynonaccstate \ar[r]^{a_6} & 
 \xyaccstate \ar[r]^{a_7} \ar@{..}[u] \ar@{..}[d]&
 \xyaccstate \ar[r] \ar@{..}[u] \ar@{..}[d] &
 {\cdots}\\
 {} & {} & {} & {} & {} & {} & {} & {} & {} 
} 
\end{equation}
An element in 
$\Fplus(\Fplusplusinf 0)$
is similarly understood, except that the initial state is regarded as a nonaccepting state.
We note that 
by its definition, 
the resulting ``decorated'' word always satisfies the B\"uchi condition.

Thus the arrows $v_1:X_1\kto \Fplus(\Fplusplusinf 0)$ and 
$v_2:X_2\kto \Fplusplusinf 0$ in (\ref{eq:fig:GFP}) are regarded as a kind of trace semantics that 
assigns each state $x\in X$ the set of infinite words accepted from $x$ 
``decorated'' with information about the corresponding accepting run.
Hence we shall call $v_1$ and $v_2$ a \emph{decorated trace semantics} for the coalgebra $c$.
The generality of the category theory allows us to define decorated trace semantics for 
systems with other transition or branching types,
e.g.\ \emph{B\"uchi tree automata} or \emph{probabilistic B\"uchi automata}.

In this paper, we also show the relationship between decorated trace semantics 
and (ordinary) trace semantics for B\"uchi automata. 
For the concrete case of B\"uchi automata sketched above, 
there exists a canonical function 
$(\myalph^+)^\omega\to \myalph^\omega$
that flattens
a sequence and hence removes the ``decorations''. 
It is easy to see that if we thus remove decorations of a decorated trace semantics
then we obtain an ordinary trace semantics.
We shall prove its categorical counterpart.

%

In fact, the framework in~\cite{urabeSH16coalgebraictrace} also covered the \emph{parity} condition,
which generalizes the B\"uchi condition.
A \emph{parity automaton} is equipped with a function $\Omega:X\to[1,2n]$ that assigns
a natural number called a \emph{priority} to each state $x\in X$.
Our new framework developed in the current paper also covers parity automata.
In order to obtain the value domain for parity automata, we repeatedly apply $(\place)^+$ and $(\place)^\plusinf$ to $F$
like $F^{+\plusinf\cdots+\plusinf}0$.

%
%
%

Compared to the existing characterization shown in (\ref{eq:fig:nestedFP}),
one of the characteristics of our new characterization as shown in (\ref{eq:fig:GFP}) is that 
information about accepting states is more explicitly captured in decorated trace semantics,
as in (\ref{eq:1801201738}).
This characteristics would be useful in categorically characterizing notions about B\"uchi or parity automata.
%
For example, we could use it for categorically characterizing
(bi)simulation notions for B\"uchi automata, e.g.\ 
\emph{delayed simulation}~\cite{etessamiWS05fairsimulation}, a simulation notion which is known to be appropriate
for state space reduction.

To summarize, our contributions in this paper are as follows:
\begin{itemize}
\item We introduce a new categorical data type $\Fplusplusinf 0$, an alternating fixed point of a functor,
 for  characterizing the B\"uchi acceptance condition.

\item Using the data type, we introduce a categorical decorated trace semantics, simply as a greatest fixed point.

\item We show the categorical relationship with ordinary trace semantics in~\cite{urabeSH16coalgebraictrace}.

\item We instantiate the framework to several types of concrete systems.

\item We extend the framework to the parity condition (in the appendix).
\end{itemize}

\subsubsection*{Related Work}
As we have mentioned, 
a categorical characterization of B\"uchi and parity conditions is also found in~\cite{CianciaV12}, but
adaptation to infinite-state or probabilistic systems seems to be difficult in their framework. 
There also exist notions which are fairly captured by their characterization 
but seem difficult to capture in the frameworks in~\cite{urabeSH16coalgebraictrace} and this paper, such as bisimilarity.

The notion of alternating fixed point of functors is also used in~\cite{ghaniHP09LMCS,ADAMEK201841}.
In~\cite{ghaniHP09LMCS} the authors characterize the set of continuous functions from $A^\omega$ to $B^\omega$ 
as an alternating fixed point $\nu X.\, \mu Y.\, (B\times X)+Y^A$ of a functor.
Although the data type and the one used in the current paper are different and incomparable,
the intuition behind them is very similar, because the former comes with a 
B\"uchi-like flavor:
if $f(a_0a_1\ldots)=b_0b_1\ldots$ then each $b_i$ should be determined by 
a \emph{finite} prefix of $a_0a_1\ldots$, and therefore $f$ is regarded 
as an \emph{infinite} sequence of such assignments.
In~\cite[\S{}7]{ADAMEK201841} a sufficient condition for the existence of such an alternating fixed point is discussed.

\subsubsection*{Organization}
\S{}\ref{sec:prelim} gives preliminaries.
In \S{}\ref{sec:altFPF} we introduce a categorical data type for 
decorated trace semantics as an alternating fixed point of functors.
In \S{}\ref{sec:lijB} we define a categorical decorated trace semantics, 
and show a relationship 
with ordinary categorical trace semantics
in~\cite{urabeSH16coalgebraictrace}.
In \S{}\ref{sec:NPTAB} we apply the framework 
to 
nondeterministic B\"uchi tree automata.
In \S{}\ref{sec:forprob}, we briefly discuss systems with other branching types.
In 
\S{}\ref{sec:conclu}, we conclude and give future work.

All the discussions in this paper also apply to the parity condition.
However, for the sake of simplicity and limited space, we mainly focus on the B\"uchi condition throughout the paper, and defer discussions about the parity condition to the appendix. 
We omit a proof if an analogous statement is proved for the parity condition in the appendix.
Some other proofs and discussions are also deferred to the appendix.

\section{Preliminaries}\label{sec:prelim}
\subsection{Notations}\label{subsec:nota}
For $m,n\in\mathbb{N}$, $[m,n]$ denotes the set $\{i\in\mathbb{N}\mid m\leq i\leq n\}$.
We write $\pi_i:\prod_{i} X_i\to X_i$ and $\kappa_i:X_i\to\coprod_{i}X_i$ 
for the canonical projection and injection respectively.
For a set $A$, $A^*$ (resp.\ $A^\omega$) denotes the set of finite (resp.\ infinite) sequences over $A$, 
 $A^\infty$ denotes $A^*\cup A^\omega$, and
 $A^+$ denotes $A^*\setminus\{\empseq\}$.
We write $\empseq$ for the empty sequence. 
For a monotone function $f:(X,\sqsubseteq)\to(X,\sqsubseteq)$, $\mu f$ (resp.\ $\nu f$) denotes its least (resp.\ greatest) fixed point (if it exists).
%
We write $\Sets$ for the category of sets and functions,
and 
$\Meas$ for the category of measurable sets and measurable functions.
For $f:X\to Y$ and $A\subseteq Y$, $f^{-1}(A)$ denotes $\{x\in X\mid f(x)\in A\}$.
%


\subsection{Fixed Point and Hierarchical Equation System}\label{subsec:HES}
In this section we review the notion of \emph{hierarchical equation system} (HES)~\cite{CleavelandKS92fmc,ArnoldNiwinski}.
It is a kind of a representation of an alternating fixed point.
\begin{mydefinition}[HES]\label{def:hes}
A \emph{hierarchical equation system} (HES for short) is a system of equations of the following form.
\[
E=
\left\{
\begin{matrix}
u_1 &=_{\eta_1} & f_1(u_1,\ldots,u_{m}) & \in (L_1,\sqsubseteq_1) \\
u_2 &=_{\eta_2} & f_2(u_1,\ldots,u_{m}) & \in (L_2,\sqsubseteq_2) \\[-1mm]
& \vdots \\[-2mm]
u_{m} &=_{\eta_m} & f_{m}(u_1,\ldots,u_{m}) & \in (L_m,\sqsubseteq_m) \\
\end{matrix}
\right.
\]
Here for each $i\in[1,m]$, $(L_i,\leq_i)$ is a complete lattice, $u_i$ is a variable that ranges over $L_i$,
$\eta_i\in\{\mu,\nu\}$ and $f_i:L_1\times\dots\times L_m\to L_i$ is a monotone function.
\end{mydefinition}

\begin{mydefinition}[solution] 
\label{def:solHES}
Let $E$ be an HES as in Def.~\ref{def:hes}.
For each $i\in[1,m]$ and $j\in[1,i]$ we inductively define $f_i^\ddagger:L_i\times\cdots\times L_m\to L_i$ and
$\lij{i}{j}:L_{i+1}\times\cdots\times L_m\to L_j$ as follows (no need to distinguish the base case from the step case):
\begin{itemize}
\item $f_i^\ddagger(u_i,\ldots,u_m):=f_i(\lij{i-1}{1}(u_i,\ldots,u_m),\ldots,\lij{i-1}{i-1}(u_i,\ldots,u_m),u_i,\ldots,u_m)$; and

\item $\lij{i}{i}(u_{i+1},\ldots,u_m):=\eta f_i^\ddagger(\place,u_{i+1},\ldots,u_m)$ where $\eta=\mu$ if $i$ is odd and $\eta=\nu$ if $i$ is even. For $j<i$, $\lij{i}{j}(u_{i+1},\ldots,u_m):=\lij{i-1}{j}(\lij{i}{i}(u_{i+1},\ldots,u_m),u_{i+1},\ldots,u_m)$.
If such a least or greatest fixed point does not exist, then it is undefined.
\end{itemize}
We call $(\lij{i}{1},\ldots,\lij{i}{i})$ the \emph{$i$-th intermediate solution}.
The \emph{solution} of the HES $E$ is a family $(u^\sol_1,\ldots,u^\sol_m)\in L_1\times\cdots\times L_m$ defined by
$u^\sol_i:=\lij{m}{i}(*)$ for each $i$.
\end{mydefinition}

\subsection{Categorical Finite and Infinitary Trace Semantics}\label{subsec:catprelim}
We review~\cite{powerT99coalgebraicfoundation,hasuoJS07generictrace,jacobs04tracesemantics,urabeH15extended} and see how finite and infinitary traces of transition systems are characterized categorically.
We assume that the readers are familiar with basic theories of categories and coalgebras.
See e.g.~\cite{borceux94handbookI,jacobs16CoalgBook} for details.

We model a system as a \emph{$(T,F)$-system}, 
a coalgebra 
$c:X\to TFX$ where $T$ is a monad representing the branching type and $F$ is an endofunctor representing the transition 
type of the system.
Here are some examples of $T$ and $F$:

\begin{mydefinition}[$\pow$, $\dist$, $\lift$ and $\giry$]\label{def:monads}
The \emph{powerset monad} is a monad $\pow=(\pow,\eta^{\pow},\mu^{\pow})$ on $\Sets$
where $\pow X:=\{A\subseteq X\}$, $\pow f(A):=\{f(x)\mid x\in A\}$, $\eta^{\pow}_X(x):=\{x\}$ and
$\mu^{\pow}_X(\Gamma):=\bigcup_{A\in\Gamma}A$.
The \emph{subdistribution monad} is a monad $\dist=(\dist,\eta^{\dist},\mu^{\dist})$ on $\Sets$
where $\dist X:=\{\delta:X\to[0,1]\mid \text{$|\{x\mid\delta(x)>0\}|$ is countable, and }\sum_{x}\delta(x)\leq 1\}$, 
$\dist f(\delta)(y):=\sum_{x\in f^{-1}(\{y\})}\delta(x)$, $\eta^{\dist}_X(x)(x')$ is $1$ if $x=x'$ and $0$ otherwise, and
$\mu^{\dist}_X(\Phi)(x):=\sum_{\delta\in\dist X}\Phi(\delta)\cdot\delta(x)$.
The \emph{lift monad} is a monad $\lift=(\lift,\eta^{\lift},\mu^{\lift})$ on $\Sets$
where $\lift X:=\{\bot\}+X$, $\lift f(a)$ is $f(a)$ if $a\in X$ and $\bot$ if $a=\bot$, $\eta^{\lift}_X(x):=x$ and
$\mu^{\lift}_X(a):=a$ if $a\in X$ and $\bot$ if $a=\bot$.
The \emph{sub-Giry monad} is a monad $\giry=(\giry,\eta^{\giry},\mu^{\giry})$ on $\Meas$
where $\giry (X,\sigalg_X)$ is carried by the set of probability measures over $(X,\sigalg_X)$,
$\giry f(\varphi)(A):=\varphi(f^{-1}(A))$, $\eta^{\giry}_X(x)(A)$ is $1$ if $x\in A$ and $0$ otherwise, and
$\mu^{\giry}_X(\Xi)(A):=\int_{\delta\in\giry X}\delta(A)d\Xi$. 
\end{mydefinition}

\begin{mydefinition}[polynomial functors]\label{def:polyfunct}
A \emph{polynomial functor $F$ on $\Sets$} is defined by the following BNF notation:
$F::=\id\mid A\mid F\times F\mid \coprod_{i\in I}F$ where $A\in\Sets$ and $I$ is countable.
A \emph{(standard Borel) polynomial functor $F$ on $\Meas$} is defined by the following BNF notation:
$F::=\id\mid A\mid F\times F\mid \coprod_{i\in I}F$ where $A\in\Meas$, $I$ is countable,
and the $\sigma$-algebras over products and coproducts are given in the standard manner (see e.g.~\cite[Def.~2.2]{urabeH15extended}).
\end{mydefinition}

A carrier of   an \emph{initial $F$-algebra} models a domain of finite traces~\cite{hasuoJS07generictrace}
while that of a \emph{final $F$-coalgebra} models a domain of infinitary traces~\cite{jacobs04tracesemantics}. 
%
For example, as we have seen in \S{}\ref{sec:intro}, for 
$F=\{\checkmark\}+\myalph\times(\place)$ on $\Sets$, the carrier set of the final $F$-coalgebra is $\myalph^\infty$
while that of the initial $F$-algebra 
is $\myalph^*$.
The situation is similar for a polynomial functor $F=(\{\checkmark\},\pow\{\checkmark\})+(\myalph,\pow\myalph)\times(\place)$ on $\Meas$. The carrier of an initial algebra is $(\myalph^*,\pow\myalph^*)$, and
that of a final coalgebra is $(\myalph^\infty,\sigalg_{\myalph^\infty})$ where 
$\sigalg_{\myalph^\omega}$ is the standard $\sigma$-algebra generated by the cylinder set.

In general, for a certain class of functors, 
an initial algebra and a final coalgebra are obtained by the following well-known construction.

\begin{mytheorem}[\cite{adamekK79leastfixed}]\label{thm:constinitfin}
\begin{enumerate}
\item\label{item:thm:constinitfin1}
Let $(A,(\pi_i:F^i0\to A)_{i\in\omega})$ be a colimit of an $\omega$-chain 
$0\xrightarrow{\initarr}F0\xrightarrow{F\initarr}F^20\xrightarrow{F^2\initarr}\dots$.
If $F$ preserves the colimit, then the unique mediating arrow $\initalg:FA\to A$ from the colimit 
 $(FA,(F\pi_i:F^{i+1}0\to FA)_{i\in\omega})$
to a cocone $(A,(\pi'_i:F^i0\to A)_{i\in\omega})$ where $\pi'_i=\pi_{i+1}$ is an initial $F$-algebra.

\item\label{item:thm:constinitfin2}
Let $(Z,(\pi_i:A\to F^i1)_{i\in\omega})$ be a limit of an $\omega^\op$-chain 
$1\xleftarrow{!}F1\xleftarrow{F!}F^21\xleftarrow{F^2!}\dots$.
If $F$ preserves the limit, then the unique mediating arrow $\fincoalg:Z\to FZ$ from 
a cone 
$(Z,(\pi'_i:A\to F^i1)_{i\in\omega})$ where $\pi'_i=\pi_{i+1}$
to the 
limit $(FZ,(F\pi_i:FZ\to F^{i+1}1)_{i\in\omega})$ is a final $F$-coalgebra.
\qed
\end{enumerate}
\end{mytheorem}

We next quickly review notions about the Kleisli category $\Kl(T)$.


\begin{mydefinition}[$\Kl(T)$, $J$, $U$ and $\overline{F}$]\label{def:klcat}
Let $T=(T,\eta,\mu)$ be a monad on $\mathbb{C}$.
The \emph{Kleisli category $\Kl(T)$} is given by $\obj{\Kl(T)}=\obj{\mathbb{C}}$ and
$\Kl(T)(X,Y)=\mathbb{C}(X,TY)$ for $X,Y\in\obj{\Kl(T)}$.
An arrow $f\in\Kl(T)(X,Y)$ is called a \emph{Kleisli arrow}, and we write $f:X\kto Y$ for distinction.
Composition of arrows $f:X\kto Y$ and $g:Y\kto Z$ is 
defined by $\mu_Z\circ Tg\circ f$, and
denoted by $g\odot f$ for distinction.
%
%
The \emph{lifting functor} $J:\mathbb{C}\to\Kl(T)$ is defined by:
$JX:= X$ and 
$J(f):=\eta_Y\circ f$
for $f:X\to Y$.
The \emph{forgetful functor} $U:\Kl(T)\to\mathbb{C}$ is defined by:
$UX:=TX$ and
$U(g):=\mu_Y\circ Tg$
for $g:X\kto Y$.
%
%
%
A functor $\overline{F}:\Kl(T)\to\Kl(T)$ is called a \emph{lifting} of $F:\mathbb{C}\to\mathbb{C}$ if 
$\overline{F}J=JF$.
\end{mydefinition}

\begin{myexample}\label{exa:1802250620}
Let $T=\pow$ and $F=\sum_{n=0}^{\omega}\Sigma_n\times(\place)^n:\Sets\to\Sets$.
A lifting $\overline{F}$ over $\Kl(T)$ is given by $\overline{F}X=FX$ for $X\in\Sets$ and 
$\overline{F}f(\sigma,x_0,\ldots,x_{n-1})=\{(\sigma,y_0,\ldots,y_{n-1})\mid \forall i.\, y_i\in f(x_i)\}$ for $f:X\kto Y$, $\sigma\in \Sigma_n$ and $x_0,\ldots,x_{n-1}\in X$.
(see e.g.~\cite{hasuoJS07generictrace}).
\end{myexample}

It is well-known that there is
a bijective correspondence between a lifting $\overline{F}$ and 
a \emph{distributive law},
a natural transformation $\lambda:FT\Rightarrow TF$  
satisfying some axioms~\cite{mulry93liftingtheorems}.
See \S{}\ref{sec:distlaws} for the details.

%



In the rest of this section, let $F$ be an endofunctor and $T$ be a monad on a category $\mathbb{C}$,
and assume that a lifting $\overline{F}:\Kl(T)\to\Kl(T)$ 
is given.

In~\cite{hasuoJS07generictrace}, a finite trace semantics of a transition system was 
characterized as the unique homomorphism to the final $\overline{F}$-coalgebra in $\Kl(T)$,
which is obtained by reversing and lifting the initial $F$-algebra in $\mathbb{C}$.

\begin{mydefinition}[$\tr(c)$]\label{def:fts}
We say $F$ and $T$ constitute a \emph{finite trace situation}  wrt.\ $\overline{F}$ 
if the following conditions are satisfied:
\begin{itemize}
\item An initial $F$-algebra $\iota^F:FA\to A$ exists.


\item 
$J(\initalg^F)^{-1}:A\kto \overline{F}A$ is a final $\overline{F}$-coalgebra.
\end{itemize}
For $c:X\kto \overline{F}X$, the unique homomorphism from $c$ to $J(\initalg^F)^{-1}$ is 
called the \emph{(coalgebraic) finite trace semantics} of $c$ and denoted by $\tr(c):X\kto A$.
\end{mydefinition}

In~\cite{hasuoJS07generictrace}, a sufficient condition for constituting a finite trace situation is given.

\begin{mytheorem}[\cite{hasuoJS07generictrace}]\label{thm:initfinal}
Assume  each homset of $\Kl(T)$ carries a partial order $\sqsubseteq$.
If the following conditions are satisfied, $F$ and $T$ constitute a finite trace situation. 
\begin{itemize}
\item The functor $F$ preserves $\omega$-colimits in $\mathbb{C}$. 


\item Each homset of $\Kl(T)$ 
constitutes an $\omega$-cpo with a bottom element $\bot$. 

\item Kleisli composition $\odot$ is monotone, and the lifting $\overline{F}$ is locally monotone,
i.e.\ $f\sqsubseteq g$ implies $\overline{F}f\sqsubseteq \overline{F}g$. 

\item Kleisli composition $\odot$ preserves $\omega$-suprema and the bottom element $\bot$. 

\end{itemize}
\end{mytheorem}
Here by Thm.~\ref{thm:constinitfin}, the first condition above implies existence of an initial algebra.

%
In~\cite{hasuoJS07generictrace} it was shown that  $T\in\{\pow,\dist,\lift\}$ and a polynomial functor $F$
satisfy the conditions in Thm.~\ref{thm:initfinal} wrt.\ some appropriate orderings and liftings, and hence constitute finite trace situations.
We can see  the result for $T=\dist$ implies $T=\giry$ and a standard Borel polynomial functor $F$
also satisfy the conditions. 

An infinitary trace semantics 
 was characterized in~\cite{jacobs04tracesemantics} as the greatest homomorphism to a 
 weakly final coalgebra obtained by lifting a final coalgebra.

\begin{mydefinition}[infinitary trace situation]\label{def:its}
We further assume that each homset of $\Kl(T)$ carries a partial order $\sqsubseteq$.
We say that $F$ and $T$ constitute an \emph{infinitary trace situation} wrt.\ $\overline{F}$ and 
$\sqsubseteq$ if the following conditions are satisfied:
\begin{itemize}
\item A final $F$-coalgebra $\fincoalg^F:Z\to FZ$ exists.


\item 
$J\fincoalg^F:Z\kto\overline{F}Z$ is a weakly final $\overline{F}$-coalgebra that admits the greatest homomorphism,
i.e.\ for an  $\overline{F}$-coalgebra $c:X\kto \overline{F}X$, there exists the greatest homomorphism 
from $c$ to $J\fincoalg^F$ wrt.\ $\sqsubseteq$.
\end{itemize}
The greatest homomorphism from $c$ to $J\fincoalg^F$ is 
called the \emph{(coalgebraic) infinitary trace semantics} of $c$ and denoted by $\trinf(c):X\kto Z$.
\end{mydefinition}
It is known that $T\in\{\pow,\dist,\lift,\giry\}$ and a polynomial functor $F$ constitute  infinitary trace situations
wrt.\ some orderings and liftings~\cite{urabeH15extended}.
Differently from finite trace situation, sufficient conditions for  infinitary trace situation are
not unified. 
In~\cite{urabeH15extended}, two families of sufficient conditions are given. One is applicable for $T=\pow$, and
the other is for $T\in\{\lift,\giry\}$. No condition is known for $T=\dist$.

\begin{myexample}\label{ex:fininfFA}
Let $T=\pow$ and $F=\{\checkmark\}+\myalph\times(\place)$. 
Then a $TF$-coalgebra $c:X\to\pow(\{\checkmark\}+\myalph\times X)$ is identified with an $\myalph$-labeled 
nondeterministic automaton whose  accepting states are given by $\{x\mid \checkmark\in c(x)\}$.
The arrow $\tr(c)$
has a type $X\kto \myalph^*$ and
assigns the set of accepted finite words to each state~\cite{hasuoJS07generictrace}:
\[
\tr(c)(x)=\left\{a_1a_2\ldots a_n\in\myalph^* \;\middle|\; \begin{aligned}
&\exists x_0,\ldots,x_n\in X.\; \forall i\in[1,n-1].\,  \\
&\qquad (a_{i+1},x_{i+1})\in c(x_i) \text{ and }\checkmark\in c(x_n)
\end{aligned}\right\}\,.
\]
In contrast, 
$\trinf(c):X\kto\myalph^\infty$ is given as follows~\cite{jacobs04tracesemantics}:
\begin{align*}
\trinf(c)(x)&=\tr(c)(x)\\&\cup\{a_1a_2\ldots \in\myalph^\omega\mid
\exists x_0,x_1,\ldots\in X.\; x=x_0, \forall i\in\omega.\;  (a_{i+1},x_{i+1})\in c(x_i) 
\}\,.
\end{align*}
\end{myexample}

\subsection{B\"uchi $(T,F)$-systems and its Coalgebraic Trace Semantics}\label{subsec:PTFSB}
The results in \S{}\ref{subsec:catprelim}
was extended for systems with the parity acceptance condition
in~\cite{urabeSH16coalgebraictrace}. 
We hereby review the results for the B\"uchi acceptance condition.
%
%
\begin{mydefinition}[B\"uchi $(T,F)$-system]\label{def:TFsysB}
Let $n\in\mathbb{N}$. A \emph{B\"uchi $(T,F)$-system} is a pair $(c,(X_1,X_{2}))$ 
of a $\overline{F}$-coalgebra $c:X\to \overline{F}X$ in $\Kl(T)$ and a partition $(X_1,X_{2})$ of $X$ (i.e.\ $X\cong X_1+X_{2}$).
For $i\in\{1,2\}$, we write $c_i$ for $c\circ\kappa_i:X_i\to \overline{F}X$.
\end{mydefinition}

Their coalgebraic trace semantics
is given by a solution of an HES.

\begin{mydefinition}[$\trB_i(c)$]\label{def:trsemTFsysB}
Assume that each homset of $\Kl(T)$ carries a partial order $\sqsubseteq$.
We say that $F$ and $T$ constitute a \emph{B\"uchi trace situation} wrt.\ $\overline{F}$ and 
$\sqsubseteq$ if they satisfy the following conditions: 
\begin{itemize}
\item A final $F$-coalgebra $\zeta:Z\to FZ$ exists.


\item For an arbitrary  B\"uchi $(T,F)$-system $\mathcal{X}=(c,(X_1,X_{2}))$, the following HES has a solution.
\begin{equation*} 
E_c=\left\{
\begin{matrix*}[l]
u_1\; &=_{\mu} \; J\fincoalg^{-1}\odot \overline{F}[u_1,u_{2}]\odot c_1 & \in (\Kl(T)(X_1,Z),\sqsubseteq_{X_1,Z})\\
u_2\; &=_{\nu} \; J\fincoalg^{-1}\odot \overline{F}[u_1,u_{2}]\odot c_2 & \in (\Kl(T)(X_2,Z),\sqsubseteq_{X_2,Z}) 
\end{matrix*}
\right.
\end{equation*}
\end{itemize}
The solution $\bigl(u_1^\sol:X_1\kto Z,u_2^\sol:X_2\kto Z\bigr)$
of $E_c$ is called the \emph{(coalgebraic) B\"uchi trace semantics} of 
$\mathcal{X}$. We write $\trB_i(c)$ for $u_i^\sol$ for each $i$ 
(see also Eq.~(\ref{eq:fig:nestedFP})). 
\end{mydefinition}

\begin{myexample}\label{ex:parityFAB}
Let $T=\pow$ and $F=\myalph\times(\place)$.
Then a B\"uchi $(T,F)$-system $(c:X\kto FX,(X_1,X_{2}))$ is identified with an $\myalph$-labeled 
B\"uchi automaton. 
Following Def.~\ref{def:solHES}
we shall sketch how the solution of the HES $E_c$ in Def.~\ref{def:trsemTFsysB} is calculated.
Note that $Z\cong \myalph^\omega$.
\begin{itemize}
\item We first calculate an intermediate solution $\lij{1}{1}(u_2):X_1\kto \myalph^\omega$ as the least fixed point of 
$u_1\mapsto J\fincoalg^{-1}\odot \overline{F}[u_1,u_{2}]\odot c_1$\,.

\item We next define $f_2^\ddagger:\Kl(T)(X_2,Z)\to\Kl(T)(X_2,Z)$ by 
$f_2^\ddagger(u_2):=J\fincoalg^{-1}\odot \overline{F}[\lij{1}{1}(u_2),u_{2}]\odot c_2$\,.

\item We calculate $\lij{2}{2}(*):X_2\kto \myalph^\omega$ as the greatest fixed point of $f_2^\ddagger$.

\item We let $\lij{2}{1}(*):=\lij{1}{1}(\lij{2}{2}):X_1\kto \myalph^\omega$.
\end{itemize}
Then for each $i$, the solution $\trB_i(c)=\lij{2}{i}(*)$
is given as follows~\cite{urabeSH16coalgebraictrace}:
\[
\trB_i(c)(x):=\left\{a_1a_2\ldots \in\myalph^\omega \;\middle|\; \begin{aligned}
&\exists x_0,x_1,\ldots\in X.\; \forall i\in\omega.\;  (a_{i+1},x_{i+1})\in c(x_i) \text{ and}\\
&\quad \text{$x_i\in X_2$ for infinitely many $i$}
\end{aligned}\right\}\,.
\]
\end{myexample}


\section{Alternating Fixed Points of Functors}\label{sec:altFPF}
\subsection{Categorical Datatypes for B\"uchi Systems}\label{subsec:refinementB}
We first introduce the categorical datatypes $\Fplus X$ and $\Fplusinf X$,
which are understood as least and greatest fixed points of a functor $F$.

\begin{mydefinition}[$\Fplus$, $\Fplusinf$]\label{def:FpFinfp}
For $F:\mathbb{C}\to\mathbb{C}$, we define functors $F^{+},\Fplusinf\colon \mathbb{C}\to\mathbb{C}$ as follows. 
Given  $X\in \mathbb{C}$, the object $F^{+}X$  is the carrier of (a choice of) an initial algebra 
$\initalg^F_X:F(F^{+}X+X)\iso F^{+}X$  for the functor $F(\place+X)$. 
Similarly, the object $\Fplusinf X$ is the carrier of a final coalgebra  $\fincoalg^F_X:\Fplusinf X\iso F(\Fplusinf X+X)$. 
For $f:X\to Y$, $\Fplus f:\Fplus X\to\Fplus Y$ is given as the unique homomorphism from $\initalg^F_X$ to 
$\initalg^F_Y\circ F(\id_{\Fplus Y}+f)$. We define $\Fplusinf f:\Fplusinf X\to\Fplusinf Y$ similarly.
\[
\small
\vcenter{  \xymatrix@R=.5em@C+.6em{
  {F(\Fplus X+X)}
  \ar@{-->}[r]_(.5){F(\Fplus f+\id)}
  \ar[dd]^{\substack{\cong}}_{\initalg^F_{X}} 
  &
  {F(F^+ Y+X)} 
  \ar[d]^{F(\id+f)}  
  \\
  &
    {F(F^+ Y+Y)}
   \ar[d]^{\substack{\cong}}_{\initalg^F_{Y}}     
%
  \\
  {F^+ X}
  \ar@{-->}[r]^(.5){\Fplus f}
  \ar@{}[ruu]|{=}
  &
  {F^+ Y}
}}
\qquad
\vcenter{  \xymatrix@R=.5em@C+.6em{
  {F(\Fplusinf X+Y)}
  \ar@{-->}[r]_(.5){F(\Fplusinf f+\id)}
  &
  {F(\Fplusinf Y+Y)} 
  \\
  {F(\Fplusinf X+X)}
  \ar[u]^{F(\id+f)}
  &
%
  \\
  {\Fplusinf X}
  \ar[u]_{\substack{\cong}}^{\fincoalg_{X}^F}  
  \ar@{-->}[r]^(.5){\Fplusinf f}
  \ar@{}[ruu]|{=}
  &
  {\Fplusinf Y}
       \ar[uu]_{\cong}^{\fincoalg_Y^F}   
}}
\]
\end{mydefinition}

\begin{myremark}\label{rem:1710141007}
 The construction $F^{+}$ resembles the \emph{free monad} $F^{*}$ over $F$. The latter is defined as follows: given $X\in \mathbb{C}$, the object $F^{*}X$ is the carrier of an initial algebra $F(F^{*}X)+X\iso F^{*}X$ for the functor $F(\place)+X$. The notations generalize the usual distinction between $*$ and $+$. Indeed, for $\mathbb{C}=\Sets$ and $F=\Sigma_{0}\times\place$ (where $\Sigma_{0}$ is an alphabet), we have $F^{+}1=\Sigma_{0}^{+}$ (the set of finite words of length $\ge 1$) and  $F^{*}1=\Sigma_{0}^{*}$ (the set of all finite words). 
 Similarly, $\Fplusinf$ resembles the \emph{free completely iterative monad}~\cite{MILIUS20051}.
\end{myremark}

\begin{myexample}\label{exa:1711141355}
For 
$F=\myalph\times(\place)$, by the construction in Thm.~\ref{thm:constinitfin},
$F^+ X\cong \myalph^+X$, $F^\plusinf X\cong \myalph^+X+\myalph^\omega$ and 
$\Fplusplusinf X\cong (\myalph^+)^+ X+(\myalph^+)^\omega$.
Especially, if we let $X=0$ then 
$\Fplusplusinf 0\cong(\myalph^+)^\omega$\,.
We identify $(a_{00}a_{01}\ldots a_{0n_0})(a_{10}a_{11}\ldots a_{1n_{1}})\ldots \in \Fplusplusinf 0\cong(\myalph^+)^\omega$
with the following ``decorated'' sequence:
\begin{equation*}
(a_{00},\accstate)(a_{01},\nonaccstate)\ldots (a_{0n_0},\nonaccstate)
(a_{10},\accstate)(a_{11},\nonaccstate)\ldots (a_{1n_1},\nonaccstate)
\ldots
\in(\myalph\times \{\nonaccstate,\accstate\})^\omega\,.
\end{equation*}
The second component of each element (i.e.\ \emph{decoration}) represents a break of a word: 
it is $2$ iff it's the beginning of a word in $A^+$. 
It is remarkable that in the sequence above, $\accstate$ always appears infinitely many times. 
Hence $w\in(\myalph^+)^\omega$ is understood as an infinite word decorated 
so that the B\"uchi condition is satisfied. 
\end{myexample}

We next define Kleisli arrows $\bi{1\,X}$ and $\bi{2\,X}$ that are used 
to define decorated trace semantics (see the diagrams in (\ref{eq:fig:GFP})).

\begin{mydefinition}\label{def:1710131524B}
We define natural transformations 
$\bi{1}:\Fplus(\Fplusplusinf+\id) \Rightarrow F(\Fplus\Fplusplusinf+\Fplusplusinf+\id)$ and
$\bi{2}:\Fplusplusinf \Rightarrow F(\Fplus\Fplusplusinf+\Fplusplusinf+\id)$ as follows.
\begin{align*}
\bi{1\,X}&:=\Bigl(&\Fplus(\Fplusplusinf X+X)\xRightarrow{(\initalg^{F}_{\Fplusplusinf X})^{-1}}
F(\Fplus\Fplusplusinf X+\Fplusplusinf X+X)\Bigr) \\
\bi{2\,X}&:=\Bigl(\Fplusplusinf X \xRightarrow{\fincoalg^{\Fplus}_X} \!\!&\Fplus(\Fplusplusinf X+X)\xRightarrow{(\initalg^{F}_{\Fplusplusinf X})^{-1}}
F(\Fplus\Fplusplusinf X+\Fplusplusinf X+X)\Bigr)
\end{align*}
\end{mydefinition}

\begin{myremark}\label{rem:1711151735}
As a final coalgebra $\fincoalg^{\Fplus}_X$ is an isomorphism, we can see from Def.~\ref{def:1710131524B} that 
$\Fplus(\Fplusplusinf X+X)\cong\Fplusplusinf X$\,.  
For $F=\myalph\times(\place)$, if we regard $\Fplusplusinf X$ as $(\myalph^+)^\omega$ as in Ex.~\ref{exa:1711141355},
$\Fplus(\Fplusplusinf X+X)$ would be understood as $\myalph^+(\myalph^+)^\omega$, which is indeed isomorphic to 
$(\myalph^+)^\omega$.
However, in this paper, mainly for the sake of simplicity of notations, 
we explicitly distinguish them and later write types of a decorated trace semantics 
of a B\"uchi $(T,F)$-system as 
$\dtr_1(c):X_1\kto \Fplus(\Fplusplusinf 0)$ and $\dtr_2(c):X_2\kto \Fplusplusinf 0$. 
%
Because of this choice, while an element in $\Fplusplusinf 0\cong(\myalph^+)^\omega$ is regarded as a decorated word
whose first letter is decorated by $\accstate$ (Ex.~\ref{exa:1711141355}),
%
an element
$a_0\ldots a_n\bigl((a_{00}a_{01}\ldots a_{0n_0})(a_{10}a_{11}\ldots a_{1n_{1}})\ldots\bigr) \in \Fplus (\Fplusplusinf 0)\cong\myalph^+(\myalph^+)^\omega$ is understood as the following decorated sequence:
\begin{multline*}
(a_{0},\nonaccstate)\ldots (a_{n},\nonaccstate)
(a_{00},\accstate)(a_{01},\nonaccstate)\ldots (a_{0n_0},\nonaccstate)
(a_{10},\accstate)(a_{11},\nonaccstate)\ldots (a_{1n_1},\nonaccstate)
\ldots.
\end{multline*}
\end{myremark}



\subsection{Natural Transformations Regarding to $\Fplus$ and $\Fplusinf$}\label{subsec:NTsB}
We introduce two natural transformations for later use. 
As  mentioned in Rem.~\ref{rem:1710141007}, $\Fplus$ resembles the free monad $F^{*}$
while $\Fplusinf$ is similar to the free completely iterative monad. 
The first natural transformation we introduce is analogous to the multiplication of those free monads.
%
%
%
\begin{mydefinition}[$\multFplusinf$]\label{def:multFplusinf}
We define a natural transformation $\multFplusinf:\Fplusinf\Fplusinf\Rightarrow \Fplusinf$ by $\multFplusinf:=(u_X\circ\kappa_1)_{X\in\mathbb{C}}$,
where $u_X$ is the unique homomorphism from $[F[\kappa_1,\kappa_2]\circ \fincoalg^{F}_{\Fplusinf X},F[\kappa_2,\kappa_3]\circ \fincoalg^F_X]$
to $\fincoalg^F_X$ (see Fig.~\ref{fig:1712121922}).
\end{mydefinition}

\begin{myexample}\label{ex:1802261909}
Let $F=\myalph\times(\place)$.
According to the characterizations in Ex.~\ref{exa:1711141355} and Rem.~\ref{rem:1711151735},
$\pij{1}{1\,X}$ has a type 
$(\myalph^+)^+(\myalph^+)^+X+(\myalph^+)^+(\myalph^+)^\omega+(\myalph^+)^\omega\to (\myalph^+)^+X+(\myalph^+)^\omega$,
and is given by the concatenating function that preserves each finite word.
\end{myexample}

The second natural transformation is for ``removing'' decorations. 
%
%
\begin{mydefinition}[$\pij{i}{j}$]\label{def:1710191132B}
We define a natural transformation $\pij{1}{1}:\Fplus\Rightarrow\Fplusinf$ so that 
$\pij{1}{1\,X}:\Fplus X\to\Fplusinf X$ is
the unique homomorphism from $J(\initalg^F_X)^{-1}$ to $J\fincoalg^F_X$.
Similarly, we define natural transformations 
$\pij{2}{1}:\Fplus(\Fplusplusinf+\id)\Rightarrow\Fplusinf$ and $\pij{2}{2}:\Fplusplusinf\Rightarrow\Fplusinf$
so that 
$[\pj{1\,X},\pj{2\,X}]:\Fplus(\Fplusplusinf X+X)+\Fplusplusinf X\to 
\Fplusinf X$ is the unique homomorphism from 
$[\bi{1\,X},\bi{2\,X}]$ to $\fincoalg^F_X$
(see Fig.~\ref{fig:1712121923B}).
\end{mydefinition}

\begin{figure}[t]
\centering
\noindent
\begin{minipage}{0.5\hsize}
\[
\small
\vcenter{  \xymatrix@R=1.4em@C=2.5em{
  {\mbox{\scriptsize $F(\Fplusinf \Fplusinf X+\Fplusinf X+X)$}}
  \ar@{-->}[r]_(.6){F(u_X+\id_X)}
  &
  {F(\Fplusinf X+X)}
  \\
  {\scriptsize\begin{aligned}
  &F(\Fplusinf\Fplusinf X+\Fplusinf X)\\[-.5em]
  &\qquad+F(\Fplusinf X+X)
  \end{aligned}}
  \ar[u]_{[F[\kappa_1,\kappa_2],F[\kappa_2,\kappa_3]]}
  &
  \\
  {\Fplusinf\Fplusinf X+\Fplusinf X}
  \ar[u]_(.4){\fincoalg^{F}_{\Fplusinf X}+\fincoalg^F_X}^{\cong}
  \ar@{-->}[r]^(.65){u_X}     
  &
  {\Fplusinf X}
  \ar[uu]^{\substack{\text{final}\\\cong}}_{\fincoalg^F_X}
}}
\]
  \vspace{-6mm}
  \captionof{figure}{the unique arrow $u_X$}
  \label{fig:1712121922}
\end{minipage}%
\begin{minipage}{0.5\hsize}
\vspace{1mm}
\[
\small
  \xymatrix@R=3.0em@C=2.5em{
 {\scriptsize\begin{aligned}&F \bigl(\Fplus(\Fplusplusinf X+X)\\[-.5em]&+\Fplusplusinf X+X\bigr)\end{aligned}} 
  \ar@{-->}[r]^(.55){\substack{F([\pij{2}{1\,X},\pij{2}{2\,X}]+\id_X)\\[1mm]\qquad\quad\mbox{}}}
 &
 {F
  (\Fplusinf X+X
 )
 } 
 \\
 {\scriptsize\begin{aligned}&\Fplus(\Fplusplusinf X+X)\\[-.5em]&+\Fplusplusinf X\end{aligned}
 } 
 \ar[u]_(.5){
 [\bi{1\,X},\bi{2\,X}]
 } 
 \ar@{-->}[r]^(.6){[\pij{2}{1\,X},\pij{2}{2\,X}]}
 &
 {\Fplusinf X}
 \ar[u]_{\fincoalg^F_X}^{\substack{\text{final}\\\cong}}
 } 
\]
  \vspace{-3mm}
  \captionof{figure}{the unique arrow $[\pij{2}{1\,X},\pij{2}{2\,X}]$}
  \label{fig:1712121923B}
\end{minipage}
\vspace{-2mm}
\end{figure}

\begin{myexample}\label{ex:1802231721}
Let $F=\myalph\times(\place)$.
According to the characterizations in Ex.~\ref{exa:1711141355} and Rem.~\ref{rem:1711151735},
$\pij{1}{1\,X}$ has a type $\myalph^+X\to\myalph^+X+\myalph^\omega$ and is given by the natural inclusion.
In contrast, $\pij{2}{1\,0}$ and $\pij{2}{2\,0}$ have types
$\myalph^+(\myalph^+)^\omega\to\myalph^\omega$ and $(\myalph^+)^\omega\to\myalph^\omega$ respectively,
and they are given by the flattening functions.
See also Prop.~\ref{prop:pijpowB}.
\end{myexample}

\subsection{Liftings $\overline{\Fplus}$ and $\overline{\Fplusinf}$ over $\Kl(T)$}\label{subsec:distlaw}
Let $\overline{F}:\Kl(T)\to\Kl(T)$ be a lifting of of a functor $F$.
We show that under certain conditions, 
it 
induces liftings $\overline{\Fplus}:\Kl(T)\to\Kl(T)$ of $\Fplus$ and $\overline{\Fplusinf}:\Kl(T)\to\Kl(T)$ of $\Fplusinf$.
%
%
Note  
that a lifting $\overline{F}$ 
induces
a lifting $\overline{F(\place+A)}:\Kl(T)\to\Kl(T)$ of $F(\place+A)$
which is defined by $\overline{F(\place+A)}f:=\overline{F}(f+\id_A)=\overline{F}([T\kappa_1,T\kappa_2]\circ(f+\eta_A))$
using the coproduct in $\Kl(T)$.

\begin{mydefinition}\label{def:liftingplus}
\begin{enumerate}
\item\label{item:prop:liftedgreatest1}
Assume  $T$ and $F$ constitute a finite trace situation.
For  $X\in\mathbb{C}$, 
we let 
$\overline{\Fplus} X:=\Fplus X$.
For $f:X\kto Y$, we define 
$\overline{\Fplus}f:\Fplus X\kto\Fplus Y$ as the unique homomorphism 
from $\overline{F}(\id_{\Fplus X}+f)\odot J(\initalg^F_X)^{-1}$ to $J(\initalg^F_Y)^{-1}$.

\item\label{item:prop:liftedgreatest2}
Assume  $T$ and $F$ constitute an infinitary trace situation.
For  $X\in\mathbb{C}$, 
we let
$\overline{\Fplusinf} X:=\Fplusinf X$.
For $f:X\kto Y$, we define 
$\overline{\Fplusinf}f:\Fplusinf X\kto\Fplusinf Y$ as the greatest homomorphism 
from $\overline{F}(\id_{\Fplusinf X}+f)\odot J\fincoalg^F_X$ to $J\fincoalg^F_Y$.
\[
\small
\vcenter{  \xymatrix@R=.7em@C+.6em{
  {F(F^+X+Y)}
  \kar@{-->}[r]_(.5){\overline{F}(\overline{\Fplus}f+\id_{\Fplus X})}
  &
  {F(F^+ Y+Y)} 
  \\
  {F(F^+ X+X)}
  \kar[u]^{\overline{F}(\id+f)}
  &
%
  \\
  {F^+ X}
  \kar[u]_{\substack{\cong}}^{J(\initalg^F_{X})^{-1}}  
  \kar@{-->}[r]^(.5){\overline{\Fplus}f}
  \ar@{}[ruu]|{=}
  &
  {F^+ Y}
       \kar[uu]_{\cong}^{J(\initalg_Y^F)^{-1}}   
}}
\qquad
\vcenter{  \xymatrix@R=.7em@C+.6em{
  {F(\Fplusinf X+Y)}
  \kar@{->}[r]_(.5){\overline{F}(\overline{\Fplusinf}f+\id_{\Fplusinf X})}
  &
  {F(\Fplusinf Y+Y)} 
  \\
  {F(\Fplusinf X+X)}
             \kar[u]^{\overline{F}(\id+f)}
  &
%
  \\
  {\Fplusinf X}
  \kar[u]_{\substack{\cong}}^{J\fincoalg_{X}^F}  
  \kar@{->}[r]^(.5){\overline{\Fplusinf}f}
  \ar@{}[ruu]|{\color{blue}=_\nu}
  &
  {\Fplusinf Y}
       \kar[uu]_{\cong}^{J\fincoalg_Y^F}   
}}
\]
\end{enumerate}
\end{mydefinition}
%
%

In the rest of this section, we check
under which conditions $\overline{\Fplus}$ and $\overline{\Fplusinf}$ are functors and 
form liftings of $\Fplus$ and $\Fplusinf$.
Functoriality of $\overline{\Fplus}$ holds iff for each $f:X\kto Y$ and $g:Y\kto W$, 
$\Fplus g\odot \Fplus f$ is the unique homomorphism from $\overline{F}(\id+g)\odot \overline{F}(\id+f)\odot J(\initalg^F_X)^{-1}$ to
$J(\initalg^F_W)^{-1}$. Similarly, functoriality of $\overline{\Fplusinf}$ holds iff 
$\Fplusinf g\odot \Fplusinf f$ is the greatest homomorphism 
from $\overline{F}(\id+g)\odot \overline{F}(\id+f)\odot J\fincoalg^F_X$ to $J\fincoalg^F_W$.

The former always holds by the finality. 
In contrast, the latter doesn't necessarily hold:
a counterexample is $T=\dist$ and $F=\{o\}\times(\place)^2$ (see Ex.~\ref{ex:cexdist} for details). 
%
%
%
%
Hence we need an extra assumption to make $\overline{\Fplusinf}$ a functor.
We hereby assume 
a stronger condition than is needed 
for the sake of discussions in \S{}\ref{sec:lijB}.

\begin{mydefinition}[$\Phi_{c,\sigma}$]\label{def:phicsigma}
Let $c:X\kto \overline{F}X$ and $\sigma:\overline{F}Y\kto Y$.
We define a function $\Phi_{c,\sigma}:\Kl(T)(X,Y)\to\Kl(T)(X,Y)$ by 
$\Phi_{c,\sigma}(f):=\sigma\odot \overline{F}f\odot c$.
\end{mydefinition}

\vspace{2mm}
\begin{wrapfigure}[4]{r}{4cm}
\vspace{-\intextsep}
\vspace{-5mm}
\[
  \xymatrix@R=1.6em@C=2.3em{
  {FX}
  \kar[r]^{\overline{F}l}
  &
  {FZ}
  \kar[r]^{\overline{F}m}
  &
  {FY}
  \kar[d]^{\sigma}
  \\
  {X}
  \kar[r]^{l}
  \kar[u]^{c}
  \ar@{}[ru]|{\color{blue}=_\nu}
  &
  {Z}
  \kar[r]^{m}
  \kar[u]^{J\fincoalg^F}_{\cong}
  \ar@{}[ru]|{\color{blue}=_\nu}  
  &
  {Y}
  }
\]
\end{wrapfigure}
\noindent\begin{minipage}{0.23\hsize}
\begin{mydefinition}\label{def:gfpprescondNew}
\end{mydefinition}
\end{minipage}
Assume that 
$T$ and $F$ constitute an infinitary trace situation.
Let $\fincoalg^F:Z\to FZ$ be a final $F$-coalgebra.
We say that $T$ and $F$ satisfy the \emph{gfp-preserving condition} 
wrt.\ an $\overline{F}$-algebra $\sigma:FY\kto Y$
if 
for each $X\in\mathbb{C}$ and $c:X\kto FX$, 
if $l:X\kto Z$ is the greatest homomorphism from $c$ to $J\fincoalg^F$
and the function $\Phi_{J\fincoalg^F,\sigma}$  has the greatest fixed point $m:Z\kto Y$,  
then $m\odot l:X\kto Y$ is the greatest fixed point of $\Phi_{c,\sigma}$.
\vspace{2mm}



We next check if $\overline{\Fplus}$ and $\overline{\Fplusinf}$ are liftings of $\Fplus$ and $\Fplusinf$.
It is immediate by definition that $\overline{\Fplus}JX=J\Fplus X$ and $\overline{\Fplusinf}JX=J\Fplusinf X$ for each $X\in\mathbb{C}$. 
Let $f:X\to Y$. 
By definition, $\overline{\Fplus}Jf=J\Fplus f$ holds iff $J\Fplus f$ is a unique homomorphism
from $\overline{F}(\id+Jf)\odot J(\initalg^F_X)^{-1}$ to $J(\fincoalg^F_Y)^{-1}$.
Similarly, $\overline{\Fplus}Jf=J\Fplus f$ holds iff $J\Fplus f$ is the greatest homomorphism
from $\overline{F}(\id+Jf)\odot J\fincoalg^F_X$ to $J\fincoalg^F_Y$.

The former is easily proved by the finality of $J(\initalg^F_Y)^{-1}$,
while the latter requires an assumption again.

\begin{mydefinition}\label{def:detgreatestcond}
Assume  $T$ and $F$ constitute an infinitary trace situation.
Let $\fincoalg^F:Z\to FZ$ be a final $F$-coalgebra.
We say that $T$ and $F$ satisfy the \emph{deterministic-greatest condition} if 
for 
$c:X\to FX$ in $\mathbb{C}$, 
if $u:X\to Z$ is the unique homomorphism from $c$ to $\fincoalg^F$ then
$Ju$ is the greatest homomorphism from 
$Jc$ to $J\fincoalg^F$.
\end{mydefinition}

Concluding the discussions so far, we obtain the following proposition.

\begin{myproposition}\label{prop:plusinffunctor}
\begin{enumerate}
\item\label{item:prop:plusinffunctor1}
If $T$ and $F(\place+A)$ constitute a finite trace situation for each $A\in\mathbb{C}$,
the operation $\overline{\Fplus}$ 
is a functor and is a lifting  of $\Fplus$. 

\item\label{item:prop:plusinffunctor2}
If $T$ and $F(\place+A)$ constitute an infinitary trace situation and 
 satisfy the gfp-preserving condition wrt.\ an arbitrary algebra and 
the deterministic-greatest condition for each $A\in\mathbb{C}$, 
then
$\overline{\Fplusinf}$ 
is a functor and is a lifting of $\Fplusinf$.
\qed
\end{enumerate}
\end{myproposition}
Hence under appropriate conditions, a lifting $\overline{F}:\Kl(T)\to\Kl(T)$ of
$F$ gives rise to liftings of $\Fplus$ and $\Fplusinf$. 
By repeating this, we can define $\overline{\Fij{i}{j}}$ for each $i$ and $j$. 

See \S{}\ref{sec:distlaws} for the distributive laws corresponding to the liftings defined above.


\begin{myexample}\label{ex:1802250612}
Let $F=\myalph\times(\place)$ and $T=\pow$. 
As we have seen in Ex.~\ref{exa:1711141355}, $\Fplusplusinf X\cong (\myalph^+)^+ X+(\myalph^+)^\omega$.
Let $\overline{F}$ be a lifting that is given as in Ex.~\ref{exa:1802250620}.
We can construct a lifting $\overline{\Fplusplusinf }$ using Prop.~\ref{prop:plusinffunctor}, and
for $f:X\kto Y$ in $\Kl(\pow)$, 
$\overline{\Fplusplusinf }f:(\myalph^+)^+ X+(\myalph^+)^\omega\kto (\myalph^+)^+ Y+(\myalph^+)^\omega$ 
is given by $\overline{\Fplusplusinf }f(w)=\{w'y\mid y\in f(x)\}$ if $w=w'x$ where $w'\in (\myalph^+)^+$ and $x\in X$, and 
$\{w\}$ if $w\in (\myalph^+)^\omega$.
\end{myexample}

\section{Decorated Trace Semantics of B\"uchi $(T,F)$-systems}\label{sec:lijB}

\subsection{Definition}\label{subsec:defdoc}
\begin{myassumption}\label{asm:assumptionB}
Throughout this section, 
let $T$ be a monad and $F$ be an endofunctor on $\mathbb{C}$, and
assume that 
each homset of $\Kl(T)$ carries a partial order $\sqsubseteq$. 
We further assume the following conditions for each $A\in\mathbb{C}$. 
\begin{enumerate}
\item\label{item:asm:assumption01B} 
$\Fplus,\Fplusplusinf:\mathbb{C}\to\mathbb{C}$ are well-defined and 
 liftings $\overline{F},\overline{\Fplus},\overline{\Fplusplusinf}:\Kl(T)\to\Kl(T)$ are given.


\item\label{item:asm:assumption1B}
$T$ and $F(\place+A)$ satisfy the conditions in Thm.~\ref{thm:initfinal} wrt.\ $\overline{F}(\place+A)$ and $\sqsubseteq$,
and hence constitute a finite trace situation.


\item\label{item:asm:assumption2B}
$T$ and $\Fplus(\place+A)$ constitute an infinitary trace situation wrt.\ 
  $\overline{\Fplus}(\place+A)$ and $\sqsubseteq$.

\item\label{item:asm:gfp-preservingB}
$T$ and $\Fplus(\place+A)$ satisfy the gfp-preserving condition wrt.\ an arbitrary $\sigma$.

\item\label{item:asm:det-greatestB}
$T$ and $\Fplus(\place+A)$ satisfy the deterministic-greatest condition.

\item\label{item:asm:assumption5B}
The liftings $\overline{\Fplus}$ and $\overline{\Fplusplusinf}$ are 
obtained from $\overline{F}$ and $\overline{\Fplus}$
using the procedure in Def.~\ref{def:liftingplus} respectively.

\item\label{item:asm:assumption6B}
$\overline{\Fplus}(\place+A)$ and $\overline{\Fplusplusinf}(\place+A)$ are locally monotone.

\item\label{item:asm:assumption101B}
$T$ and $F$ constitute a B\"uchi trace situation wrt.\ the same $\sqsubseteq$ and $\overline{F}$. 

\setcounter{asmenumi}{\value{enumi}}
\end{enumerate}
\end{myassumption}

Using the categorical data type defined in \S{}\ref{sec:altFPF}, 
we now introduce a \emph{decorated B\"uchi trace semantics} 
$\dtr_1(c):X_1\kto \Fplus(\Fplusplusinf 0)$ and $\dtr_{2}(c):X_{2}\kto \Fplusplusinf 0$.

\begin{mydefinition}[$\dtr_i(c)$]\label{def:PDTSB}
For a B\"uchi $(T,F)$-system $(c,(X_1,X_{2}))$,
the \emph{decorated B\"uchi trace semantics} is a solution
$(\dtr_1(c):X_1\kto \Fplus(\Fplusplusinf 0),\dtr_{2}(c):X_{2}\kto \Fplusplusinf 0)$
of the following HES
(see also Eq.~(\ref{eq:fig:GFP})). 
\begin{equation*} 
\left\{
\begin{matrix*}[l]
v_1\; &=_{\nu} \; J({\bi{1\;0}})^{-1}\odot \overline{F}(v_1+v_{2})\odot c_1 & \in (\Kl(T)(X_1,\Fplus(\Fplusplusinf 0)),\sqsubseteq)\\
v_2\; &=_{\nu} \; J({\bi{2\;0}})^{-1}\odot \overline{F}(v_1+v_{2})\odot c_2 & \in (\Kl(T)(X_2,\Fplusplusinf 0),\sqsubseteq) 
\end{matrix*}
\right.
\end{equation*}
\end{mydefinition}

%
Existence of a solution will be proved 
in the next section.

\subsection{Trace Semantics vs.\ Decorated Trace Semantics}\label{subsec:reltoOTB}
This section is devoted to sketching the proof of the following theorem, which
relates decorated trace semantics 
$\dtr_i(c)$ and 
B\"uchi trace semantics $\trB_i(c)$ in~\cite{urabeSH16coalgebraictrace}
via the natural transformation in Def.~\ref{def:1710191132B}.

\begin{mytheorem}\label{thm:mainthmB}
For each $i\in\{1,2\}$, $\trB_i(c)=\pij{2}{i\,0}\circ\dtr_i(c)$.
\qed
\end{mytheorem}

To prove this, we introduce Kleisli arrows $c_2^\ddagger$,
$\myl^{(1)}_1$, $\myl^{(2)}_1$ and $\myl^{(2)}_2$.
They are categorical counterparts to $f_2^\ddagger$, 
$\lij{1}{1}$, $\lij{2}{1}$ and $\lij{2}{2}$  (see Def.~\ref{def:solHES}) for the HES defining $\trB_i(c)$ (see Def.~\ref{def:trsemTFsysB}), and
 bridge the gap between $\dtr_i(c)$ and $\trB_i(c)$. 


\begin{mydefinition}[$c_2^\ddagger$, $\myl^{(1)}_1$, $\myl^{(2)}_1$, $\myl^{(2)}_2$]\label{def:1710081602NewB}
We define Kleisli arrows $\mylij{1}{1}:X_1\kto \Fplus X_2$,
$c_2^\ddagger:X_2\kto \Fplus X_2$,
$\mylij{2}{2}:X_2\kto \Fplusplusinf 0$ and 
$\mylij{2}{1}:X_1\kto \Fplusplusinf 0$ as follows:
\begin{itemize}
\item 
We define $\mylij{1}{1}:X_1\kto \Fplus X_2$ as the unique homomorphism from an $F(\place+X_2)$-coalgebra
$c_1$ to $J(\initalg^{F}_{X_2})^{-1}$ 
(see the left diagram in Eq.~(\ref{eq:1802251454}) below).

\item We define $c_2^\ddagger:X_2\kto \Fplus X_2$ by:
\[
c_2^\ddagger:=\left(
X_2\lower1ex\hbox{$\xkrightarrow{c_2}$} F(X_1+X_2)\lower1ex\hbox{$\xkrightarrow{ \overline{F}(\mylij{1}{1}+\id) }$} F(\Fplus X_2+X_2)
\lower1ex\hbox{$\xkrightarrow{J\initalg^F_{X_2}}$}  \Fplus X_2
\right)\,.
\]

\item 
We define $\mylij{2}{2}:X_2\kto \Fplusplusinf 0$ as the greatest homomorphism from
$c_2^\ddagger$ to $J\fincoalg^{\Fplus}_{0}$
(see the right diagram below). 
\[
\small
\]
\begin{equation}\label{eq:1802251454}
\small
\!\!\!\!\!\!\!\!\!\!
\vcenter{  \xymatrix@R=1.4em@C+0em{
  {F(X_1+X_2)}
  \kar@{-->}[r]^(.3){\mathrlap{\overline{F}(\mylij{1}{1}+\id)}}
  &
  {F\left(\Fplus X_2+X_2\right)}
  \\
  {X_1}
  \kar[u]_{c_1}
  \kar@{-->}[r]^{\mylij{1}{1}}
  \ar@{}[ru]|{=}
    &
  {\Fplus X_2}
  \kar[u]_(.5){J(\initalg^F_{X_2})^{-1}}^(.5){\cong}  
}}
\qquad
\vcenter{  \xymatrix@R=1.4em@C+1em{
  {\Fplus(X_2)}
  \kar@{->}[r]^(.5){\overline{\Fplus}(\mylij{2}{2})}
  &
  {\Fplus\left(\Fplusplusinf 0)
  \right)}
  \\
  {X_2}
  \kar[u]_{c_2^{\ddagger}}
  \kar[r]^{\mylij{2}{2}}
  \ar@{}[ru]|{\color{blue}=_\nu}
    &
  {\Fplusplusinf 0}
  \kar[u]_(.5){J\fincoalg^{\Fplus}_{0}}^(.5){\cong}
}}
\end{equation}

\item 
We define $\mylij{2}{1}:X_1\kto \Fplus(\Fplusplusinf 0)$ as follows:
\[
\mylij{2}{1}:=\left(
X_1\lower1ex\hbox{$\xkrightarrow{\mylij{1}{1}}$} \Fplus X_2\lower1ex\hbox{$\xkrightarrow{\overline{\Fplus}\mylij{2}{2}} $}\Fplus(\Fplusplusinf 0)
\right)\,.
\]
\end{itemize}
\end{mydefinition}

We explain an intuition why Kleisli arrows defined above bridge the gap between $\trB_i(c)$ and $\dtr_i(c)$.
One of the main differences between them 
is that  
$\trB_1(c)$ is calculated from $\lij{1}{1}(u_{2})$ which is the least fixed point of a certain function,
while $\dtr_1(c)$ is defined as the greatest homomorphism. 
%
%
The arrow $\mylij{1}{1}$ fills the gap because 
it is defined as the unique fixed point, 
which is obviously both the least and the greatest fixed point.

We shall prove Thm.~\ref{thm:mainthmB} following the intuition above.
%
%
The lemma below, which is easily proved by the finality of $a$, shows that 
not only $\mylij{1}{1}$ but also $\mylij{2}{1}$ is characterized as the unique homomorphism.
%


\vspace{1mm}
\noindent\begin{minipage}{0.45\hsize}
\begin{mylemma}\label{lem:1710161317B}
\fussy
The Kleisli arrow\linebreak $\mylij{2}{1}:X_1\kto \Fplus(\Fplusplusinf 0)$ is the unique homomorphism
from $\overline{F}(\id+\mylij{2}{2})\odot c_1$ to
$J(\initalg^{F}_{\Fplusplusinf 0})^{-1}$.
\qed
\end{mylemma}
\end{minipage}
\begin{minipage}{0.55\hsize}
\vspace{-2mm}
\[
\quad
\scriptsize
\vcenter{  \xymatrix@R=.7em@C+.1em{
  {F(X_1+\Fplusplusinf 0)}
  \kar@{-->}[r]_(.5){\overline{F}(\mylij{2}{1}+\id)}
  &
  {F(\Fplus(\Fplusplusinf 0)+\Fplusplusinf 0)} 
  \\
  {F(X_1+X_2)}
  \kar[u]^{\overline{F}(\id+\mylij{2}{2})}
  &
%
  \\
  {X_1}
  \kar[u]^{c_1}  
  \kar@{-->}[r]^(.5){\mylij{2}{1}}
  \ar@{}[ruu]|{=}
  &
  {\Fplus(\Fplusplusinf 0)}
       \kar[uu]_{\cong}^{J(\initalg_{\Fplusplusinf 0}^{F})^{-1}}   
}}
\]
\end{minipage}
\vspace{1mm}

Together with the definition of $\mylij{2}{2}$, we have the following proposition.

\begin{myproposition}\label{prop:1710160958B}
For each $i\in\{1,2\}$, $\mylij{2}{i}=\dtr_i(c)$.
\qed
\end{myproposition}
This proposition implies the
existence of a solution of the HES in Def.~\ref{def:PDTSB}.


It remains to show the relationship between the $\mylij{i}{j}$ and $\trp_i(c)$. 
%
By using that $\mylij{1}{1}$ is the unique fixed point (and hence the least fixed point), 
we can prove the following equality for an arbitrary $u_2:X_2\kto \Fplusinf 0$.
\[
\lij{1}{1}(u_2)
=\Bigl(X_1 \lower1ex\hbox{$\xkrightarrow{\mylij{1}{1}}$} \Fplus X_2 \lower1ex\hbox{$\xkrightarrow{\overline{\Fplus}u_2}$} \Fplus\Fplusinf 0
\lower1ex\hbox{$\xkrightarrow{J\pij{1}{1\,\Fplusinf 0}}$}\Fplusinf\Fplusinf 0
\lower1ex\hbox{$\xkrightarrow{\multFplusinf_0}$}\Fplusinf 0
\Bigr) 
\]
The following equalities are similarly proved using the equality above.
\begin{align*}
\lij{2}{1}(*)&=\Bigl({\scriptsize X_1 \lower1ex\hbox{$\xkrightarrow{\mylij{2}{1}}$} \Fplus\Fplusinf 0 \lower1ex\hbox{$\xkrightarrow{\overline{\Fplus\Fplusinf}\initarr_{\Fplusplusinf 0}}$} \Fplus\Fplusinf(\Fplusinf 0) 
\lower1ex\hbox{$\xkrightarrow{J\pij{2}{1\,\Fplusinf 0}}$} \Fplusinf\Fplusinf 0
\lower1ex\hbox{$\xkrightarrow{\multFplusinf_0}$} \Fplusinf 0
}\Bigr)
\quad 
\\ 
\lij{2}{2}(*)&=\Bigl({\small X_2 \lower1ex\hbox{$\xkrightarrow{\mylij{2}{2}}$} \Fplusinf 0 \lower1ex\hbox{$\xkrightarrow{\overline{\Fplusinf}\initarr_{\Fplusplusinf 0}}$} \Fplusinf(\Fplusinf 0) 
\lower1ex\hbox{$\xkrightarrow{J\pij{2}{2\,\Fplusinf 0}}$}\Fplusinf\Fplusinf 0
\lower1ex\hbox{$\xkrightarrow{\multFplusinf_0}$}\Fplusinf 0
}\Bigr) 
\end{align*}
By the definition of $\trB_i(c)$, these equalities imply the following proposition.

\begin{myproposition}\label{prop:1710191342B}
For each $i\in\{1,2\}$, $\trB_i(c)=\pij{2}{i\,0}\circ\mylij{2}{i}$.
\qed
\end{myproposition}


Prop.~\ref{prop:1710160958B} and Prop.~\ref{prop:1710191342B} 
immediately imply Thm.~\ref{thm:mainthmB}.

\section{Decorated Trace Semantics for Nondeterministic B\"uchi Tree Automata}\label{sec:NPTAB}
We apply the framework developed in \S{}\ref{sec:altFPF}--\ref{sec:lijB} to \emph{nondeterministic B\"uchi tree automata} 
(NBTA), systems that nondeterministically accept \emph{trees} wrt.\ the B\"uchi condition (see e.g.~\cite{thomas97LAL}). 
We show what datatypes $\Fplus(\Fplusplusinf0)$ and $\Fplusplusinf0$, and $\dtr_i(c)$ characterize for an NBTA.
We first review some basic notions.

\subsection{Preliminaries on B\"uchi Tree Automaton}\label{subsec:PTAB}

\begin{mydefinition}[ranked alphabet]\label{def:rankedalph}
A \emph{ranked alphabet} is a 
set $\Sigma$ equipped with an \emph{arity function} $|\place|:\Sigma\to\mathbb{N}$.
We write $\Sigma_n$ for $\{a\in\Sigma\mid|a|=n\}$.
For a set $X$, 
we regard $\Sigma+X$ as a ranked alphabet by letting $|x|=0$.
We also regard $\Sigma\times X$ as a ranked alphabet by letting $|(a,x)|=|a|$.
\end{mydefinition}

\begin{mydefinition}[$\Sigma$-labeled tree, \cite{courcelle83infinitetrees}]\label{def:tree}
A \emph{tree domain} is a set $D\subseteq\mathbb{N}^*$ s.t.: 
i) $\empseq\in D$, 
ii) for $w,w'\in\mathbb{N}^*$, $ww'\in D$ implies $w\in D$ (i.e.\ it is prefix-closed), and 
iii) for $w\in D$ and $i,j\in\mathbb{N}$, $wi\in D$ and $j\leq i$ imply $wj\in D$ (i.e.\ it is downward-closed).
A \emph{$\Sigma$-labeled (infinitary) tree} is a pair $t=(D,l)$ of a tree domain $D$ 
and a \emph{labeling function} $l:D\to \bigcup_{n\in\mathbb{N}}\Sigma_n$ s.t.\ for $w\in D$,  $|l(w)|=n$ implies
$\{i\in\mathbb{N}\mid wi\in D\}=[0,n-1]$.
A $\Sigma$-labeled tree $t=(D,l)$ is \emph{finite} if $D$ is a finite set. 
We write $\Treeinf(\Sigma)$ (resp.\ $\Treefin(\Sigma)$) for the set of $\Sigma$-labeled infinitary (resp.\ finite) trees.
For $w\in D$, the \emph{$w$-th subtree} $t_w$ of $t$  is defined by $t_w=(D_w,l_w)$ where $D_w:=\{w'\in\mathbb{N}^*\mid ww'\in D\}$ and 
$l_w(w'):=l(ww')$.
A \emph{branch} of $t$ is a possibly infinite sequence $i_1i_2\ldots\in\mathbb{N}^\infty$ s.t.\
$i_1i_2\ldots i_k\in D$ for each $k\in\mathbb{N}$, and 
if it is a finite sequence $i_1i_2\ldots i_k$ then $|l(i_0i_1\ldots i_k)|=0$.
We sometimes identify a branch  $i_0i_1\dots\in \mathbb{N}^\infty$ 
with a sequence $l(\empseq)l(i_1)l(i_1i_2)\dots\in \Sigma^\infty$.
\end{mydefinition}

%

\begin{myremark}\label{rem:1710071556}
For the sake of notational simplicity, we identify a $\Sigma$-labeled tree with a $\Sigma$-term in a natural manner.
For example, 
a $\{a,b\}$-term $(a,(b,b))$ denotes an $\{a,b\}$-labeled finite tree $t=(\{\empseq,0,1\},[\empseq\mapsto a,0\mapsto b,1\mapsto b])$.
Moreover, 
for $\{a,b,c\}$-labeled trees $t_0=(D_0,l_0)$ and $t_1=(D_1,l_1)$,
we write $(c,t_0,t_1)$ for a tree $t=(\{\empseq\cup\{0w\mid w\in D_0\}\cup\{1w\mid w\in D_1\},
[\empseq\mapsto c, 0w\mapsto l_0(w), 1w\mapsto l_1(w)])$.
\end{myremark}

\begin{mydefinition}[NBTA]\label{def:NPTAB}
A \emph{nondeterministic B\"uchi tree automaton} (NBTA) is a tuple
$\mathcal{A}=(X,\Sigma,\delta,\Acc)$
of a \emph{state space} $X$, a ranked alphabet $\Sigma$, a \emph{transition function}
$\delta:X\to\pow(\coprod_{n\in\mathbb{N}}\Sigma_n\times X^{n})$ 
and a set $\Acc\subseteq X$ of \emph{accepting states}. 
\end{mydefinition}

\begin{mydefinition}[$\langB_{\mathcal{A}}$]\label{def:langNPTAB}
\sloppy
Let $\mathcal{A}=(X,\Sigma,\delta,\Acc)$ be an NBTA. 
A \emph{run tree} 
over $\mathcal{A}$ is a $(\Sigma\times X)$-labeled tree $\rho$ such that
for each subtree $((a,x),((a_0,x_0),t_{00},\ldots,t_{0n_0}),\ldots,((a_n,x_n),t_{n0},\ldots,t_{nn_n}))$,
$(a,x_0,\ldots,x_n)\in\delta(x)$ holds.
A run tree is \emph{accepting} if for each branch $(a_0,x_0)(a_1,x_1)\ldots\in(\Sigma\times X)^\omega$,
$x_i\in\Acc$ for infinitely many $i$.
%
We write $\Run_{\mathcal{A}}(x)$ (resp.\ $\AccRun_{\mathcal{A}}(x)$) for the set of run trees (resp.\ accepting run trees) 
whose root node is labeled by $x\in X$.
For $A\subseteq X$, $\Run_{\mathcal{A}}(A)$ 
denotes $\cup_{x\in A}\Run(x)$. 
We define $\AccRun_{\mathcal{A}}(A)$ similarly.
If no confusion is likely, we omit the subscript $\mathcal{A}$.
We define $\DelSt:\Run(X)\to\Treeinf(\Sigma)$ by $\DelSt(D,l):=(D,l')$ where $l'(w):=\pi_1(l(w))$.
The \emph{language} $\langB_{\mathcal{A}}:X\to\pow\Treeinf(\Sigma)$ of $\mathcal{A}$ is defined by
$\langB_{\mathcal{A}}(x)=\DelSt(\AccRun_{\mathcal{A}}(x))$. 
\end{mydefinition}

\subsection{Decorated Trace Semantics of NPTA}\label{subsec:coalgNPTAB}
A ranked alphabet $\Sigma$ induces a  functor 
$\FSigma=\coprod_{n\in\mathbb{N}}\Sigma_n\times (\place)^n:\Sets\to\Sets$.
In~\cite{urabeSH16coalgebraictrace}, an NBTA $\mathcal{A}$ 
was modeled as a B\"uchi $(\pow,\FSigma)$-system, and it was shown that
$\langB_{\mathcal{A}}$ is characterized by a coalgebraic B\"uchi trace semantics $\trB_i(c)$. 

\begin{myproposition}[\cite{urabeSH16coalgebraictrace}]\label{prop:NPTAcoindB}
For $X,Y\in\Sets$, we define an order $\sqsubseteq$ on $\Kl(\pow)(X,Y)$ by $f\sqsubseteq g\defarrow \forall x\in X.\, f(x)\subseteq g(x)$. We define $\overline{\FSigma}:\Kl(\pow)\to\Kl(\pow)$ by $\overline{\FSigma}X:=X$ for $X\in\Kl(\pow)$ and
$\overline{\FSigma} f(a,x_1,\ldots,x_n):=\{(a,y_1,\ldots,y_n)\mid y_i\in f(x_i)\}$ for $f:X\kto Y$. 
It is easy to see that $\overline{\FSigma}$ is a lifting of $\FSigma$.
Then we have: 
\begin{enumerate}
\item\label{item:prop:NPTAcoind1} $\pow$ and $\FSigma$ constitute a B\"uchi trace situation (Def.~\ref{def:trsemTFsysB}) wrt.\ $\sqsubseteq$ and $\overline{\FSigma}$.

\item\label{item:prop:NPTAcoind2} The carrier set of the final $\FSigma$-coalgebra is isomorphic to $\Treeinf(\Sigma)$.

\item\label{item:prop:NPTAcoind3} For an NBTA $\mathcal{A}=(X,\Sigma,\delta,\Acc)$, 
we define a B\"uchi $(\pow,\FSigma)$-system $(c:X\kto\overline{\FSigma} X,(X_1,X_2))$ by 
$c:=\delta$, $X_1:=X\setminus\Acc$ and $X_2:=\Acc$.
Then we have: $[\trB_1(c),\trB_2(c)]=\langB_{\mathcal{A}}:X\to\pow\Treeinf(\Sigma)$.
\qed
\end{enumerate}
\end{myproposition}

In the rest of this section, for an NBTA $\mathcal{A}=(X,\Sigma,\delta,\Acc)$
modeled as a $(\pow,\FSigma)$-system $(c:X\to\pow\FSigma X,(X_1,X_2))$,
we describe $\dtr_i(c)$ 
and show the relationship with $\trB_i(c)$ using Thm.~\ref{thm:mainthmB}.

We first describe datatypes $\FSigmaplus(\FSigmaplusplusinf 0)$ and $\FSigmaplusplusinf 0$
referring to the construction of a final coalgebra in Thm.~\ref{thm:constinitfin}.
We can easily see that $\FSigmaplus A\cong\Treeplus(\Sigma,A) :=\Treefin(\Sigma+A)\setminus\{(x)\mid x\in A\}$.
Hence for each $i\in\omega$, by a similar characterization to Ex.~\ref{exa:1711141355}, we have:
\begin{align*}
&(\FSigmaplus(\place+0))^i 1\cong
\underbrace{\Treeplus(\Sigma,\Treeplus(\Sigma,\dots\Treeplus(\Sigma,}_{i}\{*\})\ldots))\cong\\
&\left\{{\small\begin{aligned}
&\xi\in \Treefin(\Sigma\times\{\nonaccstate,\accstate\}\\
&\qquad\qquad\qquad+\{*\})
\end{aligned}}\;\middle|\; {\small\begin{aligned}
&\text{the root node is labeled by $\accstate$, and for each branch}\\[-1mm]
&\text{whose last component is $*$, $\accstate$ appears exactly $i$-times}
\end{aligned}}\right\} \,.
\end{align*}
Therefore $\FSigmaplusplusinf 0$, a limit of the above sequence by Thm.~\ref{thm:constinitfin}, 
and $\FSigmaplus(\FSigmaplusplusinf 0)$ 
are characterized as follows:
%
%
\begin{myproposition}\label{prop:FijpowB}
We define 
$\AccTreeBj{i}(\Sigma)\subseteq\Treeinf(\Sigma\times\{\nonaccstate,\accstate\})$ by:
\[
\AccTreeBj{i}(\Sigma):=
\left\{{\small\begin{aligned}
&\xi\in \Treeinf(\Sigma\times\{\nonaccstate,\accstate\}\\
&\qquad\qquad\quad+A)
\end{aligned}}\;\middle|\; {\small\begin{aligned}
&\text{the root node is labeled by $\bullet$, and for each}\\[-1mm]
&\text{infinite branch $\accstate$ appears infinitely often}
\end{aligned}}\right\}\,.
\]
where $i\in\{1,2\}$ and $\bullet$ is $\nonaccstate$ if $i=1$ and $\accstate$ if $i=2$.
Then $\AccTreeBj{1}(\Sigma)\cong\FSigmaplus(\FSigmaplusplusinf 0)$ and $\AccTreeBj{2}(\Sigma,A)\cong\FSigmaplusplusinf 0$.
\qed
\end{myproposition}

%

We now show what $\dtr_i(c)$ characterizes for an NBTA wrt.\ the characterization in Prop.~\ref{prop:FijpowB}.
Firstly, the assumptions in the previous section are satisfied.
\begin{myproposition}\label{prop:asmsatisfiedB}
Asm.~\ref{asm:assumptionB} is satisfied by 
$(T,F)=(\pow,\FSigma)$. 
\qed
\end{myproposition}

\fussy
By Prop.~\ref{prop:FijpowB}, for $i\in\{1,2\}$,
$\bi{i\,0}$ (see Def.~\ref{def:1710131524B}) has a type 
\[
\bi{i\,0}:\AccTreeBj{i}(\Sigma)\to \textstyle{\coprod_{n\in\omega}}\Sigma_n\times(\AccTreeBj{1}(\Sigma)+\AccTreeBj{2}(\Sigma))\,,
\]
and is given by $\bi{i\,A}(\xi)=(a,\xi_0,\ldots,\xi_{n-1})$ if the root  of $\xi$ is labeled by $(a,\bullet)\in\Sigma_n\times \{\nonaccstate,\accstate\}$.
Using this, we can show the following characterization of $\dtr_i(c)$.

\begin{myproposition}\label{prop:dtrpowB}
Let $\mathcal{A}=(X,\Sigma,\delta,\Acc)$ be an NBTA. 
We define $\Omega:\Run(X)\to\Treeinf(\Sigma\times\{\nonaccstate,\accstate\})$ 
by $\Omega(D,l):=(D,l')$ where for $w\in D$ s.t.\ $l(w)=(a,x)$, $l'(w):=(a,\nonaccstate)$ if $x\notin Acc$ and $(a,\accstate)$ if $x\in Acc$.
We define a B\"uchi $(\pow,\FSigma)$-system $(c:X\kto\overline{\FSigma} X,(X_1,X_{2}))$ as in 
Prop.~\ref{prop:NPTAcoindB}.\ref{item:prop:NPTAcoind3}. Then for $i\in[1,2n]$ and $x\in X_i$, 
\begin{align*}
\dtr_i(c)(x)=\bigl\{\Omega(\rho)\in\AccTreeBj{i}(\Sigma)\mid \rho\in\AccRun_{\mathcal{A}}(x)\bigr\}\,.
\tag*{\qed}
\end{align*}
\end{myproposition}

We conclude this section by instantiating $\pij{2}{i\,A}$ (Def.~\ref{def:1710191132B}) for NBTAs.

\begin{myproposition}\label{prop:pijpowB}
We overload $\DelSt$ and define $\DelSt:\AccTreeBj{1}(\Sigma)+\AccTreeBj{2}(\Sigma)\to\Treeinf(\Sigma)$ by 
$\DelSt(D,l):=(D,l')$ where $l'(w):=\pi_1(l(w))$.
Then with respect to the isomorphism in Prop.~\ref{prop:FijpowB}, 
$\DelSt(\xi)=\pij{2}{i\,A}(\xi)$ for each $i\in\{1,2\}$ and $\xi\in \AccTreeBj{i}(\Sigma)$.
\qed
\end{myproposition}

Hence Thm.~\ref{thm:mainthmB} results in the following (obvious) equation  for NBTAs:
\[
\bigl\{\DelSt(\Omega(\rho))\mid \rho\in\AccRun_{\mathcal{A}}(x)\bigr\}=\langB_{\mathcal{A}}(x)\,.
\]

%
%


\section{Systems with Other Branching Types}\label{sec:forprob}
In this section we briefly discuss other monads than $T=\pow$.
As we have discussed in~\S{}\ref{subsec:distlaw},
the framework does not apply to $T=\dist$.

Let $T=\lift$ and $F=\FSigma$. A B\"uchi $(\lift,\FSigma)$-system $(c:X\kto\overline{\FSigma} X,(X_1,\ldots,X_{2n}))$ is understood as a 
$\Sigma$-labeled deterministic B\"uchi tree automaton with an exception. 
In a similar manner to $T=\pow$ we can prove that they satisfy Asm.~\ref{asm:assumptionB}. 
The resulting decorated trace semantics has a type  $\dtr_i(c):X_i\to\{\bot\}+\AccTreeBj{i}(\Sigma)$.
Note that once $x\in X$ is fixed, either of the following occurs: a decorated tree is determined according to $c$; or $\bot$ is reached at some point.
The function $\dtr_i(c)$ assigns $\bot$ to $x\in X_i$ iff $\bot$ is encountered from $x$ or the resulting decorated tree
does not satisfy the B\"uchi condition: 
otherwise, the generated tree is assigned to $x$.
%
See \S{}\ref{sec:dtslift} for detailed discussions, which includes the case of parity automata. 

We next let $T=\giry$. A B\"uchi $(\giry,\FSigma)$-system is understood as a \emph{probabilistic B\"uchi tree automaton}.
In fact, it is open if $T=\giry$ and $F=\FSigma$ satisfy Asm.~\ref{asm:assumptionB}. 
The challenging part is the gfp-preserving condition (Asm.~\ref{asm:assumptionB}.\ref{item:asm:gfp-preservingB}).
However, by carefully checking the proofs of the lemmas and the propositions where the gfp-preserving condition is used
(i.e. Prop.~\ref{prop:plusinffunctor}, Lem.~\ref{lem:1710161317B} and Prop.~\ref{prop:1710191342B}), we can show that 
Asm.~\ref{asm:assumptionB}.\ref{item:asm:gfp-preservingB} can be relaxed to the following weaker but more complicated 
conditions:
\begin{enumerate}
\item[\ref{item:asm:gfp-preservingB}'-1.] $T$ and $\Fplus(\place+A)$ satisfy the gfp-preserving condition wrt.\ an algebra 
$\overline{\Fplus}(\Fplusplusinf B+A)\xkrightarrow{\overline{\Fj{i}}(\id+f)}\overline{\Fplus}(\Fplusplusinf B+B)\xkrightarrow{J(\fincoalg^{\Fplus}_B)^{-1}}\Fplusplusinf B$ for each $f:A\kto B$;

\item[\ref{item:asm:gfp-preservingB}'-2.] $T$ and $\Fplus(\place+A)$ satisfy the gfp-preserving condition wrt.\ an algebra
$\Fplus (F^{\plusinf\plusinf} A+A)\xkrightarrow{J\plustoinf}\Fplusinf (F^{\plusinf\plusinf} A+A)\xkrightarrow{J(\fincoalg^{\Fplusinf}_A)^{-1}}F^{\plusinf\plusinf}A$ where $\plustoinf$ is the unique homomorphism from $(\initalg^{F}_{F^{\plusinf\plusinf} A+A})^{-1}$ to $\fincoalg^{F}_{F^{\plusinf\plusinf} A+A}$; and

\item[\ref{item:asm:gfp-preservingB}'-3.] $T$ and $F(\place+A)$ satisfy the gfp-preserving condition wrt.\ an algebra
$F(\Fplusinf A+\Fplusinf A+A)\xkrightarrow{JF([\id,\id]+\id)}F(\Fplusinf A+A)\xkrightarrow{J(\fincoalg^{F}_A)^{-1}}\Fplusinf A$. 
\end{enumerate}
In fact, only the first condition is sufficient to prove Prop.~\ref{prop:plusinffunctor} and Lem.~\ref{lem:1710161317B}.

We can show that $T=\giry$ and $F=\FSigma$ on $\Meas$ satisfy the above weakened  gfp-preserving condition, and
hence we can consider a decorated trace semantics $\dtr_i(c)$ for a B\"uchi $(\giry,\FSigma)$-system $(c:X\kto\overline{\FSigma} X,(X_1,X_2))$ and use Thm.~\ref{thm:mainthmB}.

Assume $X$ is a countable set equipped with a discrete $\sigma$-algebra for simplicity.
Then the resulting decorated trace semantics $\dtr_i(c)$ has a type $X_i\to\giry(\AccTreeBj{i}(\Sigma),\sigalg_{\AccTreeBj{i}(\Sigma)})$ where 
$\sigalg_{\AccTreeBj{i}(\Sigma)}$ is the standard $\sigma$-algebra generated by cylinders. The probability measure assigned to $x\in X_i$ by $\dtr_i(c)$ is 
defined in a similar manner to the probability measure over the set of run trees generated by a probabilistic B\"uchi tree automaton 
(see e.g.\ \cite{urabeH15extended}).

The situation is similar for parity $(\giry,\FSigma)$-systems.
See \S{}\ref{sec:dtsgiry} for the details.


\section{Conclusions and Future Work}\label{sec:conclu}
We have introduced 
a categorical data type for capturing behavior of systems with B\"uchi acceptance conditions. 
The data type was defined as an alternating fixed point of a functor, 
which is understood as the set of traces decorated with priorities. 
We then defined a notion of coalgebraic decorated trace semantics, and compared it with the coalgebraic trace semantics in~\cite{urabeSH16coalgebraictrace}.
%
We have applied our framework for nondeterministic B\"uchi tree automata,
and showed that decorated trace semantics is concretized to a function that assigns 
a set of trees decorated with priorities so that the B\"uchi condition is satisfied in every branch.
We have focused on the B\"uchi acceptance condition for simplicity, but all the results can be extended to 
the parity acceptance condition
(see \S{}\ref{sec:parity}).

\subsubsection*{Future Work} 
There are some directions for future work.
In this paper we focused on systems with a simple branching type like nondeterministic  or probabilistic. 
Extending this so that we can deal with systems with more complicated branching type like \emph{two-player games} (systems with two kinds of nondeterministic branching) or 
\emph{Markov decision processes} (systems with both nondeterministic and probabilistic branching) 
is a possible direction of future work.


Another direction would be to use the framework developed here to categorically generalize a verification method.
For example, using the framework of coalgebraic trace semantics in~\cite{urabeSH16coalgebraictrace},
a simulation notion for B\"uchi automata is generalized in~\cite{urabeH17fairsimJourn}.
Searching for an existing verification method that we can successfully generalize in our framework would be interesting.

Finally, 
it was left open in \S{}\ref{sec:forprob} if Asm.~\ref{asm:assumptionB}.\ref{item:asm:gfp-preserving} 
is satisfied by $T=\giry$ and $F=\FSigma$. Investigating this
is clearly a future work.


\subsubsection*{Acknowledgments}  
We thank Kenta Cho, Shin'ya Katsumata and the anonymous referees for useful comments.
The authors are supported by JST ERATO
HASUO Metamathematics for Systems Design Project (No.\ JPMJER1603), \pagebreak
and JSPS KAKENHI Grant Numbers 15KT0012 \& 15K11984.
Natsuki Urabe is supported by JSPS KAKENHI Grant Number 16J08157.

\bibliographystyle{splncs03}
\bibliography{myref}

\newpage
\appendix

\noindent{\Large\bf Appendix}
\section{Categorical Parity Conditions via Alternating Fixed Points of Functors}\label{sec:parity}
\begin{wrapfigure}[4]{r}[5pt]{3.6cm}
\small
\vspace*{-1.0cm}
\hspace{-4mm}
\begin{tikzpicture}[node distance=1.0cm, initial text=, bend angle=30, accepting/.style={double distance=2.0pt}, every loop/.style={min distance=5mm,looseness=5}]
\node [state, initial,   label={above:$x$}, minimum size=0pt] (x)                  {1};
\node [state, label={above:$y$},   minimum size=0pt] (y) [right of = x] {2};
\node [state, label={above:$z$},  label={[label distance=1.0mm]60:$\mathcal{A}'$}, minimum size=0pt] (z) [right of = y] {3};
\path [->] (x) edge [bend left] node             [above]  {$b$}  (y)
               edge [in=310, out=230, loop] node [below]  {$a$}  ()
           (y) edge [bend left] node             [below]  {$a$}  (x)
               edge [bend left] node             [above]  {$c$}  (z)
               edge [in=310, out=230, loop] node [below]  {$b$}  ()
           (z) edge [bend left] node             [below]  {$b$}  (y)
               edge [in=310, out=230, loop] node [below]  {$c$}  ();               
\end{tikzpicture}
\end{wrapfigure}
The \emph{parity condition} is a generalization of the B\"uchi condition.
A \emph{parity automaton} is equipped with a priority function $\Omega:X\to[1,2n]$ that assigns
a natural number called a \emph{priority} to each state $x\in X$.
An infinite run $x_{0}\xrightarrow{a_{0}}x_{1}\xrightarrow{a_{1}}\cdots$ satisfies the \emph{parity} condition if 
$\limsup_{i\to\infty}\Omega(x_i)$ is even.
For example, the parity automaton on the right above accepts an infinite word iff it contains infinitely many $b$ 
but only finitely many $c$.

\subsection{Parity $(T,F)$-systems and its Coalgebraic Trace Semantics}\label{subsec:PTFS}
%
Analogous results to those in 
\S{}\ref{subsec:catprelim}
hold for the
parity acceptance condition~\cite{urabeSH16coalgebraictrace}. 
In~\cite{urabeSH16coalgebraictrace} a priority function of a parity automaton was captured by a partition $X=X_1+\cdots+X_{2n}$ of a state space $X$ 
so that $X_i$ collects states with priority $i$,
and trace semantics was modeled by a solution of a hierarchical equation system that is similar to (\ref{eq:fig:nestedFP})
but consists of $2n$ diagrams.
\begin{mydefinition}[parity $(T,F)$-system]\label{def:TFsys}
Let $n\in\mathbb{N}$. A \emph{parity $(T,F)$-system} is a pair $(c,(X_1,\ldots,X_{2n}))$ 
of a $\overline{F}$-coalgebra $c:X\to \overline{F}X$ in $\Kl(T)$ and a partition $(X_1,\ldots,X_{2n})$ of $X$ (i.e.\ $X\cong X_1+\cdots+X_{2n}$).
For $i\in[1,2n]$, we write $c_i$ for $c\circ\kappa_i:X_i\to \overline{F}X$.
\end{mydefinition}

Their coalgebraic trace semantics
is given by a solution of an HES.

\begin{mydefinition}[$\trp_i(c)$]\label{def:trsemTFsys}
Assume that each homset of $\Kl(T)$ carries a partial order $\sqsubseteq$.
We say that $F$ and $T$ constitute a \emph{parity trace situation} wrt.\ $\overline{F}$ and 
$\sqsubseteq$ if they satisfy the following conditions: 
\begin{itemize}
\item A final $F$-coalgebra $\zeta:Z\to FZ$ exists.


\item For an arbitrary  parity $(T,F)$-system $\mathcal{X}=(c,(X_1,\ldots,X_{2n}))$, the following HES has a solution.
\begin{equation*} 
E_c=\left\{
\begin{matrix*}[l]
u_1\; &=_{\mu} \; J\fincoalg^{-1}\odot \overline{F}[u_1,\ldots,u_{2n}]\odot c_1 & \in (\Kl(T)(X_1,Z),\sqsubseteq_{X_1,Z})\\
u_2\; &=_{\nu} \; J\fincoalg^{-1}\odot \overline{F}[u_1,\ldots,u_{2n}]\odot c_2 & \in (\Kl(T)(X_2,Z),\sqsubseteq_{X_2,Z}) \\[-1.2mm]
& \qquad\vdots \\[-1.2mm]
u_{2n}\; &=_{\nu} \; J\fincoalg^{-1}\odot \overline{F}[u_1,\ldots,u_{2n}]\odot c_{2n} & \in (\Kl(T)(X_{2n},Z),\sqsubseteq_{X_{2n},Z})
\end{matrix*}
\right.
\end{equation*}
\end{itemize}
The solution $(u_i^\sol:X_i\kto Z)_{1\leq i\leq 2n}$ of $E_c$ is called the \emph{(coalgebraic) parity trace semantics} of 
$\mathcal{X}$. We write $\trp_i(c)$ for $u_i^\sol$ for each $i\in[1,2n]$
(see also Eq.~(\ref{eq:fig:nestedFP})). 
\end{mydefinition}

Note that the notions of B\"uchi $(T,F)$-system and B\"uchi trace semantics are 
special cases of those of parity $(T,F)$-system and parity trace semantics respectively.

\begin{myexample}\label{ex:parityFA}
Let $T=\pow$ and $F=\myalph\times(\place)$.
Then a parity $(T,F)$-system $(c:X\kto FX,(X_1,\ldots,X_{2n}))$ is identified with an $\myalph$-labeled 
parity automaton. 
Each $\trp_i(c)$ has a type $X_i\kto  \myalph^\omega$,
and
it is given as follows~\cite{urabeSH16coalgebraictrace}:
\[
\trp_i(c)(x):=\left\{a_1a_2\ldots \in\myalph^\omega \;\middle|\; \begin{aligned}
&\exists x_0,x_1,\ldots\in X.\; \forall i\in\omega.\;  x_i\in X_{p_i},\\
&\quad  (a_{i+1},x_{i+1})\in c(x_i) \text{ and }\textstyle{\limsup_{i\to\infty}p}_i\text{ is even}
\end{aligned}\right\}\,.
\]
\end{myexample}

\subsection{Categorical Datatypes for parity Systems}\label{subsec:refinement}
By repeatedly applying the operations $(\place)^+$ and $(\place)^\plusinf$ to $F$,
we can obtain functors $F^{(+\plusinf)^i}$ and $F^{(+\plusinf)^i+}$ where $(+\plusinf)^i$ denotes $i$-repetition of $+\plusinf$.
We introduce notations for them for simplicity.
\begin{mydefinition}[$\Fj{i}$]\label{def:Fj}
For $i\in\mathbb{N}$, we define $\Fj{i}:\mathbb{C}\to\mathbb{C}$ by 
$\Fj{i}:=F^{(+\plusinf)^l}$ if $i=2l$ and $\Fj{i}:=F^{(+\plusinf)^l+}$ if $i=2l+1$.
\end{mydefinition}
%
%

\begin{myexample}\label{exa:1710071248}
We continue Ex.~\ref{exa:1711141355}. 
In general, 
for  $i>0$,
\begin{align}
\Fj{i} X
&\cong 
\bigl\{
(a_0,p_0)\ldots (a_{k},p_{k})x\in (\myalph\times [1,i])^+X\mid k\in\mathbb{N}, p_0=i
\bigr\} \notag\\
&\cup\bigl\{
(a_0,p_0)(a_1,p_1)\ldots\in (\myalph\times [1,i])^\omega\mid p_0=i, \textstyle{\limsup_{i\to\infty}}p_i\text{ is even}
\bigl\}
\,. \label{eq:1712121834}
\end{align}
\end{myexample}

For B\"uchi $(T,F)$-systems, we have distinguished the following datatypes that are isomorphic to each other, and 
we wrote types of decorated trace semantics as $\dtr_2(c):X_2\kto\Fplusplusinf0$ and $\dtr_1(c):X_1\kto \Fplus\Fplusplusinf0$
for the sake of simplicity (Rem.~\ref{rem:1711151735}):
\[
\Fplusplusinf X\;\cong\; \Fplus(\Fplusplusinf X+X)\;\cong\;
F(\Fplus(\Fplusplusinf X+X)+\Fplusplusinf X+X)\,.
\]
Similarly, for parity $(T,F)$-systems, we distinguish the following datatypes.
\begin{multline*}
F^{(+\plusinf)^{n}} X\;\cong\; F^{(+\plusinf)^{n-1}+}(F^{(+\plusinf)^{n}} X+X)\;\\
\cong\;
F^{(+\plusinf)^{n-1}}(F^{(+\plusinf)^{2n-1}+}(F^{(+\plusinf)^{n}} X+X)+F^{(+\plusinf)^{n}} X+X)\;\cong\cdots\,.
\end{multline*}
We hereby introduce a short notation for each of the above.

\begin{mydefinition}[$\Fij{i}{j}, \aij{i}{j},\bij{i}{j}$]\label{def:1710131524}
For $i\in\mathbb{N}$ and $j\in[0,i]$,
we inductively define a functor 
$\Fij{i}{j}:\mathbb{C}\to\mathbb{C}$ 
as follows:
i) $\Fij{i}{i}:=\Fj{i}$; and
%
ii) $\Fij{i}{j}:=\Fj{j}\bigl(\coprod_{k=j+1}^{i}\Fij{i}{k}(\place)+\place\bigr)$ for $j<i$. 
Moreover, for $i\in\mathbb{N}$ and $j\in[1,i]$, we define a natural transformation
$\aij{i}{j}:\Fij{i}{j}\Rightarrow\Fij{i}{j-1}$ by 
$\aij{i}{j}:=\bigl(\initalg^{\Fj{j-1}}_{\coprod_{k=j+1}^{i}\Fij{i}{k}X+X}\bigr)^{-1}$ if $j$ is odd and
$\fincoalg^{\Fj{j-1}}_{\coprod_{k=j+1}^{i}\Fij{i}{k}X+X}$ if $j$ is even.
By its definition, each $\aij{i}{j}$ is an isomorphism. 
Furthermore, we define a natural transformation $\bij{i}{j}:\Fij{i}{j}\Rightarrow \Fij{i}{0}$ by:
\[
\bij{i}{j}\;:=\;\Bigl(\Fij{i}{j}\xRightarrow{\aij{i}{j}}\Fij{i}{j-1}\xRightarrow{\aij{i}{j-1}}\cdots\xRightarrow{\aij{i}{2}}\Fij{i}{1}\xRightarrow{\aij{i}{1}}\Fij{i}{0}\Bigr)\,.
\]
\end{mydefinition}

\subsection{Natural Transformation $\pij{i}{j}$}\label{subsec:NTs}
We introduce a natural transformation $\pij{i}{j}$, which is a generalization of the transformations introduced in
Def.~\ref{def:1710191132B} and removes decorations.
\begin{mydefinition}[$\pij{i}{j}$]\label{def:1710191132}
For $i\in\mathbb{N}$ and $j\in[1,i]$, we define a natural transformation 
$\pij{i}{j}:\Fij{i}{j}\Rightarrow\Fplusinf$ so that 
$[\pij{i}{1,X},\ldots,\pij{i}{i,X}]:\coprod_{j=1}^{i}\Fij{i}{j}(X)\to 
\Fplusinf X$ is the unique homomorphism from $[\bij{i}{1}_{\!\!\!\! X},\ldots,\bij{i}{i}_{\!\!\!\!X}]$ to $\fincoalg^F_X$.
\[
\small
  \xymatrix@R=3.0em@C=2.5em{
 {F (\mbox{\scriptsize$\smash{\displaystyle\coprod_{j=1}^{i}}$}\Fij{i}{j}X+X)} 
  \ar@{-->}[r]^(.55){\substack{F([\pij{i}{1\,X},\ldots,\pij{i}{i\,X}]+\id_X)\\\qquad\quad\mbox{}}}
 &
 {F
  (\Fplusinf X+X
 )
 } 
 \\
 {\qquad
 \mbox{\scriptsize$\smash{\displaystyle\coprod_{j=1}^{i}}$}\Fij{i}{j}X
 } 
 \ar[u]_(.5){
 [\bij{i}{1\,X},\ldots,\bij{i}{i\,X}]
 } 
 \ar@{-->}[r]^(.55){[\pij{i}{1\,X},\ldots,\pij{i}{i\,X}]}
 &
 {\Fplusinf X}
 \ar[u]_{\fincoalg^F_X}^{\substack{\text{final}\\\cong}}
 } 
\]
\end{mydefinition}


\subsection{Decorated Trace Semantics for Parity $(T,F)$-systems}\label{subsec:dtrP}
In order to deal with parity $(T,F)$-systems, we modify
Asm.~\ref{asm:assumptionB} as follows.

\begin{myassumption}\label{asm:assumption}
Throughout this section, 
let $T$ be a monad and $F$ be an endofunctor on $\mathbb{C}$, and
assume that 
each homset of $\Kl(T)$ carries a partial order $\sqsubseteq$. 
We further assume the following conditions for each $n\in\mathbb{N}$ and $A\in\mathbb{C}$. 
\begin{enumerate}
\item\label{item:asm:assumption01} $\Fj{n}:\mathbb{C}\to\mathbb{C}$ is well-defined and a lifting $\overline{\Fj{n}}:\Kl(T)\to\Kl(T)$ of $\Fj{n}$ is given.


\item\label{item:asm:assumption1}
If $n$ is even, $T$ and $\Fj{n}(\place+A)$ satisfy the conditions in Thm.~\ref{thm:initfinal}.

\item\label{item:asm:assumption2}
If $n$ is odd,
$T$ and $\Fj{n}(\place+A)$ constitute an infinitary trace situation wrt.\ 
  $\overline{\Fj{n}}$ and $\sqsubseteq$.



\item\label{item:asm:gfp-preserving}
If $n$ is odd,
$T$ and $\Fj{n}(\place+A)$ satisfy the gfp-preserving condition wrt.\ an arbitrary $\sigma$.

\item\label{item:asm:det-greatest}
If $n$ is odd,
$T$ and $\Fj{n}(\place+A)$ satisfy the deterministic-greatest condition.
%

\item\label{item:asm:assumption5}
The lifting $\overline{\Fj{n+1}}$ is obtained from $\overline{\Fj{n}}$ using the procedure in Def.~\ref{def:liftingplus}.

\item\label{item:asm:assumption6}
For $n\in\mathbb{N}$ and $A\in\mathbb{C}$, 
$\overline{\Fj{n}}(\place+A)$ is locally monotone.

\item\label{item:asm:assumption101}
$T$ and $F$ constitute a parity trace situation wrt.\ the same $\sqsubseteq$ and $\overline{F}$. 

\setcounter{asmenumi}{\value{enumi}}
\end{enumerate}
\end{myassumption}




\begin{mydefinition}[$\dtr_i(c)$]\label{def:PDTS}
For a parity $(T,F)$-system $(c,(X_1,\ldots,X_{2n}))$,
the \emph{decorated parity trace semantics} is a solution
$(\dtr_i(c):X_i\kto \Fij{2n}{i}0)_{1\leq i\leq 2n}$
of the following HES, all of whose equal symbols are subscripted by $\nu$.
\begin{equation*} 
\left\{
\begin{matrix*}[l]
v_1\; &=_{\nu} \; J({\bij{2n}{1\;0}})^{-1}\odot \overline{F}(v_1+\cdots+v_{2n})\odot c_1 & \in (\Kl(T)(X_1,\Fij{2n}{1}0),\sqsubseteq)\\
v_2\; &=_{\nu} \; J({\bij{2n}{2\;0}})^{-1}\odot \overline{F}(v_1+\cdots+v_{2n})\odot c_2 & \in (\Kl(T)(X_2,\Fij{2n}{2}0),\sqsubseteq) \\[-1mm]
& \qquad\vdots \\[-1mm]
v_{2n}\; &=_{\nu} \; J({\bij{2n}{2n\;0}})^{-1}\odot \overline{F}(v_1+\cdots+v_{2n})\odot c_{2n} & \in (\Kl(T)(X_{2n},\Fij{2n}{2n}0),\sqsubseteq)
\end{matrix*}
\right.
\end{equation*}
\end{mydefinition}

\subsection{Trace Semantics vs.\ Decorated Trace Semantics: Parity Case}\label{subsec:reltoOT}
We shall prove the following theorem, which generalizes Thm.~\ref{thm:mainthmB}.

\begin{mytheorem}\label{thm:mainthm}
For each $i\in[1,2n]$, $\trp_i(c)=\pij{2n}{i\,0}\circ\dtr_i(c)$.
\end{mytheorem}

We prove the above theorem in a way that was sketched in \S{}\ref{subsec:reltoOTB} for B\"uchi $(T,F)$-systems.
Def.~\ref{def:1710081602New},  Lem.~\ref{lem:1710161317},
Prop.~\ref{prop:1710160958} and Prop.~\ref{prop:1710191342} in the below generalize
Def.~\ref{def:1710081602NewB},  Lem.~\ref{lem:1710161317B},
Prop.~\ref{prop:1710160958B} and Prop.~\ref{prop:1710191342B} respectively.

\begin{mydefinition}[$c_i^\ddagger$, $\myl^{(i)}_j$]\label{def:1710081602New}
For $i\in[1,2n]$ and $j\in[1,i]$, 
we inductively define Kleisli arrows $c_i^\ddagger:X_i\kto \Fj{i-1}(X_{i}+\cdots+X_{2n})$ 
and 
$\myl^{(i)}_j:X_j\kto \Fij{i}{j}(X_{i+1}+\cdots+X_{2n})$ as follows 
(no need to distinguish the base case from the step case):
\begin{itemize}
\item $c_i^\ddagger:X_i\kto \Fj{i-1}(X_{i}+\cdots+X_{2n})$ is defined by:
\[
c_i^\ddagger:=\left(
{\small
\begin{aligned}
&X_i\xkrightarrow{c_i} F(X_1+\cdots+X_{2n})
\lower1ex\hbox{$\xkrightarrow{\overline{F}(\coprod_{j=1}^{i-1}\myl_j^{(i-1)}+\id_{X_i+\cdots+X_{2n}})}$} \\
&\quad F(\textstyle{\coprod_{j=1}^{i-1}}\Fij{i-1}{j}(X_i+\cdots+X_{2n})+X_i+\cdots+X_{2n}) \\[-1mm]
&\quad =\Fij{i-1}{0}(X_i+\cdots+X_{2n})\!\!\!\lower1ex\hbox{$\xkrightarrow{(\bij{i-1}{i-1}_{X_{i}+\cdots+X_{2n}})^{-1}}$}
\Fj{i-1}(X_{i}+\cdots+X_{2n})
\end{aligned}
}
\right)\,.
\]

\item $\myl^{(i)}_i:X_i\kto \Fij{i}{i}(X_{i+1}+\cdots+X_{2n})$ is defined as follows:
\begin{itemize}
\item If $i=2k+1$,
we define $\myl^{(i)}_i:X_i\kto F^{(+\plusinf)^{k}+}(X_{i+1}+\cdots+X_{2n})$ as the unique homomorphism from
$c_i^\ddagger$ to $J(\initalg^{F^{(+\plusinf)^{k-1}+}}_{X_{i}+\cdots+X_{2n}})^{-1}$. 
\[
\small
\!\!\!\!\!\!\!\!\!\!
\vcenter{  \xymatrix@R=1.4em@C+1em{
  {F^{(+\plusinf)^{k}}(X_{i}+\cdots+X_{2n})}
  \kar@{-->}[r]_(.3){\mathrlap{\overline{F^{(+\plusinf)^{k}}}(\myl^{(i)}_i+\id)}}
  &
  {F^{(+\plusinf)^{k}}\left(\begin{aligned}
  &F^{(+\plusinf)^{k}+}(X_{i+1}+\cdots+X_{2n})\\[-1mm]
  &\quad+X_{i+1}+\cdots+X_{2n}
  \end{aligned}
  \right)}
  \\
  {X_i}
  \kar[u]_{c_i^\ddagger}
  \kar@{-->}[r]^{\myl^{(i)}_i}
  \ar@{}[ru]|{=}
    &
  {F^{(+\plusinf)^{k}+}(X_{i+1}+\cdots+X_{2n})}
  \kar[u]_(.4){J(\initalg^{F^{(+\plusinf)^{k-1}+}}_{X_{i}+\cdots+X_{2n}})^{-1}}^(.4){\cong}  
  \mathrlap{\,.}
}}
\]

\item If $i=2k$,
we define $\myl^{(i)}_i:X_i\kto F^{(+\plusinf)^{k}}(X_{i+1}+\cdots+X_{2n})$ as the greatest homomorphism from
$c_i^\ddagger$ to $J\fincoalg^{F^{(+\plusinf)^{k-1}}}_{X_{i}+\cdots+X_{2n}}$. 
\[
\small
\vcenter{  \xymatrix@R=1.4em@C+1em{
  {F^{(+\plusinf)^{k-1}+}(X_{i}+\cdots+X_{2n})}
  \kar@{->}[r]_(.5){\overline{F^{(+\plusinf)^{k-1}+}}(\myl^{(i)}_i+\id)}
  &
  {F^{(+\plusinf)^{k-1}+}\left(\begin{aligned}
  &F^{(+\plusinf)^{k}}(X_{i+1}+\cdots+X_{2n})\\[-1mm]
  &\quad+X_{i+1}+\cdots+X_{2n}
  \end{aligned}
  \right)}
  \\
  {X_i}
  \kar[u]_{c_i^{\ddagger}}
  \kar[r]^{\myl^{(i)}_i}
  \ar@{}[ru]|{\color{blue}=_\nu}
    &
  {F^{(+\plusinf)^{k}}(X_{i+1}+\cdots+X_{2n})}
  \kar[u]_(.4){J\fincoalg^{F^{(+\plusinf)^{k-1}}}_{X_{i}+\cdots+X_{2n}}}^(.4){\cong}
 \mathrlap{\enspace.} 
}}
\]

\end{itemize}

\item For $j<i$, $\myl^{(i)}_j:X_j\kto \Fij{i}{j}(X_{i+1}+\cdots+X_{2n})$ is defined by:
\[
\myl^{(i)}_j:=\left({\small\begin{aligned}
&X_j\lower1ex\hbox{$\xkrightarrow{\myl^{(i-1)}_j}$}
\Fij{i-1}{j}(X_i+X_{i+1}+\cdots+X_{2n})
\lower1ex\hbox{$\xkrightarrow{\overline{\Fij{i-1}{j}}(\myl^{(i)}_i+J\id)}$}\\
&\quad \Fij{i-1}{j}(\Fij{i}{i}(X_{i+1}+\cdots+X_{2n})+X_{i+1}+\cdots+X_{2n})\\
&\quad =\Fij{i}{j}(X_{i+1}+\cdots+X_{2n})
\end{aligned}}\right)\,.
\]
\end{itemize}
\end{mydefinition}
Here the last equality is by the following lemma.
%
\begin{mylemma}\label{lem:1710151826}
For $i\in\mathbb{N}$ and $j\in[0,i]$,
$\Fij{i}{j}(\Fij{i+1}{i+1}(\place)+\place)=\Fij{i+1}{j}$.
\end{mylemma}

\begin{myproof} 
We prove the statement by the induction on $j$.

If $j=i$ then the statement is immediate by definition. 

If $j=2l-1<i$, we have:
\begin{align*}
&\Fij{i}{j}(\Fij{i+1}{i+1}(\place)+\place)\\
&=F^{(+\plusinf)^{l-1}+}\bigl(\coprod_{k=2l}^{i}\Fij{i}{k}(\Fij{i+1}{i+1}(\place)+\place)+\Fij{i+1}{i+1}(\place)+\place\bigr) \tag*{(\text{by definition})}\\
&=F^{(+\plusinf)^{l-1}+}\bigl(\coprod_{k=2l}^{i}\Fij{i+1}{k}(\place)+\Fij{i+1}{i+1}(\place)+\place\bigr)  \tag*{(\text{by IH})}\\
&=F^{(+\plusinf)^{l-1}+}\bigl(\coprod_{k=2l}^{i+1}\Fij{i+1}{k}(\place)+\place\bigr)  \\
&=\Fij{i+1}{j} \tag*{(\text{by definition}).}
\end{align*}
We can similarly prove the statement when $j=2l-2<i$.
\qed
\end{myproof}

%
%

The lemma below shows that if $j$ is odd (resp.\ even), 
not only $\mylij{j}{j}$ but also $\mylij{i}{j}$ with $i>j$ is characterized as the least (resp.\ greatest) homomorphism.
%
\begin{mylemma}\label{lem:1710161317}
Let $i\in[1,2n]$ and $j\in[1,i]$.
For simplicity, we write $\Xij{i}{j}$ for 
$\coprod_{k=j+1}^{i}\Fij{i}{k}(X_{i+1}+\cdots+X_{2n})+X_{i+1}+\cdots+X_{2n}$.
\begin{enumerate}
\item\label{item:lem:17101613171}
If $j$ is odd,  
$\mylij{i}{j}:X_j\kto \Fij{i}{j}(X_{i+1}+\cdots+X_{2n})$ is the unique homomorphism
from $\overline{\Fj{j-1}}(\id_{X_j}+\coprod_{k=j+1}^{i}\mylij{i}{k}+\id_{X_{i+1}+\cdots+X_{2n}})\odot c^\ddagger_j$ to
$J(\initalg^{\Fj{j-1}}_{\Xij{i}{j}})^{-1}$.

\item\label{item:lem:17101613172}
If $j$ is even, 
$\mylij{i}{j}:X_j\kto \Fij{i}{j}(X_{i+1}+\cdots+X_{2n})$ is the greatest homomorphism
from $\overline{\Fj{j-1}}(\id_{X_j}+\coprod_{k=j+1}^{i}\mylij{i}{k}+\id_{X_{i+1}+\cdots+X_{2n}})\odot c^\ddagger_j$ to
$J\fincoalg^{\Fj{j-1}}_{\Xij{i}{j}}$.
\end{enumerate}
\end{mylemma}
%

\begin{myproof} 
Item~\ref{item:lem:17101613171} is easily proved by the finality of $J(\initalg^{\Fj{j-1}}_{X_{i+1}+\cdots+X_{2n}})^{-1}$.
We prove Item~\ref{item:lem:17101613172} by the induction on $i$.

If $i=j$, then the statement is immediate by the definition of 
$\myl^{(j)}_j:X_j\kto \Fij{j}{j}(X_{j+1}+\cdots+X_{2n})$ (Def.~\ref{def:1710081602New}).

Let $i>j$ and assume that 
$\myl^{(i-1)}_j:X_j\kto \Fij{i-1}{j}(X_{i}+\cdots+X_{2n})$ is the greatest homomorphism from 
 $\overline{\Fj{j-1}}(\id_{X_j}+\coprod_{k=j+1}^{i-1}\mylij{i-1}{k}+\id_{X_{i}+\cdots+X_{2n}})\odot c^\ddagger_j$ to
$J\fincoalg^{\Fj{j-1}}_{\Xij{i-1}{j}}$.
By the definition of $\Fij{i-1}{j}$, we have the following equation.
\[
\overline{\Fij{i-1}{j}}(\myl^{(i)}_i+\id) 
=
\overline{\Fj{j}}
\bigl(
\coprod_{k=j+1}^{i-1}
\overline{\Fij{i-1}{k}}(\mylij{i}{i}+\id)+\mylij{i}{i}+\id\bigr)
\] 
By the definition of a lifting $\overline{\Fj{j}}$, 
this means that 
$\overline{\Fij{i-1}{j}}(\myl^{(i)}_i+\id)$ is the greatest homomorphism from 
$\overline{\Fj{j-1}}\bigl(\id+\coprod_{k=j+1}^{i-1}\Fij{i}{k}(\mylij{i}{i}+\id)+\mylij{i}{i}+\id\bigr)\odot J\fincoalg^{\Fj{j-1}}_{\Xij{i-1}{j}}$ to $J\fincoalg^{\Fj{j-1}}_{\Xij{i}{j}}$.
Hence by the gfp-preserving condition, $\mylij{i}{j}=\overline{\Fij{i-1}{j}}(\mylij{i}{i}+\id)\odot\mylij{i-1}{j}$ 
is the greatest homomorphism from 
$\overline{\Fj{j-1}}\bigl(\id+\coprod_{k=j+1}^{i-1}\overline{\Fij{i}{k}}(\mylij{i}{i}+\id)+\mylij{i}{i}+\id\bigr)\odot\overline{\Fj{j-1}}(\id+\coprod_{k=j+1}^{i-1}\mylij{i-1}{k}+\id)\odot c^\ddagger_j$ to $J\fincoalg^{\Fj{j-1}}_{\Xij{i}{j}}$.
As we have
\begin{align*}
&\overline{\Fj{j-1}}\bigl(\id+\coprod_{k=j+1}^{i-1}\overline{\Fij{i}{k}}(\mylij{i}{i}+\id)+\mylij{i}{i}+\id\bigr)\odot\overline{\Fj{j-1}}(\id+\coprod_{k=j+1}^{i-1}\mylij{i-1}{k}+\id)\odot c^\ddagger_j\\
&=\overline{\Fj{j-1}}\bigl(\id+\bigl(\coprod_{k=j+1}^{i-1}\overline{\Fij{i}{k}}(\mylij{i}{i}+\id)\odot \mylij{i-1}{k}\bigr)+\mylij{i}{i}+\id\bigr)\odot c^\ddagger_j\\
&=\overline{\Fj{j-1}}(\id+\coprod_{k=j+1}^{i}\mylij{i}{k}+\id)\odot c^\ddagger_j\,,
\end{align*}
by the definition of $\mylij{i}{k}$,
the statement is proved.
See also Fig.~\ref{fig:1711052057}.
\qed
\end{myproof}

\afterpage{%
\begin{landscape}
\begin{figure}[p]
\[
\hspace{-3cm}
\vcenter{  \xymatrix@R=4.0em@C+.6em{ 
  {\scriptsize\Fj{j-1}\left(\begin{aligned}
  &X_j+\\ 
  &\textstyle{\coprod_{k=j+1}^{i}}\Fij{i}{k}(X_{i+1}+\cdots+X_{2n})\\
  &+X_{i+1}+\cdots+X_{2n}\end{aligned}
  \right)}
  \kar@{->}[r]^(.55){\overline{\Fj{j-1}}(\myl^{(i-1)}_j+\id)}
  &
  {\scriptsize\Fj{j-1}\left(\begin{aligned}
  &\Fij{i-1}{j}(X_{i+1}+\cdots+X_{2n})+\\ 
  &\textstyle{\coprod_{k=j+1}^{i}}\Fij{i}{k}(X_{i+1}+\cdots+X_{2n})\\
  &+X_{i+1}+\cdots+X_{2n}
  \end{aligned}
  \right)}
    \kar@{->}[r]
  &
  {\scriptsize\Fj{j-1}\left(\begin{aligned}
  &\Fij{i}{j}(X_{i+1}+\cdots+X_{2n})+\\ 
  &\textstyle{\coprod_{k=j+1}^{i}}\Fij{i}{k}(X_{i+1}+\cdots+X_{2n})\\
  &+X_{i+1}+\cdots+X_{2n}
  \end{aligned}
  \right)}
  \\
  {\scriptsize\Fj{j-1}\left(\begin{aligned}
  &X_j+\\ 
  &\textstyle{\coprod_{k=j+1}^{i-1}}\Fij{i-1}{k}(X_{i}+\cdots+X_{2n})\\
  &+X_{i}+\cdots+X_{2n}\end{aligned}
  \right)}
  \kar@{->}[r]^(.55){\overline{\Fj{j-1}}(\myl^{(i-1)}_j+\id)}
  \kar[u]^(.5){\overline{\Fj{j-1}}\bigl(\id+\coprod_{k=j+1}^{i-1}\overline{\Fij{i}{k}}(\mylij{i}{i}+\id)+\mylij{i}{i}+\id\bigr)} 
  &
  {\scriptsize\Fj{j-1}\left(\begin{aligned}
  &\Fij{i-1}{j}(X_{i}+\cdots+X_{2n})+\\ 
  &\textstyle{\coprod_{k=j+1}^{i-1}}\overline{\Fij{i}{k}}(X_{i}+\cdots+X_{2n})\\
  &+X_{i}+\cdots+X_{2n}
  \end{aligned}
  \right)}
  \kar[u]^(.5){\overline{\Fj{j-1}}\bigl(\id+\coprod_{k=j+1}^{i-1}\Fij{i}{k}(\mylij{i}{i}+\id)+\mylij{i}{i}+\id\bigr)}   
  &
  {}
  \\
  {\Fj{j-1}(X_{j}+\cdots+X_{2n})}
  \kar[u]^(.4){\overline{\Fj{j-1}}(\id+\coprod_{k=j+1}^{i-1}\mylij{i-1}{k}+\id)}
  &
  {}
  \\
  {X_j}
  \kar[u]^{c^\ddagger_j}
  \kar@{->}[r]^{\myl^{(i-1)}_j}
  \ar@{}[ruu]|{\color{blue}=_\nu}
  &
    {\Fij{i-1}{j}(X_{i}+\cdots+X_{2n})
  }
  \kar[uu]_(.4){J\fincoalg^{\Fj{j-1}}_{\Xij{i-1}{j}}}^(.4){\substack{\cong}}  
    \kar@{->}[r]^{\overline{\Fij{i-1}{j}}(\myl^{(i)}_i+\id)}  
    &
  {\Fij{i}{j}(X_{i+1}+\cdots+X_{2n})
  }
  \kar[uuu]_(.4){J\fincoalg^{\Fj{j-1}}_{\Xij{i}{j}}}^(.4){\substack{\cong}}  
}}
\]
\caption{$\mylij{i}{j}:=\overline{\Fij{i-1}{j}}(\myl^{(i)}_i+\id)\odot \mylij{i-1}{j}$ is the greatest homomorphism}
\label{fig:1711052057}
\end{figure}
\end{landscape}
}


%





\mbox{}

\begin{myproposition}\label{prop:1710160958}
For each $i\in[1,2n]$, $\myl^{(2n)}_i=\dtr_i(c)$.
\end{myproposition}

\begin{myproof} 
Assume that $j$ is odd.
By Lem.~\ref{lem:1710161317}.\ref{item:lem:17101613171}, 
$\mylij{2n}{j}$ is the unique homomorphism from
$\overline{\Fj{j-1}}(\myl^{(i)}_j+\id)\odot c^\ddagger_j$ to 
$J(\initalg^{\Fj{j-1}}_{\coprod_{k=j+1}^{2n}\Fij{2n}{k}0})^{-1}$.
This means that it is the greatest homomorphism.
\[
\vcenter{  \xymatrix@R=1.2em@C+.6em{ 
  {\scriptsize\Fj{j-1}(X_j+\textstyle{\coprod_{k=j+1}^{2n}}\Fij{i}{k}0
  )}
  \kar@{->}[r]^(.55){\overline{\Fj{j-1}}(\myl^{(2n)}_j+\id)}
  &
  {\scriptsize\Fj{j-1}(\Fij{2n}{j}0+\textstyle{\coprod_{k=j+1}^{2n}}\Fij{2n}{k}0
  )}
  \\
  {\Fj{j-1}(X_j+\cdots+X_{2n})}
  \kar[u]^(.5){\overline{\Fj{j-1}}(\id+\coprod_{k=j+1}^{2n}\mylij{2n}{k})}
  &
  \\
  {X_j}
  \kar[u]_{c_j^\ddagger}
  \kar@{->}[r]^{\myl^{(2n)}_j}
  \ar@{}[ruu]|{\color{blue}=_\nu}
    &
  {\Fij{2n}{j}0
  }
  \kar[uu]_(.4){J(\initalg^{\Fj{j-1}}_{\coprod_{k=j+1}^{2n}\Fij{2n}{k}0})^{-1}}^(.4){\substack{\cong}}  
}}
\]

By the definition of $c_j^\ddagger$ (Def.~\ref{def:1710081602New}), 
this means that $\mylij{2n}{j}$ is the greatest fixed point of the following function.
\begin{align*}
f\mapsto &J\initalg^{\Fj{j-1}}_{\coprod_{k=j+1}^{2n}\Fij{2n}{k}0}\odot \overline{\Fj{j-1}}(f+\mylij{2n}{j+1}+\cdots+\mylij{2n}{2n})
\\&\qquad\odot
J(\bij{j-1}{j-1}_{X_{i}+\cdots+X_{2n}})^{-1}\odot \overline{F}(\mylij{i-1}{1}+\cdots+\mylij{i-1}{i-1}+\id)\odot c_j
\end{align*}
Note here that the right hand side can be transformed as follows: 
\begin{align*}
&J\initalg^{\Fj{j-1}}_{\coprod_{k=j+1}^{2n}\Fij{2n}{k}0}\odot \overline{\Fj{j-1}}(f+\mylij{2n}{j+1}+\cdots+\mylij{2n}{2n})\odot
J(\bij{j-1}{j-1}_{\coprod_{k=j}^{2n}X_k})^{-1}\\
&\qquad\odot \overline{F}(\mylij{j-1}{1}+\cdots+\mylij{j-1}{j-1}+\id)\odot c_j \\
%
%
&=J\initalg^{\Fj{j-1}}_{\coprod_{k=j+1}^{2n}\Fij{2n}{k}0}\odot J(\bij{j-1}{j-1}_{\coprod_{k=j}^{2n}\Fij{2n}{k}0})^{-1}\odot \overline{\Fij{j-1}{0}}(f+\mylij{2n}{j+1}+\cdots+\mylij{2n}{2n})\\
&\qquad\odot 
\overline{F}(\mylij{j-1}{1}+\cdots+\mylij{j-1}{j-1}+\id)
\odot  c_j 
\tag*{(\text{by naturality of $\bij{j-1}{j-1}$})} \\
&=J\initalg^{\Fj{j-1}}_{\coprod_{k=j+1}^{2n}\Fij{2n}{k}0}\odot J(\bij{j-1}{j-1}_{\coprod_{k=j}^{2n}\Fij{2n}{k}0})^{-1}\\
&\qquad\odot 
\overline{F}(\mylij{2n}{1}+\cdots+\mylij{2n}{j-1}+f+\mylij{2n}{j+1}+\cdots+\mylij{2n}{2n})
\odot  c_j 
\tag*{(\text{by Def.~\ref{def:1710081602New}})} \\
&=J\initalg^{\Fj{j-1}}_{\coprod_{k=j+1}^{2n}\Fij{2n}{k}0}\odot 
J(\bij{2n}{j-1}_{0})^{-1}\\
&\qquad\odot 
\overline{F}(\mylij{2n}{1}+\cdots+\mylij{2n}{j-1}+f+\mylij{2n}{j+1}+\cdots+\mylij{2n}{2n})
\odot  c_j 
 \tag*{(\text{by Def.~\ref{def:1710131524}})} \\
&=J(\bij{2n}{j}_{0})^{-1}\odot 
\overline{F}(\mylij{2n}{1}+\cdots+\mylij{2n}{j-1}+f+\mylij{2n}{j+1}+\cdots+\mylij{2n}{2n})
\odot  c_j 
\tag*{(\text{by Def.~\ref{def:1710131524}})\,.} 
\end{align*}
Hence $\mylij{2n}{j}$ is the greatest fixed point of the following function:
\[
f\mapsto 
J({\bij{2n}{j}}_{0})^{-1}\odot \overline{F}(\mylij{2n}{1}+\cdots+\mylij{2n}{j-1}+f+\mylij{2n}{j+1}+\cdots+\mylij{2n}{2n})\odot c_j\,.
\]
We can similarly prove the same statement when $j$ is even.
Hence $(\mylij{2n}{1},\ldots,\mylij{2n}{2n})$ is the solution of the HES in Def.~\ref{def:PDTS}, 
and this concludes the proof.
\qed
\end{myproof}

This proposition implies the
existence of a solution of the HES in Def.~\ref{def:PDTS}.


\begin{mylemma}\label{lem:1710170852}
For the HES in Def.~\ref{def:trsemTFsys}, we define
the intermediate solution 
$l^{(i)}_j:\Kl(T)(X_{i+1},\Fplusinf 0)\times\cdots\times\Kl(T)(X_{2n},\Fplusinf 0)\to\Kl(T)(X_i,\Fplusinf 0)$ 
as in Def.~\ref{def:solHES} (note that $Z$ in Def.~\ref{def:trsemTFsys} is $\Fplusinf 0$).
Then for 
$(u_k:X_k\kto \Fplusinf0)_{k\in[i+1,2n]}$, 
we have: 

\noindent\begin{minipage}{0.95\hsize}
\begin{equation}\label{eq:1711181620}
l^{(i)}_j(u_{i+1},\ldots,u_{2n})
=J\multFplusinf_0\odot J\pij{i}{j\,\Fplusinf0}\odot\overline{\Fij{i}{j}}[u_{i+1},\ldots,u_{2n}] \odot\mylij{i}{j}\,.
\end{equation}
\end{minipage}
\begin{minipage}{0.05\hsize}
\vspace{6mm}
\qed
\end{minipage}
\vspace{2mm}
\end{mylemma}

The proof of the above lemma is very long, so we defer the proof to \S{}\ref{subsec:proof:lem:1710170852}.


\begin{myproposition}\label{prop:1710191342}
For each $i\in[1,2n]$, $\trp_i(c)=\pij{2n}{i\,0}\circ\myl^{(2n)}_i$.
\end{myproposition}
\begin{mylemma}\label{lem:1712021619}
For each $j\in[1,i]$, $\multFplusinf_0\circ \pij{i}{j\,\Fplusinf0}\circ \Fij{i}{j}\initarr_{\Fplusinf0}=\pij{i}{j}_0$\,.
\end{mylemma}

\begin{myproof} 
By definition, it suffices to prove that the following arrow is a homomorphism from 
$[\bij{i}{1}_0,\ldots,\bij{i}{i}_0]$ to $\fincoalg^F_0$.
\[
\bigl[
\multFplusinf_0\circ \pij{i}{1\,\Fplusinf0}\circ \Fij{i}{1}\initarr_{\Fplusinf0},\ldots,
\multFplusinf_0\circ \pij{i}{i\,\Fplusinf0}\circ \Fij{i}{i}\initarr_{\Fplusinf0}
\bigr]:\coprod_{j=1}^{i}\Fij{i}{j}0\to 
\Fplusinf0
\]
We have:
\begin{align*}
&\fincoalg^F_0\circ\bigl[
\multFplusinf_0\circ \pij{i}{1\,\Fplusinf0}\circ \Fij{i}{1}\initarr_{\Fplusinf0},\ldots,
\multFplusinf_0\circ \pij{i}{i\,\Fplusinf0}\circ \Fij{i}{i}\initarr_{\Fplusinf0}
\bigr] \\
&=F[\multFplusinf_0,\id_{\Fplusinf0}]\circ\fincoalg^F_{\Fplusinf0}\circ 
\bigl[
\pij{i}{1\,\Fplusinf0}\circ \Fij{i}{1}\initarr_{\Fplusinf0},\ldots,
\pij{i}{i\,\Fplusinf0}\circ \Fij{i}{i}\initarr_{\Fplusinf0}
\bigr]
\tag*{(\text{by Def.~\ref{def:multFplusinf}})} \\
&=F[\multFplusinf_0,\id_{\Fplusinf0}]\circ
F\bigl([\pij{i}{1\,\Fplusinf0},\ldots,\pij{i}{i\,\Fplusinf0}]+\id_{\Fplusinf0}\bigr)\\
&\qquad\circ
[\bij{i}{1\,\Fplusinf0}\circ \Fij{i}{1}\initarr_{\Fplusinf0},\ldots,\bij{i}{i\,\Fplusinf0}\circ\Fij{i}{i}\initarr_{\Fplusinf0}] 
\tag*{(\text{by Def.~\ref{def:1710191132}})} \\
&=F[\multFplusinf_0,\id_{\Fplusinf0}]\circ
F\bigl([\pij{i}{1\,\Fplusinf0},\ldots,\pij{i}{i\,\Fplusinf0}]+\id_{\Fplusinf0}\bigr)\circ
\Fij{i}{0}\initarr_{\Fplusinf0}\circ [\bij{i}{1}_0,\ldots,\bij{i}{i}_0] 
\tag*{(\text{by naturality})} \\
&=F\Bigl(\bigl[
\multFplusinf_0\circ \pij{i}{1\,\Fplusinf0}\circ \Fij{i}{1}\initarr_{\Fplusinf0},\ldots,
\multFplusinf_0\circ \pij{i}{i\,\Fplusinf0}\circ \Fij{i}{i}\initarr_{\Fplusinf0}
\bigr]\Bigr)\circ [\bij{i}{1}_0,\ldots,\bij{i}{i}_0]\,.
\end{align*}
This concludes the proof.
\qed
\end{myproof}

\begin{myproof}[Prop.~\ref{prop:1710191342}]
Immediate by Lem.~\ref{lem:1710170852} and Lem.~\ref{lem:1712021619}
\qed
\end{myproof}

\begin{myproof}[Thm.~\ref{thm:mainthm}]
Immediate from Prop.~\ref{prop:1710160958} and Prop.~\ref{prop:1710191342}.
\qed
\end{myproof}


\subsection{Decorated Trace Semantics for Nondeterministic Parity Tree Automata}\label{subsec:NPTA}
We extend the discussions in \S{}\ref{sec:NPTAB} for 
nondeterministic \emph{parity} tree automata (NPTA).

\begin{mydefinition}[NPTA, see e.g.~\cite{thomas97LAL}]\label{def:NPTA}
A \emph{nondeterministic parity tree automaton} (NPTA for short) is a quadruple
$\mathcal{A}=(X,\Sigma,\delta,\Omega)$
of a set $X$ of \emph{states}, a ranked alphabet $\Sigma$, a \emph{transition function}
$\delta:X\to\pow(\coprod_{n\in\mathbb{N}}\Sigma_n\times X^{n})$ 
and a \emph{priority function} $\Omega:X\to[1,2n]$.
For $i\in[1,2n]$, we write $X_i$ for $\{x\in X\mid \Omega(x)=i\}$.
\end{mydefinition}

\begin{mydefinition}[$\langp_{\mathcal{A}}$]\label{def:langNPTA}
Let $\mathcal{A}=(X,\Sigma,\delta,\Omega)$ be an NPTA where $\Omega:X\to[1,2n]$.
A \emph{run tree} 
over $\mathcal{A}$ is 
is defined in a similar manner to that over an NBTA (Def.~\ref{def:langNPTAB}).
A run tree is \emph{accepting} if for each branch $(a_0,x_0)(a_1,x_1)\ldots\in(\Sigma\times X)^\omega$,
$\limsup_{k\to\infty}\Omega(x_k)$ is even.
%
We write $\Run_{\mathcal{A}}(x)$ (resp.\ $\AccRun_{\mathcal{A}}(x)$) for the set of run trees (resp.\ accepting run trees) 
whose root node is labeled by $x\in X$.
For $A\subseteq X$, $\Run_{\mathcal{A}}(A)$ 
denotes $\cup_{x\in A}\Run(x)$. 
We define $\AccRun_{\mathcal{A}}(A)$ similarly.
If no confusion is likely, we omit the subscript $\mathcal{A}$.
We define $\DelSt:\Run(X)\to\Treeinf(\Sigma)$ 
in a similar manner to that for NBTA.
%
The \emph{language} $\langp_{\mathcal{A}}:X\to\pow\Treeinf(\Sigma)$ of $\mathcal{A}$ is defined by
$\lang_{\mathcal{A}}(x)=\DelSt(\AccRun_{\mathcal{A}}(x))$. 
\end{mydefinition}

\begin{myproposition}[\cite{urabeSH16coalgebraictrace}]\label{prop:NPTAcoind}
We assume the situation in Prop.~\ref{prop:NPTAcoindB}.
\begin{enumerate}
\item[\ref{item:prop:NPTAcoind1}'] $\pow$ and $\FSigma$ constitute a parity trace situation (Def.~\ref{def:trsemTFsys}) wrt.\ $\sqsubseteq$ and $\overline{\FSigma}$.

\item[\ref{item:prop:NPTAcoind3}']
For an NPTA $\mathcal{A}=(X,\Sigma,\delta,\Omega)$ where $\Omega:X\to[1,2n]$,
we define a parity $(\pow,\FSigma)$-system $(c:X\kto\overline{\FSigma} X,(X_1,\ldots,X_{2n}))$ by 
$c:=\delta$ and $X_i:=\{x\in X\mid \Omega(x)=i\}$.
Then we have: $[\trp_1(c),\ldots,\trp_{2n}(c)]=\langp_{\mathcal{A}}:X\to\pow\Treeinf(\Sigma)$.
\qed
\end{enumerate}
\end{myproposition}

In the rest of this section, 
as in \S{}\ref{subsec:coalgNPTAB},
we describe $\Fij{i}{j}0$ and $\dtr_i(c)$ for $F=\FSigma$, and show the relationship with $\trp_i(c)$ in accordance with Thm.~\ref{thm:mainthm}
for an NPTA $\mathcal{A}=(X,\Sigma,\delta,\Omega)$
modeled as a $(\pow,\FSigma)$-system $(c:X\to\pow\FSigma X,(X_1,\ldots,X_{2n}))$.

The following proposition generalizes Prop.~\ref{prop:FijpowB}. It is proved in a similar manner to Ex.~\ref{exa:1710071248}.

\begin{myproposition}\label{prop:Fijpow}
We define a set
$\AccTreeij{i}{j}(\Sigma,A)\subseteq\Treeinf(\Sigma\times[1,i]+A)$ by:
\[
\AccTreeij{i}{j}(\Sigma,A):=
\left\{\begin{aligned}
&\xi\in \Treeinf(\Sigma\times[1,i]\\
&\qquad\qquad+A)
\end{aligned}
\;\middle|\; {\small\begin{aligned}
&\text{the root node is labeled by $j$, and for}\\[-1mm]
&\text{each infinite branch, the maximum}\\[-1mm]
&\text{priority appearing infinitely is even}
\end{aligned}}\right\}\,.
\]
Moreover, we define a function
\[
\decompij{i}{j}:
\AccTreeij{i}{j}(\Sigma,A)\to 
\AccTreeij{i}{j-1}(\Sigma,A)
\]
by 
$\decompij{i}{j}(D,l):= (D,l')$ where 
\[
l'(w):=\begin{cases}
(a,j-1) & (\text{$w=\empseq$ and $l(\empseq)=(a,j)$}) \\
l(w) & (\text{otherwise}).
\end{cases}
\]
Then $\AccTreeij{i}{j}(\Sigma,A)\cong\FSigmaij{i}{j}A$,
and 
\begin{equation*}
\Bigl(\decompij{i}{j}
:
\AccTreeij{i}{j}(\Sigma,A)\to
\AccTreeij{i}{j-1}(\Sigma,A)
\Bigr)
\cong
\Bigl(
\aij{i}{j\,A}
:\FSigmaij{i}{j}A\to \FSigmaij{i}{j-1}A
\Bigr)
\end{equation*}
where $\aij{i}{j}$ is defined as in Def.~\ref{def:1710131524}
and $\AccTreeij{i}{0}(\Sigma,A)$ is defined as follows:
\[
\AccTreeij{i}{0}(\Sigma,A)
:=
\left\{\begin{aligned}
&\xi\in \Treeinf(\Sigma\times[0,i]\\
&\qquad\qquad+A)
\end{aligned}
\;\middle|\; {\small\begin{aligned}
&\text{only the root node is labeled by $0$, and for}\\[-1mm]
&\text{each infinite branch, the maximum}\\[-1mm]
&\text{priority appearing infinitely is even}
\end{aligned}}\right\}\,.
\]
\qed
\end{myproposition}

By using the characterizations in the proposition above,
we can concretely prove that 
the assumptions required in the previous sections are satisfied by 
$\pow$ and $\FSigma$.
\begin{myproposition}\label{prop:asmsatisfied}
Asm.~\ref{asm:assumption} is satisfied by 
$(T,F)=(\pow,\FSigma)$. 
\end{myproposition}

\begin{myproof} 
Cond.~\ref{item:asm:assumption1} is proved in a similar manner to~\cite{hasuoJS07generictrace}.
Cond.~\ref{item:asm:assumption2} is proved in a similar manner to~\cite{urabeH15extended}
using Prop.~\ref{prop:Fijpow}..

We prove that Cond.~\ref{item:asm:gfp-preserving} is satisfied.
Let $c:X\kto\FSigmaj{i}(X+A)$ and $\sigma: \FSigmaj{i}(Y+A)\kto Y$. Let $l:X\kto (\FSigmaj{i})^\plusinf A$ be the greatest homomorphism from $c$ to $J\fincoalg^{\FSigmaj{i}}_A$,
and $m:(\FSigmaj{i})^\plusinf A\kto Y$ be the greatest fixed point of $\Phi_{J\fincoalg^{\FSigmaj{i}}_A,\sigma}$.
It is easy to see that $m\odot l$ is a fixed point of $\Phi_{c,\sigma}$.
We show that it is the greatest fixed point.
Let $t:X\kto (\FSigmaj{i})^\plusinf A$ be a fixed point of $\Phi_{c,\sigma}$.

For each $k\in\omega$, we inductively define $\pi_k:(\FSigmaj{i})^\plusinf A\to(\FSigmaj{i}(\place+A))^k1$ as follows:
$\pi_0:=!_{(\FSigmaj{i})^\plusinf A}$ and $\pi_{k+1}:=(\FSigmaj{i}(\place+\id_A))^k!\circ \FSigmaj{i}(\pi_k+\id_A)\circ \fincoalg^{\FSigmaj{i}}_A$.
By Thm.~\ref{thm:constinitfin}, $((\FSigmaj{i})^\plusinf A,(\pi_k)_{k\in\omega})$ is a limit over a final sequence 
$1\xleftarrow{!}\FSigmaj{i}(1+A)\xleftarrow{\FSigmaj{i}(!+\id_A)}\FSigmaj{i}(\FSigmaj{i}(1+A)+A)\xleftarrow{\FSigmaj{i}(\FSigmaj{i}(!+\id_A)+\id_A)}\ldots$.
Hence we can identify $(\FSigmaj{i})^\plusinf A$ with the following set:
\[
\Bigl\{(z_k\in\FSigmaj{i}(\place+A)^k1)_{k\in\omega} 
\mid 
\forall k\in\omega.\; \Fj{i}(\place+\id_A)^k!(z_{k+1})=z_k
\Bigr\}\,.
\]

For each $k\in\omega$, we inductively define $t_k:X\kto \Fj{i}(\place+A)^k1$ as follows:
i) $t_0:=J!\odot t$ and ii) $t_{k+1}:=\FSigmaj{i}(t_k+A)\odot c$.
Moreover
we define a function $l':X\kto (\FSigmaj{i})^\plusinf A$ as follows:
\[
l'(x):=\{(z_k\in\FSigmaj{i}(\place+A)^k1)_{k\in\omega}\in (\FSigmaj{i})^\plusinf A\mid \forall k\in\omega.\; z_k\in t_k(x) \}\,.
\]
Then $l'$ is a homomorphism from $c$ to $J\fincoalg^{\FSigmaj{i}}_A$, and moreover $m\odot l'=t$.
The former implies $l'\sqsubseteq l$. Hence we have $t=m\odot l'\sqsubseteq m\odot l$.
Therefore $m\odot l$ is the greatest fixed point of $\Phi_{c,\sigma}$. 
Hence Cond.~\ref{item:asm:gfp-preserving} is satisfied.

We prove that Cond.~\ref{item:asm:det-greatest} is satisfied.
Let $c:X\to \FSigmaj{n}(X+A)$ be an $\FSigmaj{n}(\place+A)$-coalgebra and 
$u:X\to (\FSigmaj{n})^{\plusinf}A$ be the unique homomorphism from $c$ to $\fincoalg^{\FSigmaj{n}}_A$.
Let $f:X\kto (\FSigmaj{n})^{\plusinf}A$ be a homomorphism from $Jc$ to $J\fincoalg^{\FSigmaj{n}}_A$ and $x\in X$,
and assume that $t\in f(x)$. 
Then as $f$ and $Ju$ are homomorphism from $Jc$ to $J\fincoalg^{\FSigmaj{n}}_A$, 
we can prove $t=u(x)$ by the induction on the structure of $t$.
Hence $Ju$ is the greatest homomorphism.

By Cond.~\ref{item:asm:gfp-preserving}--\ref{item:asm:det-greatest}, we can inductively define 
a lifting $\overline{\FSigmaj{n}}:\Kl(T)\to\Kl(T)$ for each $n\in\mathbb{N}$.
Then Cond.~\ref{item:asm:assumption01} and Cond.~\ref{item:asm:assumption5} are  satisfied. 

It is proved in a similar manner to \cite{hasuoJS07generictrace,urabeH15extended,urabeSH16coalgebraictrace} that
Cond.~\ref{item:asm:assumption6} is satisfied using Prop.~\ref{prop:Fijpow}.

Using Prop.~\ref{prop:Fijpow}, Cond.~\ref{item:asm:assumption101} is proved in a similar manner to~\cite{urabeSH16coalgebraictrace}.
\qed
\end{myproof}

We now show what $\dtr_i(c)$ characterizes for an NPTA wrt.\ the characterization in Prop.~\ref{prop:Fijpow}.
By Prop.~\ref{prop:Fijpow},
$\bij{i}{j\,A}$ is isomorphic to the following type 
\[
\bij{i}{j\,A}:\AccTreeij{i}{j}(\Sigma,A)\to \textstyle{\coprod_{n\in\omega}}\Sigma_n\times(\textstyle{\coprod_{k=1}^i}\AccTreeij{i}{k}(\Sigma,A)+A)\,,
\]
and is given by $\bij{i}{j\,A}(\xi)=(a,\xi_0,\ldots,\xi_{n-1})$ if the root node of $\xi$ is labeled by $(a,j)\in\Sigma_n\times [1,i]$.
We write $\AccTreeij{i}{j}(\Sigma)$ for $\AccTreeij{i}{j}(\Sigma,\emptyset)$.

\begin{myproposition}\label{prop:dtrpow}
Let $\mathcal{A}=(X,\Sigma,\delta,\Omega)$ be an NPTA where $\Omega:X\to[1,2n]$.
We overload $\Omega$ and define $\Omega:\Run(X)\to\Treeinf(\Sigma\times[1,2n])$ 
by $\Omega(D,l):=(D,l')$ where $l'(w):=(a,\Omega(x))$ if $l(w)=(a,x)$.
We define a parity $(\pow,\FSigma)$-system $(c:X\kto\overline{\FSigma} X,(X_1,\ldots,X_{2n}))$ as in 
Prop.~\ref{prop:NPTAcoind}.\ref{item:prop:NPTAcoind3}'. Then for $i\in[1,2n]$ and $x\in X_i$, 
\begin{align*}
\dtr_i(c)(x)=\{\Omega(\rho)\in\AccTreeij{2n}{i}(\Sigma)\mid \rho\in\AccRun_{\mathcal{A}}(x)\}\,.
\end{align*}
\end{myproposition}

\begin{myproof} 
%
By Prop.~\ref{prop:Fijpow} and the definition of $\AccRun_{\mathcal{A}}(x)$, it is easy to see that 
$\{\Omega(\rho)\mid \rho\in\AccRun_{\mathcal{A}}(x)\}\subseteq \Fij{2n}{i}0$.
For each $i\in[1,2n]$, we define $f_i:X_i\to \pow\AccTreeij{2n}{i}(\Sigma)$ by 
$f_i(x):= \{\Omega(\rho)\mid \rho\in\AccRun_{\mathcal{A}}(x)\}$.
We show that a family $(f_i)_{i\in[1,2n]}$ is the solution of the HES as in Def.~\ref{def:PDTS}.

We first prove 
$f_i= J(\bij{2n}{i\,0})^{-1}\odot \overline{\FSigma}(f_1+\cdots+f_{2n})\odot c_i$ for each $i$.
For each $x\in X_i$, we have:
\begin{align*}
&J(\bij{2n}{i\,0})^{-1}\odot \overline{\FSigma}(f_1+\cdots+f_{2n})\odot c_i(x)\\
&= J(\bij{2n}{i\,0})^{-1}\odot \overline{\FSigma}(f_1+\cdots+f_{2n})\bigl(\{(a,x_0,\ldots,x_{m-1})\in \tau(x)\}\bigr) \\
&= J(\bij{2n}{i\,0})^{-1}\Bigl(\bigl\{(a,\Omega(\rho_0),\ldots,\Omega(\rho_{m-1})) \in\Sigma_m\times \bigl(\coprod_{k=1}^{2n}\AccTreeij{2n}{k}(\Sigma)\bigr)^m\\
&\qquad\qquad\qquad\qquad
\mid (a,x_0,\ldots,x_{m-1})\in \tau(x), \rho_t\in \AccRun_{\mathcal{A}}(x_t)\text{ for each $t\in[0,m-1]$}\bigr\}\Bigr) \\
&= \bigl\{((a,i),\Omega(\rho_0),\ldots,\Omega(\rho_{m-1})) \in\AccTreeij{2n}{i}(\Sigma)\\
&\qquad\qquad\qquad\qquad
\mid (a,x_0,\ldots,x_{m-1})\in \tau(x), \rho_t\in \AccRun_{\mathcal{A}}(x_t)\text{ for each $t\in[0,m-1]$}\bigr\} \\
&=\bigl\{\Omega(\rho)\in\AccTreeij{2n}{i}(\Sigma)\mid \rho\in\AccRun_{\mathcal{A}}(x)\bigr\} \\
&=f_i(x)\,.
\end{align*}

We next show that it is the greatest fixed point.
Let $(g_i:X_i\to\pow\AccTreeij{2n}{i}(\Sigma))_{1\leq i\leq 2n}$ be a family of functions such that 
$g_i= J(\bij{2n}{i\,0})^{-1}\odot \overline{\FSigma}(g_1+\cdots+g_{2n})\odot c_i$ for each $i$.
It suffices to show that $g_i(x)\subseteq f_i(x)$ for each $i\in[1,2n]$ and $x\in X_i$.

Let $i\in[1,2n]$ and $x\in X_i$, and assume that $\xi=(D,l)\in g_i(x)$.
We write $l_1(w)$ and $l_2(w)$ for $\pi_1(l(w))$ and $\pi_2(l(w))$ respectively.
We hereby define a function $l':D\to \Sigma\times X$ by $l'(w):=(l_1(w),l'_2(w))$,
where $l'_2(w)\in X$ is inductively defined as follows so that the following condition is satisfied:
%
for each $w\in D$, 
$\xi_w\in [g_1,\ldots,g_{2n}](l'_2(w))$ 
(recall that $\xi_w$ denotes the $w$-th subtree of $\xi$).
%
\begin{itemize}
\item For $w=\empseq$, we let $l'_2(\empseq)=x$. 

\item Let $w\in D$, $\pi_1(l(w))=m$. Assume that we have fixed $l'_2(w)$ so that 
the condition above is satisfied. Assume $l'_2(w)\in X_i$.
Assume that the root node of the $w$-th subtree $\xi_w$ of $\xi$ is labeled by $a\in\Sigma_m$.
Then $\xi_w$ has a shape $((a,i),\xi_{w0},\ldots,\xi_{w(m-1)})$. 
Here by the assumption on $g_i$,  we have:
\allowdisplaybreaks[1]
\begin{align*}
\xi_w &=((a,i),\xi_{w0},\ldots,\xi_{w(m-1)}) \\
&\in g_i(l'_2(w))\\
&= J(\bij{2n}{i\,0})^{-1}\odot \overline{\FSigma}(g_1+\cdots+g_{2n})\odot c_i(x')  \\
&= J(\bij{2n}{i\,0})^{-1}\odot \overline{\FSigma}(g_1+\cdots+g_{2n})\Bigl(\bigl\{(a',x_0,\ldots,x_{m'-1})\in \tau(l'_2(w))\bigr\}\Bigr)  \\
&= J(\bij{2n}{i\,0})^{-1}\Bigl(\bigl\{(a',\xi'_0,\ldots,\xi'_{m'01})\in\Sigma_{m'}\times \bigl(\coprod_{k=1}^{2n}\AccTreeij{2n}{k}(\Sigma)\bigr)^{m'}\\
&\qquad\qquad\qquad\quad\mid (a',x_0,\ldots,x_{m'-1})\in \tau(l'_2(w)), \\
&\qquad\qquad\qquad\qquad\xi'_t\in [g_1,\ldots,g_{2n}](x_t)\text{ for each $t\in[0,m'-1]$}\bigr\}\Bigr)  \\
&=\bigl\{((a',i),\xi'_0,\ldots,\xi'_{m'-1})\in\AccTreeij{2n}{i}(\Sigma) \\
&\qquad\quad\mid (a',x_0,\ldots,x_{m'-1})\in \tau(l'_2(w)), \\
&\qquad\qquad\xi'_t\in [g_1,\ldots,g_{2n}](x_t)\text{ for each $t\in[0,m'-1]$}\bigr\}\,.
\end{align*}
This means that there exists $(a,x_0,\ldots,x_{m-1})\in\tau(l'_2(w))$ such that 
$\xi_{wt}\in [g_1,\ldots,g_{2n}](x_t)$ for each $t\in[0,m-1]$.
We let $l'_2(tw):=x_t$ for each $t$.
\end{itemize}

Let $\rho=(D,l')$. By its construction, we can easily see that $\Omega(\rho)=\xi$.

In contrast, by the construction, $\rho$ is a run tree over $\mathcal{A}$, and moreover, as $\xi\in\AccTreeij{2n}{i}(\Sigma)$,
$\rho$ is accepting. Therefore by the definition of $f_i$, we have $\Omega(\rho)\in f_i(x)$.
This concludes the proof.
\qed  
\end{myproof}

The following proposition generalizes Prop.~\ref{prop:pijpowB}.

\begin{myproposition}\label{prop:pijpow}
We define $\DelStij{i}{j}:\AccTreeij{i}{j}(\Sigma,A)\to\Treeinf(\Sigma+A)$ by 
$\DelStij{i}{j}(D,l):=(D,l')$ where $l'(w):=\pi_1(l(w))$.
Then with respect to the isomorphism in Prop.~\ref{prop:Fijpow}, 
$\DelStij{i}{j}(\xi)=\pij{i}{j\,A}(\xi)$.
\end{myproposition}

\begin{myproof}[Prop.~\ref{prop:pijpow}]
It is easy to see that $\DelStij{i}{j}:\AccTreeij{i}{j}(\Sigma,A)\to\Treeinf(\Sigma+A)$
satisfies the following equality for each $\xi=((a,i),(\xi_0,\ldots,\xi_{m-1}))\in\AccTreeij{i}{j}(\Sigma,A)$.
\[
\DelStij{i}{j}(\xi)=
\Bigl(a,\bigl([\DelStij{i}{1},\ldots,\DelStij{i}{i}](\xi_0),\ldots,[\DelStij{i}{1},\ldots,\DelStij{i}{i}](\xi_{m-1})\bigr)\Bigr)\,.
\]
By the characterization of $\FSigmaplusinf A$ and $\fincoalg^F_A$ 
in \S{}\ref{subsec:coalgNPTAB} and by Prop.~\ref{prop:Fijpow},
the equation above means that $[\DelStij{i}{1},\ldots,\DelStij{i}{i}]$
is a homomorphism from $[\bij{i}{1\,A},\ldots,\bij{i}{i}_A]$ to $\fincoalg^F_A$.
Therefore immediate by the definition of $\pij{i}{j}$.
\qed
\end{myproof}

Hence Thm.~\ref{thm:mainthm} results in the following equation, which is again obvious.
\[
\bigl\{[\DelStij{2n}{1},\ldots,\DelStij{2n}{2n}](\Omega(\rho))\mid \rho\in\AccRun_{\mathcal{A}}(x)\bigr\}=\langp(x)\,.
\]

\section{Omitted Proofs}\label{sec:omittedproofs}
\subsection{Proof of Prop.~\ref{prop:plusinffunctor}}


\begin{mylemma}\label{lem:gfpprescondOld}
Assume that $T$ and $F$ constitute an infinitary trace situation
and satisfy the gfp-preserving condition (Def.~\ref{def:gfpprescondNew}).
For each $X,A,B\in\mathbb{C}$, $c:X\kto F(X+A)$ and $f:A\kto B$,
if $l:X\kto \Fplusinf A$ is the greatest homomorphism from $c$ to $J\fincoalg^F_A$,
then $\overline{\Fplusinf}f\odot l:X\kto \Fplusinf B$ is the greatest homomorphism from 
$\overline{F}(\id_X+f)\odot c$ to $J\fincoalg^F_B$.
\vspace{-2mm}
\[
  \xymatrix@R=1.6em@C=4.3em{
  {F(X+B)}
  \kar[r]^{\overline{F}(l+\id)}
  &
  {F(\Fplusinf A+B)}
  \kar[r]^{\overline{F}(\overline{\Fplusinf} f+\id)}
  &
  {F(\Fplusinf B+B)}
  \\
  {F(X+A)}
  \kar[r]^{\overline{F}(l+\id)}
  \kar[u]^{\overline{F}(\id+f)}
  &
  {F(\Fplusinf A+A)}
  \kar[u]^{\overline{F}(\id+f)}
  &
  \\
  {X}
  \kar[r]^{l}
  \kar[u]^{c}
  \ar@{}[ru]|{\color{blue}=_\nu}
  &
  {\Fplusinf A}
  \kar[r]^{\overline{\Fplusinf} f}
  \kar[u]^{J\fincoalg^F_A}_{\cong}
  \ar@{}[ruu]|{\color{blue}=_\nu}  
  &
  {\Fplusinf B}
  \kar[uu]^{J\fincoalg^F_B}_{\cong}
  }
\]
\end{mylemma}

\begin{myproof}
It is easy to see that $\overline{\Fplusinf}f$ is the greatest fixed point of 
$\Phi_{J\fincoalg^F_A,J(\fincoalg^F_B)^{-1}\odot\overline{F}(\id+f)}$,
and the greatest homomorphism from $\overline{F}(\id_X+f)\odot c$ to $J\fincoalg^F_B$ is the greatest fixed point of
$\Phi_{c,J(\fincoalg^F_B)^{-1}\odot\overline{F}(\id+f)}$.
\[
  \xymatrix@R=1.6em@C=4.3em{
 {F(X+A)}
  \kar[r]^{\overline{F}(l+\id)}
  &
  {F(\Fplusinf A+A)}
    \kar[r]^{\overline{F}(m+\id)}
  &
  {F(\Fplusinf B+A)}
    \kar[d]^{\overline{F}(\id+f)}
  \\
  &
  &
  {F(\Fplusinf B+B)}
  \kar[d]^{J(\fincoalg^F_B)^{-1}}_{\cong}
  \\
  {X}
  \kar[r]^{l}
  \kar[uu]^{c}
  \ar@{}[ruu]|{\color{blue}=_\nu}
  &
  {\Fplusinf A}
  \kar[r]^{\overline{\Fplusinf} f}
  \kar[uu]^{J\fincoalg^F_A}_{\cong}
  \ar@{}[ruu]|{\color{blue}=_\nu}  
  &
  {\Fplusinf B}
  }
\]
Hence immediate by the gfp-preserving condition.
\end{myproof}

\begin{myproof}[Prop.~\ref{prop:plusinffunctor}]
Item.~\ref{item:prop:plusinffunctor1} is immediate by the finality.
Item.~\ref{item:prop:plusinffunctor2} is easily proved by the gfp-preserving condition,
the deterministic-greatest condition and Lem.~\ref{lem:gfpprescondOld}.
\qed
\end{myproof}




\subsection{Proof of Lem.~\ref{lem:1710170852}}\label{subsec:proof:lem:1710170852}
\begin{mysublemma}\label{sublem:1712011751}
For $A\in\mathbb{C}$,
the unique homomorphism from $[F[\kappa_1,\kappa_2]\circ \fincoalg^{F}_{\Fplusinf A},F[\kappa_2,\kappa_3]\circ \fincoalg^F_A]$
to $\fincoalg^F_A$
is given by $[\multFplusinf_A,\id_A]$.
\end{mysublemma}

\begin{myproof}
Let $u:\Fplusinf\Fplusinf A+\Fplusinf A\to\Fplusinf A$ be the unique homomorphism from $[F[\kappa_1,\kappa_2]\circ \fincoalg^{F}_{\Fplusinf A},F[\kappa_2,\kappa_3]\circ \fincoalg^F_A]$ to $\fincoalg^F_A$.

Note that $u=[u\circ\kappa_1,u\circ\kappa_2]$.
By Def.~\ref{def:multFplusinf}, $u\circ\kappa_1=\multFplusinf_A$.
We shall show that $u\circ\kappa_2=\id_A$

It is easy to see that $\kappa_2:\Fplusinf A\to\Fplusinf\Fplusinf A+\Fplusinf A$ is a homomorphism from 
$\fincoalg^F_A$ to $[F[\kappa_1,\kappa_2]\circ \fincoalg^{F}_{\Fplusinf A},F[\kappa_2,\kappa_3]\circ \fincoalg^F_A]$.
Therefore $u\circ \kappa_2$ is a homomorphism from $\fincoalg^F_A$ to itself, on the one hand.
On the other hand, $\id_A$ is also a homomorphism from $\fincoalg^F_A$.
Hence by the finality of $\fincoalg^F_A$, we have $u\circ\kappa_2=\id_{\Fplusinf A}$.
\[
\vcenter{  \xymatrix@R=1.4em@C+.1em{
  {F(\Fplusinf A+A)}
    \ar@{->}[r]_(.45){F(\kappa_2+\id_A)}
  &
  {F(\Fplusinf \Fplusinf A+\Fplusinf A+A)}
  \ar@{-->}[r]_(.55){F(u+\id_A)}
  &
  {F(\Fplusinf A+A)}
  \\
  &
  {F(\Fplusinf\Fplusinf A+\Fplusinf A)+F(\Fplusinf A+A)}
  \ar[u]_{[F[\kappa_1,\kappa_2],F[\kappa_2,\kappa_3]]}
  &
  \\
  {\Fplusinf A}
  \ar[uu]^{\substack{\cong}}_{\fincoalg^F_A}
  \ar@{->}[r]^(.45){\kappa_2}
  &
  {\Fplusinf\Fplusinf A+\Fplusinf A}
  \ar[u]_{\fincoalg^{F}_{\Fplusinf A}+\fincoalg^F_A}^{\cong}
  \ar@{-->}[r]^(.55){u}     
  &
  {\Fplusinf A}
  \ar[uu]^{\substack{\text{final}\\\cong}}_{\fincoalg^F_A}
}}
\]
Hence we have $u=[u\circ\kappa_1,u\circ\kappa_2]=[\multFplusinf_A,\id_A]$.
\qed
\end{myproof}

\begin{mysublemma}\label{sublem:1712021229}
\sloppy
We define an $F$-coalgebra $\gamma_i:\coprod_{j=1}^{i-1}\Fij{i}{j}\Fplusinf0+\Fplusinf0\to F(\coprod_{j=1}^{i-1}\Fij{i}{j}\Fplusinf0+\Fplusinf0)$ as follows:
\begin{align*}
\gamma_i:=&\textstyle{\coprod_{j=1}^{i-1}}\Fij{i}{j}\Fplusinf0+\Fplusinf0
\xrightarrow{[\bij{i}{1}_{\Fplusinf0},\ldots,\bij{i}{i-i}_{\Fplusinf0},F\kappa_{i+1}\circ\fincoalg^F_0]} 
F(\textstyle{\coprod_{j=1}^{i}}\Fij{i}{j}\Fplusinf0+\Fplusinf0) \\
&
\xrightarrow{F(\id_{\Fij{i}{1}\Fplusinf0}+\cdots+\id_{\Fij{i}{i-1}\Fplusinf0}+[\multFplusinf_0\circ\pij{i}{i},\id_{\Fplusinf0}])} 
F(\textstyle{\coprod_{j=1}^{i-1}}\Fij{i}{j}\Fplusinf0+\Fplusinf0)\,.
\end{align*}
Then the unique homomorphism from $\gamma_i$ 
to $\fincoalg^F_0:\Fplusinf0\to F\Fplusinf0$
is given by the following arrow:
\[
[\multFplusinf_0\circ\pij{i}{1\,\Fplusinf0},\ldots,
\multFplusinf_0\circ\pij{i}{i-1\,\Fplusinf0},\id_{\Fplusinf0}]:
\coprod_{j=1}^{i-1}\Fij{i}{j}0+\Fplusinf0\to\Fplusinf0\,.
\]
\end{mysublemma}

\begin{myproof}
It suffices to show that it is a homomorphism.
We have:
\allowdisplaybreaks[1]
\begin{align*}
&\fincoalg^F_0\circ[\multFplusinf_0\circ\pij{i}{1\,\Fplusinf0},\ldots,\multFplusinf_0\circ\pij{i}{i-1\,\Fplusinf0},\id_{\Fplusinf0}] \\
&= \fincoalg^F_0\circ[\multFplusinf_0,\id_{\Fplusinf0}]\circ([\pij{i}{1\,\Fplusinf0},\ldots,\pij{i}{i-1\,\Fplusinf0}]+\id_{\Fplusinf0}) 
\\
&= F[\multFplusinf_0,\id_{\Fplusinf0}]\circ[\fincoalg^{F}_{\Fplusinf0},F\kappa_2\circ \fincoalg^F_0] \circ([\pij{i}{1\,\Fplusinf0},\ldots,\pij{i}{i-1\,\Fplusinf0}]+\id_{\Fplusinf0}) 
\tag*{(\text{by Sublem.~\ref{sublem:1712011751}})}\\
&= F[\multFplusinf_0,\id_{\Fplusinf0}]\circ[F([\pij{i}{1\,\Fplusinf0},\ldots,\pij{i}{i\,\Fplusinf0}]+\id_{\Fplusinf0})\circ[\bij{i}{1\,\Fplusinf0},\ldots,\bij{i}{i-1\,\Fplusinf0}],\\
&\qquad\quad F\kappa_2\circ \fincoalg^F_0]  
\tag*{(\text{by Def.~\ref{def:1710191132}})}\\
&=F[\multFplusinf_0\circ\pij{i}{1\,\Fplusinf0},\ldots,
\multFplusinf_0\circ\pij{i}{i\,\Fplusinf0},\id_{\Fplusinf0}]\circ[\bij{i}{1\,\Fplusinf0},\ldots,\bij{i}{i-1\,\Fplusinf0},F\kappa_{i}\circ\fincoalg^F_0]\\
&=F[\multFplusinf_0\circ\pij{i}{1\,\Fplusinf0},\ldots,
\multFplusinf_0\circ\pij{i}{i-1\,\Fplusinf0},\id_{\Fplusinf0}] \\
&\qquad \circ \bigl(F(\id_{\Fij{i}{1}\Fplusinf0}+\cdots+\id_{\Fij{i}{i-1}\Fplusinf0}+[\multFplusinf_0\circ\pij{i}{i},\id_{\Fplusinf0}])\\
&\qquad\qquad \circ[\bij{i}{1\,\Fplusinf0},\ldots,\bij{i}{i-1\,\Fplusinf0},F\kappa_{i+1}\circ\fincoalg^F_0]\bigr)\,.
\end{align*}
This concludes the proof.
\qed
\end{myproof}

\begin{mysublemma}\label{sublem:1712011520}
For each $j\in[1,i-1]$, we have the following equality.
\[
\multFplusinf_0\circ\pij{i}{j\,\Fplusinf0}=\multFplusinf_0\circ\pij{i-1}{j\,\Fplusinf0}\circ \Fij{i-1}{j}[\multFplusinf_{0}\circ \pij{i}{i\,\Fplusinf0},\id_{\Fplusinf0}]\,.
\]
\end{mysublemma}

\begin{myproof}
By Sublem.~\ref{sublem:1712021229},
it suffices to show that the following arrow is a homomorphism from 
$\gamma_i$ to $\fincoalg^F_0$.
\begin{multline*}
\bigl[
\multFplusinf_0\circ\pij{i-1}{1\,\Fplusinf0}\circ \Fij{i-1}{1}[\multFplusinf_{0}\circ \pij{i}{i\,\Fplusinf0},\id_{\Fplusinf0}],\ldots, 
\\
\multFplusinf_0\circ\pij{i-1}{i-1\,\Fplusinf0}\circ \Fij{i-1}{i-1}[\multFplusinf_{0}\circ \pij{i}{i\,\Fplusinf0},\id_{\Fplusinf0}],
\id_{\Fplusinf0}
\bigr]
\end{multline*}
\fussy
We have:
\allowdisplaybreaks[1]
\begin{align*}
&\fincoalg^F_{0}\circ\bigl[\multFplusinf_0\circ\pij{i-1}{1\,\Fplusinf0}\circ \Fij{i-1}{1}[\multFplusinf_{0}\circ \pij{i}{i\,\Fplusinf0},\id_{\Fplusinf0}],\ldots, \\
&\qquad\quad\multFplusinf_0\circ\pij{i-1}{i-1\,\Fplusinf0}\circ \Fij{i-1}{i-1}[\multFplusinf_{0}\circ \pij{i}{i\,\Fplusinf0},\id_{\Fplusinf0}],\id_{\Fplusinf0}
\bigr] \\
&=\fincoalg^F_{0}\circ[\multFplusinf_0,\id_{\Fplusinf0}]\circ\Bigl(\bigl[\pij{i-1}{1\,\Fplusinf0}\circ \Fij{i-1}{1}[\multFplusinf_{0}\circ \pij{i}{i\,\Fplusinf0},\id_{\Fplusinf0}],\ldots, \\
&\qquad\qquad\qquad\qquad\qquad\qquad\pij{i-1}{i-1\,\Fplusinf0}\circ \Fij{i-1}{i-1}[\multFplusinf_{0}\circ \pij{i}{i},\id_{\Fplusinf0}]\bigr]+\id_{\Fplusinf0}\Bigr) \\
&=F[\multFplusinf_0,\id_{\Fplusinf0}]\circ[\fincoalg^{F}_{\Fplusinf0},F\kappa_2\circ \fincoalg^F_0] \\
&\qquad\circ\Bigl(\bigl[\pij{i-1}{1\,\Fplusinf0}\circ \Fij{i-1}{1}[\multFplusinf_{0}\circ \pij{i}{i\,\Fplusinf0},\id_{\Fplusinf0}],\ldots, \\
&\qquad\qquad\pij{i-1}{i-1\,\Fplusinf0}\circ \Fij{i-1}{i-1}[\multFplusinf_{0}\circ \pij{i}{i\,\Fplusinf0},\id_{\Fplusinf0}]\bigr]+\id_{\Fplusinf0}\Bigr) 
\tag*{(\text{by Sublem.~\ref{sublem:1712011751}})}\\
&=F[\multFplusinf_0,\id_{\Fplusinf0}] \\
&\qquad\circ\Bigl[\fincoalg^F_{\Fplusinf0}\circ[\pij{i-1}{1}_A,\ldots,\pij{i-1}{i-1}_A]
\\
 &\qquad\qquad
 \circ\bigl(
 \Fij{i-1}{1}[\multFplusinf_{\Fplusinf0}\circ \pij{i}{i\,\Fplusinf0},\id_{\Fplusinf0}]+\cdots+\Fij{i-1}{i-1}[\multFplusinf_{\Fplusinf0}\circ \pij{i}{i\,\Fplusinf0},\id_{\Fplusinf0}]\bigr),\\
 &\qquad\qquad
F\kappa_2\circ \fincoalg^F_0\Bigr] \\
&=F[\multFplusinf_0,\id_{\Fplusinf0}] \\
&\qquad\circ\Bigl[F\bigl([\pij{i-1}{1\,\Fplusinf0},\ldots,\pij{i-1}{i-1\,\Fplusinf0}]+\id_{\Fplusinf0}\bigr)\circ [\bij{i-1}{1\,\Fplusinf0},\ldots,\bij{i-1}{i-1\,\Fplusinf0}]
\\
 &\qquad\qquad
 \circ\bigl(
 \Fij{i-1}{1}[\multFplusinf_{\Fplusinf0}\circ \pij{i}{i\,\Fplusinf0},\id_{\Fplusinf0}]+\cdots+\Fij{i-1}{i-1}[\multFplusinf_{\Fplusinf0}\circ \pij{i}{i\,\Fplusinf0},\id_{\Fplusinf0}]\bigr),\\
 &\qquad\qquad
F\kappa_2\circ \fincoalg^F_0\Bigr]
\tag*{(\text{by Def.~\ref{def:1710191132}})} \\
&=F[\multFplusinf_0,\id_{\Fplusinf0}] \\
&\qquad\circ\Bigl[F\bigl([\pij{i-1}{1\,\Fplusinf0},\ldots,\pij{i-1}{i-1\,\Fplusinf0}]+\id_{\Fplusinf0}\bigr)\\
&\qquad\qquad\quad\circ 
F\bigl(\textstyle{\coprod_{j=1}^{i-1}}\Fij{i-1}{j}[\multFplusinf_{\Fplusinf0}\circ \pij{i}{i\,\Fplusinf0},\id_{\Fplusinf0}]
+[\multFplusinf_{\Fplusinf0}\circ \pij{i}{i\,\Fplusinf0},\id_{\Fplusinf0}]
\bigr) \\
&\qquad\qquad\quad\circ [\bij{i-1}{1\,\Fij{i}{i}\Fplusinf0+\Fplusinf0},\ldots,\bij{i-1}{i-1\,\Fij{i}{i}\Fplusinf0+\Fplusinf0}],\\
 &\qquad\qquad
F\kappa_2\circ \fincoalg^F_0\Bigr]
\tag*{(\text{by naturality})} \\
&=F[\multFplusinf_0,\id_{\Fplusinf0}] \\
&\qquad\circ\Bigl[F\bigl([\pij{i-1}{1\,\Fplusinf0},\ldots,\pij{i-1}{i-1\,\Fplusinf0}]+\id_{\Fplusinf0}\bigr)\\
&\qquad\qquad\quad\circ 
F\bigl(\textstyle{\coprod_{j=1}^{i-1}}\Fij{i-1}{j}[\multFplusinf_{\Fplusinf0}\circ \pij{i}{i\,\Fplusinf0},\id_{\Fplusinf0}]
+[\multFplusinf_{\Fplusinf0}\circ \pij{i}{i\,\Fplusinf0},\id_{\Fplusinf0}]
\bigr) \\
&\qquad\qquad\quad\circ [\bij{i}{1\,\Fplusinf0},\ldots,\bij{i}{i-1\,\Fplusinf0}],\\
 &\qquad\qquad
F\kappa_2\circ \fincoalg^F_0\Bigr]
\tag*{(\text{by Def.~\ref{def:1710131524}})} \\
&=F[\multFplusinf_0,\id_{\Fplusinf0}] \circ
F \bigl([\pij{i-1}{1\,\Fplusinf0},\ldots,\pij{i-1}{i-1\,\Fplusinf0}]+\id_{\Fplusinf0}\bigr)\\
&\qquad\circ 
F\bigl(\textstyle{\coprod_{j=1}^{i-1}}\Fij{i-1}{j}[\multFplusinf_{\Fplusinf0}\circ \pij{i}{i\,\Fplusinf0},\id_{\Fplusinf0}]
+[\multFplusinf_{\Fplusinf0}\circ \pij{i}{i\,\Fplusinf0},\id_{\Fplusinf0}]
\bigr) \\
&\qquad\circ [\bij{i}{1\,\Fplusinf0},\ldots,\bij{i}{i-1\,\Fplusinf0},F\kappa_{i+1}\circ\fincoalg^F_0] \\
&=F\bigl[
\multFplusinf_{\Fplusinf0}\circ\pij{i-1}{1\,\Fplusinf0}\circ \Fij{i-1}{1}[\multFplusinf_A\circ \pij{i}{i\,\Fplusinf0},\id_{\Fplusinf0}],
\ldots, \\
&\qquad\qquad
\multFplusinf_{\Fplusinf0}\circ\pij{i-1}{i-1\,\Fplusinf0}\circ \Fij{i-1}{i-1}[\multFplusinf_{\Fplusinf0}\circ \pij{i}{i\,\Fplusinf0},\id_{\Fplusinf0}]\bigr] \\
&\qquad\circ \bigl(F(\id_{\Fij{i}{1}\Fplusinf0}+\cdots+\id_{\Fij{i}{i-1}\Fplusinf0}+[\multFplusinf_0\circ\pij{i}{i},\id_{\Fplusinf0}])\\
&\qquad \circ[\bij{i}{1\,\Fplusinf0},\ldots,\bij{i}{i-1\,\Fplusinf0},F\kappa_{i+1}\circ\fincoalg^F_0]\bigr)\,.
\end{align*}
This concludes the proof.
\qed
\end{myproof}

\begin{mysublemma}\label{sublem:1712012042}
The unique homomorphism from an $F$-coalgebra 
\[
[\bij{i}{1\,\Fplusinf0},\ldots,\bij{i}{i\,\Fplusinf0},F\kappa_{i+1}\circ\fincoalg^F_0]:\coprod_{j=1}^i\Fij{i}{j}\Fplusinf0+\Fplusinf0\to F(\coprod_{j=1}^i\Fij{i}{j}\Fplusinf0+\Fplusinf0)
\] 
to $\fincoalg^F_0:\Fplusinf0\to F\Fplusinf 0$
is given by the following arrow:
\[
[\multFplusinf_0\circ\pij{i}{1\,\Fplusinf0},\ldots,
\multFplusinf_0\circ\pij{i}{i\,\Fplusinf0},\id_{\Fplusinf0}]:
\coprod_{j=1}^i\Fij{i}{j}0+\Fplusinf0\to\Fplusinf0\,.
\]
\end{mysublemma}

\begin{myproof}
It suffices to that the arrow is a homomorphism.
We have:
\allowdisplaybreaks[1]
\begin{align*}
&\fincoalg^F_0\circ[\multFplusinf_0\circ\pij{i}{1\,\Fplusinf0},\ldots,\multFplusinf_0\circ\pij{i}{i\,\Fplusinf0},\id_{\Fplusinf0}] \\
&= \fincoalg^F_0\circ[\multFplusinf_0,\id_{\Fplusinf0}]\circ([\pij{i}{1\,\Fplusinf0},\ldots,\pij{i}{i\,\Fplusinf0}]+\id_{\Fplusinf0}) 
\\
&= F[\multFplusinf_0,\id_{\Fplusinf0}]\circ[\fincoalg^{F}_{\Fplusinf0},F\kappa_2\circ \fincoalg^F_0] \circ([\pij{i}{1\,\Fplusinf0},\ldots,\pij{i}{i\,\Fplusinf0}]+\id_{\Fplusinf0}) 
\tag*{(\text{by Sublem.~\ref{sublem:1712011751}})}\\
&= F[\multFplusinf_0,\id_{\Fplusinf0}]\circ[F([\pij{i}{1\,\Fplusinf0},\ldots,\pij{i}{i\,\Fplusinf0}]+\id_{\Fplusinf0})\circ[\bij{i}{1\,\Fplusinf0},\ldots,\bij{i}{i\,\Fplusinf0}],\\
&\qquad\quad F\kappa_2\circ \fincoalg^F_0]  
\tag*{(\text{by Def.~\ref{def:1710191132}})}\\
&=F[\multFplusinf_0\circ\pij{i}{1\,\Fplusinf0},\ldots,
\multFplusinf_0\circ\pij{i}{i\,\Fplusinf0},\id_{\Fplusinf0}]\circ[\bij{i}{1\,\Fplusinf0},\ldots,\bij{i}{i\,\Fplusinf0},F\kappa_{i+1}\circ\fincoalg^F_0]\,.
\end{align*}
This concludes the proof.
\qed
\end{myproof}

\begin{mysublemma}\label{sublem:1711301734}
Let $i>0$ and $A\in\mathbb{C}$. 
We define an $\Fj{i-1}(\place+\Fplusinf 0)$-algebra $\sigma_i:\Fj{i-1}(\Fplusinf 0+\Fplusinf 0)\to \Fplusinf 0$
as follows:
\begin{align*}
&\sigma_i:=\Fj{i-1}(\Fplusinf 0+\Fplusinf 0)
\xrightarrow{\Fj{i-1}[\id_{\Fplusinf 0},\id_{\Fplusinf 0}]}
\Fj{i-1}\Fplusinf 0
\xrightarrow{\bij{i-1}{i-1\,\Fplusinf 0}} \\
&\qquad\qquad F(\textstyle{\coprod_{j=1}^{i-1}}\Fij{i-1}{j}\Fplusinf 0+\Fplusinf 0)
\xrightarrow{F[\multFplusinf_0\circ\pij{i-1}{1\,\Fplusinf0},\ldots,\multFplusinf_0\circ\pij{i-1}{i-1\,\Fplusinf0},\id_{\Fplusinf 0}]} \\
&\qquad\qquad F\Fplusinf 0 \xrightarrow{(\fincoalg^F_0)^{-1}} \Fplusinf 0\,.
\end{align*}
Then if $i$ is even (resp.\ odd), 
$J\multFplusinf_0\odot J\pij{i}{i\,\Fplusinf0}:\Fj{i}\Fplusinf 0\kto \Fplusinf0$
is the greatest fixed point of $\Phi_{J\fincoalg^{\Fj{i-1}}_{\Fplusinf0},J\sigma_i}$ 
(resp.\ $\Phi_{J(\initalg^{\Fj{i-1}}_{\Fplusinf0})^{-1},J\sigma_i}$).
\end{mysublemma}

\begin{myproof}
Assume that $i$ is even. We write $\Phi$ for $\Phi_{J\fincoalg^{\Fj{i-1}}_{\Fplusinf0},J\sigma_i}$ for simplicity.
%
For $f:\Fj{i}\Fplusinf 0\kto \Fplusinf0$, we have:
\allowdisplaybreaks[4]
\begin{align*}
&\Phi(f)\\
&=J\sigma_i\odot \overline{\Fj{i-1}}(f+\id_{\Fplusinf0})\odot J\fincoalg^{\Fj{i-1}}_{\Fplusinf0} 
\tag*{(by Def.~\ref{def:gfpprescondNew})} \\
&=J(\fincoalg^F_0)^{-1}\odot JF[\multFplusinf_0\circ\pij{i-1}{1\,\Fplusinf0},\ldots,\multFplusinf_0\circ\pij{i-1}{i-1\,\Fplusinf0},\id_{\Fplusinf 0}]
\odot J\bij{i-1}{i-1\,\Fplusinf 0}\\
&\qquad\odot \overline{\Fj{i-1}}[\id_{\Fplusinf 0},\id_{\Fplusinf 0}]\odot  \overline{\Fj{i-1}}(f+\id_{\Fplusinf0})\odot J\fincoalg^{\Fj{i-1}}_{\Fplusinf0} 
\tag*{(by definition)} \\
&=J(\fincoalg^F_0)^{-1}\odot JF[\multFplusinf_0\circ\pij{i-1}{1\,\Fplusinf0},\ldots,\multFplusinf_0\circ\pij{i-1}{i-1\,\Fplusinf0},\id_{\Fplusinf 0}]
\odot  \overline{\Fij{i-1}{0}}[f,\id_{\Fplusinf 0}] \\
&\qquad \odot  J\bij{i-1}{i-1\,\Fj{i}\Fplusinf0+\Fplusinf0} \odot J\fincoalg^{\Fj{i-1}}_{\Fplusinf0} 
\tag*{(by the naturality of $\bij{i-1}{i-1}$)} \\
&=J(\fincoalg^F_0)^{-1}\odot JF[\multFplusinf_0\circ\pij{i-1}{1\,\Fplusinf0},\ldots,\multFplusinf_0\circ\pij{i-1}{i-1\,\Fplusinf0},\id_{\Fplusinf 0}]
\odot  \overline{\Fij{i-1}{0}}[f,\id_{\Fplusinf 0}] \\
&\qquad \odot  J\bij{i}{i\,\Fplusinf0}
\tag*{(by Def.~\ref{def:1710131524})} \\
&=J(\fincoalg^F_0)^{-1}\odot \overline{F}\bigl[J\multFplusinf_0\odot J\pij{i-1}{1\,\Fplusinf0}\odot\overline{\Fij{i-1}{1}}[f,\id_{\Fplusinf 0}],\ldots,
 \\
&\qquad\qquad\qquad\qquad J\multFplusinf_0\odot J\pij{i-1}{i-1\,\Fplusinf0}\odot\overline{\Fij{i-1}{i-1}}[f,\id_{\Fplusinf 0}],f,\id_{\Fplusinf 0}\bigr]\odot  J\bij{i}{i\,\Fplusinf0} 
\tag*{(by Def.~\ref{def:1710131524})\,.\quad\raisebox{1.3em}{\smash{\begin{minipage}{1em}\begin{equation}\label{eq:1712012129}\end{equation}\end{minipage}}}} 
\end{align*}
We now show that $J\multFplusinf_0\odot\pij{i}{i\,\Fplusinf0}$ is a fixed point of $\Phi$.
By Eq.~(\ref{eq:1712012129}) above, we have:
\begin{align*}
&\Phi(J\multFplusinf_0\odot\pij{i}{i\,\Fplusinf0})\\
&=J(\fincoalg^F_0)^{-1}\odot \overline{F}\bigl[J\multFplusinf_0\odot J\pij{i-1}{1\,\Fplusinf0}\odot\overline{\Fij{i-1}{1}}[J\multFplusinf_0\odot\pij{i}{i\,\Fplusinf0},\id_{\Fplusinf 0}],\ldots,
 \\
&\qquad\qquad\qquad J\multFplusinf_0\odot J\pij{i-1}{i-1\,\Fplusinf0}\odot\overline{\Fij{i-1}{i-1}}[J\multFplusinf_0\odot\pij{i}{i\,\Fplusinf0},\id_{\Fplusinf 0}],\pij{i}{i\,\Fplusinf0},\id_{\Fplusinf 0}\bigr]\\
&\qquad\odot  J\bij{i}{i\,\Fplusinf0}  \\
&=J(\fincoalg^F_0)^{-1}\odot \overline{F}\bigl[J\multFplusinf_0\odot J\pij{i}{1\,\Fplusinf0},\ldots,J\multFplusinf_0\odot J\pij{i}{1\,\Fplusinf0},J\multFplusinf_0\odot\pij{i}{i\,\Fplusinf0},\id_{\Fplusinf 0}\bigr]
 \\
&\qquad\quad \odot  J\bij{i}{i\,\Fplusinf0}  
\tag*{(\text{by Sublem.~\ref{sublem:1712011520}})} \\
&=J\multFplusinf_0\odot J\pij{i}{j\,\Fplusinf0}
\tag*{(\text{by Sublem.~\ref{sublem:1712012042}})\,.}
\end{align*}
Hence $J\multFplusinf_0\odot J\pij{i}{i\,\Fplusinf0}$ is a fixed point of $\Phi$.

It remains to show that it is the greatest fixed point.
Let $f$ be a fixed point of $\Phi$.
For each $j\in[1,i-1]$, we have:
\begin{align*}
&J\fincoalg^F_0\odot (J\multFplusinf_0\odot J\pij{i-1}{j}\odot\overline{\Fij{i-1}{j}}[f,\id_{\Fplusinf 0}]) \\
&=JF[\multFplusinf_0\circ\pij{i-1}{1\,\Fplusinf0},\ldots,
\multFplusinf_0\circ\pij{i-1}{i-1\,\Fplusinf0},\id_{\Fplusinf0}]\odot J\bij{i-1}{j\,\Fplusinf0}
\odot\overline{\Fij{i-1}{j}}[f,\id_{\Fplusinf 0}]
\tag*{(by Sublem.~\ref{sublem:1712012042})} \\
&=JF[\multFplusinf_0\circ\pij{i-1}{1\,\Fplusinf0},\ldots,
\multFplusinf_0\circ\pij{i-1}{i-1\,\Fplusinf0},\id_{\Fplusinf0}]
\odot\overline{\Fij{i-1}{0}}[f,\id_{\Fplusinf 0}] \\
&\qquad\quad\odot J\bij{i-1}{j\,\Fij{i}{i}\Fplusinf0+\Fplusinf0}
\tag*{(by the naturality of $\bij{i-1}{j}$)} \\
&=JF[\multFplusinf_0\circ\pij{i-1}{1\,\Fplusinf0},\ldots,
\multFplusinf_0\circ\pij{i-1}{i-1\,\Fplusinf0},\id_{\Fplusinf0}]
\odot\overline{\Fij{i-1}{0}}[f,\id_{\Fplusinf 0}] \odot J\bij{i}{j\,\Fplusinf0}
\tag*{(by Def.~\ref{def:1710131524})} \\
&= \overline{F}\bigl[J\multFplusinf_0\odot J\pij{i-1}{1\,\Fplusinf0}\odot\overline{\Fij{i-1}{1}}[f,\id_{\Fplusinf 0}],\ldots,
 \\
&\qquad\quad J\multFplusinf_0\odot J\pij{i-1}{i-1\,\Fplusinf0}\odot\overline{\Fij{i-1}{i-1}}[f,\id_{\Fplusinf 0}],f,\id_{\Fplusinf 0}\bigr]\odot  J\bij{i}{j\,\Fplusinf0} 
\tag*{(by Def.~\ref{def:1710131524})\,.} 
\end{align*}
Therefore, together with Eq.~(\ref{eq:1712012129}), we can see that the following arrow is a homomorphism from 
$[\bij{i}{1\,\Fplusinf0},\ldots,\bij{i}{i\,\Fplusinf0},F\kappa_{i+1}\circ\fincoalg^F_0]$ to $\fincoalg^F_0$.
\begin{multline*}
\bigl[J\multFplusinf_0\odot J\pij{i-1}{1}\odot\overline{\Fij{i-1}{1}}[f,\id_{\Fplusinf 0}],\ldots,
 \\
 J\multFplusinf_0\odot J\pij{i-1}{i-1}\odot\overline{\Fij{i-1}{i-1}}[f,\id_{\Fplusinf 0}],f,\id_{\Fplusinf 0}\bigr]
\end{multline*}
Hence by Sublem.~\ref{sublem:1712012042} and the deterministic-greatest condition (Asm.~\ref{asm:assumption}.\ref{item:asm:det-greatest}), we have:
\begin{multline*}
\bigl[J\multFplusinf_0\odot J\pij{i-1}{1\,\Fplusinf0}\odot\overline{\Fij{i-1}{1}}[f,\id_{\Fplusinf 0}],\ldots,
 \\
 J\multFplusinf_0\odot J\pij{i-1}{i-1\,\Fplusinf0}\odot\overline{\Fij{i-1}{i-1}}[f,\id_{\Fplusinf 0}],f,\id_{\Fplusinf 0}\bigr]\\
 \sqsubseteq J[\multFplusinf_0\circ\pij{i}{1\,\Fplusinf0},\ldots,
\multFplusinf_0\circ\pij{i}{i\,\Fplusinf0},\id_{\Fplusinf0}]\,.
\end{multline*}
This immediately implies $f\sqsubseteq J\multFplusinf_0\odot J\pij{i}{i\,\Fplusinf0}$.
Hence $J\multFplusinf_0\odot J\pij{i}{i\,\Fplusinf0}$ is the greatest fixed point of $\Phi$.

The proof when $i$ is odd is similar. 
\qed
\end{myproof}

\vspace{2mm}
\begin{wrapfigure}[4]{r}{4cm}
\vspace{-\intextsep}
\vspace{-5mm}
\[
  \xymatrix@R=1.6em@C=2.3em{
  {FX}
  \kar@{-->}[r]^{\overline{F}u}
  &
  {FA}
  \kar[r]^{\overline{F}m}
  &
  {FY}
  \kar[d]^{\sigma}
  \\
  {X}
  \kar@{-->}[r]^{u}
  \kar[u]^{c}
  &
  {A}
  \kar[r]^{m}
  \kar[u]^{J(\initalg^F)^{-1}}_{\cong}
  \ar@{}[ru]|{=}  
  &
  {Y}
  }
\]
\end{wrapfigure}
\noindent\begin{minipage}{0.24\hsize}
\begin{mysublemma}\label{sublem:lfpprescondNew}
\end{mysublemma}
\end{minipage}
Let $T$ be a monad and $F$ be an endofunctor on $\mathbb{C}$.
Assume that a lifting $\overline{F}:\Kl(T)\to\Kl(T)$ of $F$ is given and
each homset of $\Kl(T)$ carries a partial order $\sqsubseteq$.
Assume further that they satisfy the conditions in Thm.~\ref{thm:initfinal}.
Let $\initalg^F:FA\to A$ be an initial $F$-algebra.
For each $X,Y\in\mathbb{C}$, $c:X\kto FX$ and $\sigma:FY\kto Y$,
if $u:X\kto A$ is the unique homomorphism from $c$ to $J(\initalg^F)^{-1}$
and a function $\Phi_{J(\initalg^F),\sigma}$ (see Def.~\ref{def:gfpprescondNew}) has a fixed point $m:A\kto Y$,  
then $m\odot u:X\kto Y$ is the least fixed point of $\Phi_{c,\sigma}$.
\vspace{2mm}

\begin{myproof}
It is easy to see that $m\odot u$ is a fixed point of $\Phi_{c,\sigma}$.
We shall show that it is the least fixed point.
Let $f:X\kto Y$ be a fixed point of $\Phi_{c,\sigma}$.

By the conditions in Thm.~\ref{thm:initfinal},
a homset $\Kl(T)(X,A)$ is $\omega$-complete and has the least element $\bot$, and 
a function $\Phi_{c,J\initalg}:\Kl(T)(X,A)\to\Kl(T)(X,A)$ is monotone and $\omega$-continuous.
Hence by the Kleene fixed point theorem, 
the unique fixed point $u:X\kto A$ of $\Phi_{c,J\initalg}$ is given by $\bigsqcup_{i\in\omega}\Phi^i_{c,J\initalg}(\bot)$.

We now prove $m\odot \Phi^i_{c,J\initalg}(\bot)\sqsubseteq f$ by the induction on $i$.
\begin{itemize}
\item If $i=0$, by the conditions in Thm.~\ref{thm:initfinal}, we have:
\[
m\odot \Phi^i_{c,J\initalg}(\bot)=m\odot\bot=\bot\sqsubseteq f\,.
\]

\item Assume that $m\odot \Phi^i_{c,J\initalg}(\bot)\sqsubseteq f$.
Then  we have:
\begin{align*}
&m\odot \Phi^{i+1}_{c,J\initalg}(\bot)\\
&=m\odot J\initalg^F \odot F(\Phi^{i}_{c,J\initalg}(\bot)) \odot c \tag*{(\text{by definition})}\\
&=\sigma\odot Fm\odot J(\initalg^F)^{-1}\odot J\initalg^F \odot F(\Phi^{i}_{c,J\initalg}(\bot)) \odot c \tag*{(\text{$m$ is a fixed point})}\\
&=\sigma\odot F(m\odot \Phi^{i}_{c,J\initalg}(\bot)) \odot c \\
&\sqsubseteq \sigma\odot Ff \odot c\tag*{(\text{by IH})} \\
&=f \tag*{(\text{$f$ is a fixed point}) \,.}
\end{align*}
\end{itemize}
Hence we have $m\odot \Phi^i_{c,J\initalg}(\bot)\sqsubseteq f$ for each $i\in\omega$.
Therefore we have:
\[
m\odot  u= m\odot \bigsqcup_{i\in\omega}\Phi^i_{c,J\initalg}(\bot) =\bigsqcup_{i\in\omega}(m\odot \Phi^i_{c,J\initalg}(\bot))\sqsubseteq f\,.
\]
Hence $m\odot u$ is the least fixed point.
\qed
\end{myproof}

\begin{myproof}[Lem.~\ref{lem:1710170852}]
We prove Eq.~(\ref{eq:1711181620}) by the induction on $i$. 
We don't have to distinguish the base case from the step case.  

We first prove Eq.~(\ref{eq:1711181620}) for $j=i$. 
Assume that $i$ is even. 
By the definition of intermediate solutions (Def.~\ref{def:solHES}), it suffices to show that
$J\multFplusinf_0\odot J\pij{i}{i\,\Fplusinf0}\odot\overline{\Fij{i}{i}}[u_{i+1},\ldots,u_{2n}] \odot\mylij{i}{i}$ is 
the greatest fixed point of the following function:
\begin{align*}
f\;\mapsto \;
&J(\fincoalg^F_0)^{-1}\odot
\overline{F}\bigl[l^{(i-1)}_1(f,u_{i+1},\ldots,u_{2n}),\ldots,
\lij{i-1}{i-1}(f,u_{i+1},\ldots,u_{2n}),\\
&\qquad\qquad\qquad f,u_{i+1},\ldots,u_{2n}\bigr] \odot c_i\,.
\end{align*}
Here the right hand side can be deformed as follows:
\allowdisplaybreaks[4]
\begin{align*}
&J(\fincoalg^F_0)^{-1}\odot
\overline{F}\bigl[l^{(i-1)}_1(f,u_{i+1},\ldots,u_{2n}),\ldots,
\lij{i-1}{i-1}(f,u_{i+1},\ldots,u_{2n}),\\
&\qquad\qquad\qquad f,u_{i+1},\ldots,u_{2n}\bigr] \odot c_i \\
&=J(\fincoalg^F_0)^{-1}\odot
\overline{F}\bigl[J\multFplusinf_0\odot J\pij{i-1}{1}\odot\overline{\Fij{i-1}{1}}[f,u_{i+1},\ldots,u_{2n}] \odot\mylij{i-1}{1},\ldots,\\
&\qquad\qquad\qquad\qquad
J\multFplusinf_0\odot J\pij{i-1}{i-1}\odot\overline{\Fij{i-1}{i-1}}[f,u_{i+1},\ldots,u_{2n}] \odot\mylij{i-1}{i-1},\\
&\qquad\qquad\qquad\qquad f, u_{i+1},\ldots, u_{2n}\bigr] \odot c_i 
\tag*{(\text{by IH})} \\
&=J(\fincoalg^F_0)^{-1}\odot
\overline{F}\bigl[J\multFplusinf_0\odot J\pij{i-1}{1}\odot\overline{\Fij{i-1}{1}}[f,u_{i+1},\ldots,u_{2n}],\ldots,\\
&\qquad\qquad\qquad\qquad
J\multFplusinf_0\odot J\pij{i-1}{i-1}\odot\overline{\Fij{i-1}{i-1}}[f,u_{i+1},\ldots,u_{2n}] ,\\
&\qquad\qquad\qquad\qquad f,u_{i+1},\ldots,u_{2n}\bigr] \odot J\bij{i-1}{i-1\,X_{i}+\cdots+X_{2n}}\odot c^\ddagger_i 
\tag*{(\text{by the definition of $c_i^\ddagger$})} \\
&=J(\fincoalg^F_0)^{-1}\odot
\overline{F}\bigl[J\pij{i-1}{1},\ldots,J\pij{i-1}{i-1},\id_{\Fplusinf0}\bigr] \\
&\qquad\odot
\overline{F}\bigl(\textstyle{\coprod_{j=1}^{i-1}} \overline{\Fij{i-1}{j}}[f,u_{i+1},\ldots,u_{2n}]+[f,u_{i+1},\ldots,u_{2n}]\bigr)\\
&\qquad
 \odot J\bij{i-1}{i-1\,X_{i}+\cdots+X_{2n}}\odot c^\ddagger_i \\
&=J(\fincoalg^F_0)^{-1}\odot
\overline{F}\bigl[J\pij{i-1}{1},\ldots,J\pij{i-1}{i-1},\id_{\Fplusinf0}\bigr]
\odot\overline{\Fij{i-1}{0}}[f,u_{i+1},\ldots,u_{2n}]\\
&\qquad
 \odot J\bij{i-1}{i-1\,X_{i}+\cdots+X_{2n}}\odot c^\ddagger_i 
\tag*{(\text{by the definition of $\Fij{i-1}{0}$})} \\
&=J(\fincoalg^F_0)^{-1}\odot
JF\bigl[\multFplusinf_0\circ\pij{i-1}{1},\ldots,\multFplusinf_0\circ\pij{i-1}{i-1},\id_{\Fplusinf0}\bigr]
\odot J\bij{i-1}{i-1\,\Fplusinf 0}
\\
&\qquad \odot \overline{\Fj{i-1}}[f,u_{i+1},\ldots,u_{2n}]
 \odot c^\ddagger_i 
\tag*{(\text{by naturality of $\bij{i-1}{i-1}$})}\\
&=J(\fincoalg^F_0)^{-1}\odot
JF\bigl[\multFplusinf_0\circ\pij{i-1}{1},\ldots,\multFplusinf_0\circ\pij{i-1}{i-1},\id_{\Fplusinf0}\bigr]
\odot J\bij{i-1}{i-1\,\Fplusinf 0}
\\
&\qquad \odot J\Fj{i-1}[\id_{\Fplusinf0},\id_{\Fplusinf0}]\odot\overline{\Fj{i-1}}(f+\id_{\Fplusinf0})\odot
 \overline{\Fj{i-1}}(\id_{X_i}+[u_{i+1},\ldots,u_{2n}])\odot c^\ddagger_i \\
&= J\sigma_i\odot \overline{\Fj{i-1}}(f+\id_{\Fplusinf0})\odot \bigl(
\overline{\Fj{i-1}}(\id_{X_i}+[u_{i+1},\ldots,u_{2n}])
 \odot c^\ddagger_i\bigr)
\tag*{(by the definition of $\sigma_i$)\,.
\quad\raisebox{1.3em}{\smash{\begin{minipage}{1em}\begin{equation}\label{eq:1712021714}\end{equation}\end{minipage}}}}  
\end{align*}
Therefore we have to show that $J\multFplusinf_0\odot J\pij{i}{i\,\Fplusinf0}\odot\overline{\Fij{i}{i}}[u_{i+1},\ldots,u_{2n}] \odot\mylij{i}{i}$ is the greatest fixed point of $\Phi_{\overline{\Fj{i-1}}(\id_{X_i}+[u_{i+1},\ldots,u_{2n}])
 \odot c^\ddagger_i,J\sigma_i}$ (see also Fig.~\ref{fig:1711151943New3}). 
 
By definition, $\mylij{i}{i}$ is the greatest homomorphism from $c^\ddagger_i$ to $J\fincoalg^F_{X_{i+1}+\cdots+X_{2n}}$,
and
$\overline{\Fij{i}{i}}[u_{i+1},\ldots,u_{2n}]$ is the greatest homomorphism from 
$\overline{\Fj{i-1}}(\id_{X_{i+1}+\cdots+X_{2n}}+[u_{i+1},\ldots,u_{2n}])\odot J\fincoalg^{\Fj{i-1}}_{X_{i+1}+\cdots+X_{2n}}$ to
$J\fincoalg^{\Fj{i-1}}_{\Fplusinf0}$. 
Therefore by Lem.~\ref{lem:gfpprescondOld}, 
$\overline{\Fij{i}{i}}[u_{i+1},\ldots,u_{2n}] \odot\mylij{i}{i}$ is the greatest homomorphism from 
$\overline{\Fj{i-1}}(\id_{X_i}+[u_{i+1},\ldots,u_{2n}]) \odot c^\ddagger_i$ to $J\fincoalg^{\Fj{i-1}}_{\Fplusinf0}$.

By Sublem.~\ref{sublem:1711301734}, $J\multFplusinf_0\odot J\pij{i}{i\,\Fplusinf0}$ is the greatest fixed point of 
$\Phi_{J\fincoalg^{\Fj{i-1}}_{\Fplusinf0},J\sigma_i}$.
Therefore by the gfp-preserving condition (Asm.~\ref{asm:assumption}.\ref{item:asm:gfp-preserving}), 
$J\multFplusinf_0\odot J\pij{i}{i\,\Fplusinf0}\odot\overline{\Fij{i}{i}}[u_{i+1},\ldots,u_{2n}] \odot\mylij{i}{i}$ is the greatest fixed point of $\Phi_{\overline{\Fj{i-1}}(\id_{X_i}+[u_{i+1},\ldots,u_{2n}])
 \odot c^\ddagger_i,J\sigma_i}$.

Eq.~(\ref{eq:1711181620}) is similarly proved when $i$ is odd, except that we use 
the finality instead of Lem.~\ref{lem:gfpprescondOld}, and 
Sublem.~\ref{sublem:lfpprescondNew} instead of the gfp-preserving condition respectively.

It remains to prove Eq.~(\ref{eq:1711181620}) for $j<i$.
We have:
\allowdisplaybreaks[4]
\begin{align*}
&\lij{i}{j}(u_{i+1},\ldots,u_{2n}) \\
&=\lij{i-1}{j}\bigl(\lij{i}{i}( u_{i+1},\ldots, u_{2n}), u_{i+1},\ldots, u_{2n}\bigr) \tag*{(by definition)}\\
&=\lij{i-1}{j}\bigl(J\multFplusinf_0\odot J\pij{i}{i\,\Fplusinf0}\odot\overline{\Fij{i}{i}}[u_{i+1},\ldots,u_{2n}] \odot\mylij{i}{i}, u_{i+1},\ldots, u_{2n}\bigr) \tag*{(by the discussion above)}\\
&=J\multFplusinf_0\odot J\pij{i-1}{j\,\Fplusinf0}\\
&\qquad\odot\overline{\Fij{i-1}{j}}\bigl[J\multFplusinf_0\odot J\pij{i}{i\,\Fplusinf0}\odot\overline{\Fij{i}{i}}[u_{i+1},\ldots,u_{2n}] \odot\mylij{i}{i},u_{i+1},\ldots,u_{2n}\bigr] \odot\mylij{i-1}{j} \tag*{(by IH)}\\
&=J\multFplusinf_0\odot J\pij{i-1}{j\,\Fplusinf0}\\
&\qquad\odot\overline{\Fij{i-1}{j}}\bigl[J\multFplusinf_0\odot J\pij{i}{i\,\Fplusinf0}\odot\overline{\Fij{i}{i}}[u_{i+1},\ldots,u_{2n}],u_{i+1},\ldots,u_{2n}\bigr] \odot\mylij{i}{j} \tag*{(by Def.~\ref{def:1710081602New})}\\
&=J\multFplusinf_0\odot J\pij{i-1}{j\,\Fplusinf0}\odot \overline{\Fij{i-1}{j}}[J\multFplusinf_0\odot J\pij{i}{i\,\Fplusinf0},\id_{\Fplusinf0}]\\
&\qquad\odot \overline{\Fij{i}{j}}[u_{i+1},\ldots,u_{2n}] \odot\mylij{i}{j} \tag*{(by Lem.~\ref{lem:1710151826})}\\
&=J\multFplusinf_0\odot J\pij{i}{j\,\Fplusinf0}\odot\overline{\Fij{i}{j}}[u_{i+1},\ldots,u_{2n}] \odot\mylij{i}{j}\tag*{(by Sublem.~\ref{sublem:1712011520})\,.}
\end{align*}
This concludes the proof. 
\qed
\end{myproof}

\afterpage{%
\begin{landscape}
\begin{figure}[p]
\[
\hspace{-3cm}
\vcenter{ \xymatrix@R=4.0em@C+1.6em{ 
  \\
  {\Fj{i-1}(
  X_i+ 
  \Fplusinf0
  )}
  \kar@{->}[r]^(.4){\overline{\Fj{i-1}}(\myl^{(i)}_i+\id)}
  &
  {\Fj{i-1}\left(\begin{aligned}
  &\Fj{i}(X_{i+1}+\cdots+X_{2n})\\ 
  &+\Fplusinf0 
  \end{aligned}
  \right)}
    \kar@{->}[r]^{\substack{\overline{\Fj{i-1}}\bigl(\overline{\Fj{i}}[u_{i+1},\ldots,u_{2n}]\\+\id\bigr)\\\mbox{}}} 
  &
  {\Fj{i-1}(
  \Fj{i}\Fplusinf 0+\Fplusinf 0
  )}
    \kar@{->}[r]^(.5){\overline{\Fj{i-1}}(J\pij{i}{i}+\id)}  
  &
  {\Fj{i-1}(\Fplusinf\Fplusinf0+\Fplusinf0)}
     \kar@{->}[r]^(.5){\overline{\Fj{i-1}}(J\multFplusinf_0+\id)}  
  &
    {\Fj{i-1}(\Fplusinf0+\Fplusinf0)}
   \kar[dd]^(.5){J\sigma_i}
  \\
  {\Fj{i-1}(X_i+\cdots+X_{2n})}
  \kar@{->}[r]^(.45){\overline{\Fj{i-1}}(\mylij{i}{j}+\id)}
  \kar[u]_(.5){\overline{\Fj{i-1}}(\id+[u_{i+1},\ldots,u_{2n}])}
  &
  {\Fj{i-1}\left(\begin{aligned}
  &\Fj{i}(X_{i+1}+\cdots+X_{2n})\\ 
  &+X_{i+1}+\cdots+X_{2n}
  \end{aligned}
  \right)}
  \kar[u]_(.5){\overline{\Fj{i-1}}(\id+[u_{i+1},\ldots,u_{2n}])}
  &
  {}
  &
  {}
  &
  {}
  \\
  {X_i}
  \kar[u]^{c^\ddagger_i}
  \kar@{->}[r]^(.4){\mylij{i}{i}}
  \ar@{}[ru]|{\color{blue}=_\nu}
  &
  {\Fj{i}(X_{i+1}+\cdots+X_{2n})}
  \kar[u]_(.5){J\fincoalg^{\Fj{i-1}}_{X_{i+1}+\cdots+X_{2n}}}
  ^(.5){\substack{\cong}}  
  \kar@{->}[r]^(.6){\overline{\Fj{i}}[u_{i+1},\ldots,u_{2n}]}  
  \ar@{}[ruu]|{\color{blue}=_\nu}  
    &
  {\Fj{i}\Fplusinf 0}
  \kar[uu]_(.5){J\fincoalg^{\Fj{i-1}}_{\Fplusinf 0}}^(.5){\substack{\cong}}  
    \kar@{->}[r]^{J{\pij{i}{i\,\Fplusinf0}}}
  \ar@{}[rruu]|{\color{blue}=_\nu}      
    &
  {\Fplusinf\Fplusinf 0}
  \kar@{->}[r]^{J\multFplusinf_0}
  &
  {\Fplusinf 0}
}}
\]
\caption{$J\multFplusinf_0\odot J\pij{i}{j\,\Fplusinf0}\odot\overline{\Fij{i}{j}}[u_{i+1},\ldots,u_{2n}] \odot\mylij{i}{j}$ is the greatest fixed point.}
\label{fig:1711151943New3}
\end{figure}
\end{landscape}
}






\section{Omitted Example}\label{sec:omittedex}
\begin{myexample}\label{ex:cexdist}
We define $F:\Sets\to\Sets$ by $F=\{o\}\times(\place)\times(\place)$.
Let $X=\{x\}$ and $Y=\{y_1,y_2\}$, and define $f:X\kto Y$ and $g:Y\kto X$ in $\Kl(\dist)$ by 
$f(x)=[y_1\mapsto \frac{1}{2},y_2\mapsto\frac{1}{2}]$ and $f(y_1)=f(y_2)=[x\mapsto 1]$.
In a similar manner to Ex.~\ref{exa:1711141355}, 
we can show that $\Fplusinf X$ is identified with the set of 
infinitary binary trees whose depth is greater than $1$,
nodes are labeled with $o$ and leaves are labeled with $x$.
A set $\Fplusinf Y$ is similarly characterized.
Let $t_X\in\Fplusinf X$ be an element identified with a tree
$o(x,o(x,o(x,\ldots)))$.
For each $t_Y\in\Fplusinf Y$,
\[
\overline{\Fplusinf}f(t_X)(t_Y) 
=J(\fincoalg^F_Y)^{-1} \odot \overline{F}(\overline{\Fplusinf}f+\id_Y) \odot \overline{F}(\id+f)\odot \fincoalg^F_X(t_X)(t_Y)  
=\frac{1}{2}\cdot \overline{\Fplusinf}f(t_X)(t_Y)\,.
\]
This implies $\overline{\Fplusinf}f(t_X)(t_Y)=0$,
and therefore $\overline{\Fplusinf}g\odot \overline{\Fplusinf}f(t_X)(t_X)=0$.
In contrast, $\id_X:X\kto X$ is a homomorphism from $\overline{F}(\id+g)\odot\overline{F}(\id+f)\odot J\fincoalg^F_X=J\fincoalg^F_X$ to itself,
and $\id_X(t_X)(t_X)=1\neq 0$. 
Hence $\overline{\Fplusinf}(g\odot f)(t_X)(t_X)\geq 1$, and this means that the operation $\overline{\Fplusinf}$ does not satisfy the functoriality.
\end{myexample}

\section{Distributive Laws from $T$ to $\Fplus$ and $\Fplusinf$}\label{sec:distlaws}
\begin{mydefinition}\label{def:distlaw}
A \emph{distributive law} from $T$ to $F$ is a natural transformation $\lambda:FT\Rightarrow TF$ that makes the following diagrams 
commute for each $X$.

\noindent
\begin{minipage}{0.4\hsize}
\begin{equation}\label{eq:distlaw1}
\xymatrix@R=1.6em@C=2.7em{
 {FTX} \ar[r]^{\lambda_X}   & 
 {TF X}  \\
 {F X} \ar[u]^{F\eta_{X}} \ar[ur]_{\eta_{FX}} & 
 }
 \end{equation}
\end{minipage}
\begin{minipage}{0.6\hsize}
\begin{equation}\label{eq:distlaw2}
\xymatrix@R=1.6em@C=2.7em{
 {FTX} \ar[rr]^{\lambda_X}   & 
 {} &
 {TF X}  \\
 {F T^2X} \ar[u]^{F\mu_X} \ar[r]^{\lambda_{TX}} & 
 {TFTX} \ar[r]^{T\lambda_{X}} &
 {T^2 FX} \ar[u]^{\mu_{FX}}
 }
 \end{equation}
\end{minipage}
\end{mydefinition}

\begin{myproposition}\label{prop:1710031711New} 
Let $\lambda:FT\Rightarrow TF$ be a distributive law from $T$ to $F$.
For  $A,X\in\mathbb{C}$, we write $\dot{\lambda}_{A,X}$ for 
$\lambda_{A+X}\circ F[T\kappa_1\circ\eta_{A},T\kappa_2]:F(A+TX)\to TF(A+X)$.
\begin{enumerate}
\item\label{item:lem:17100317111New}
Assume $T$ and $F(\place+A)$ constitute a finite trace situation for each $A\in\mathbb{C}$.
For $X\in\mathbb{C}$ we define $\lambda_{+,X}:F^+TX\to TF^+X$ as the unique 
homomorphism from 
$\dot{\lambda}_{\Fplus TX,X}
\odot J(\initalg^F_{TX})^{-1}$ 
to $J(\initalg_X^F)^{-1}$.
Then $\lambda_+:=(\lambda_{+,X})_{X\in\mathbb{C}}$ is a natural transformation $F^+T\Rightarrow TF^+$, and 
is a distributive law from $T$ to $F^+$.

\item\label{item:lem:17100317112New}
Assume  
$T$ and $F(\place+A)$ constitute an infinitary trace situation
 and 
 satisfy the gfp-preserving condition and 
the deterministic-greatest condition for each $A\in\mathbb{C}$.
For  $X\in\mathbb{C}$,  let $\lambda_{\plusinf,X}:\Fplusinf TX\to T\Fplusinf X$ be the greatest 
homomorphism from 
$\dot{\lambda}_{\Fplusinf TX,X}\odot J\fincoalg^F_{TX}$ 
to $J\fincoalg^F_{X}$.
Then 
$\lambda_\plusinf:=(\lambda_{\plusinf,X})_{X\in\mathbb{C}}$ is a natural transformation $\Fplusinf T\Rightarrow T\Fplusinf$, and 
is a distributive law from $T$ to $\Fplusinf$.
\end{enumerate}
\[
\vcenter{  \xymatrix@R=.8em@C+.6em{
  {F(F^+ TX+X)}
  \kar@{-->}[r]^(.5){\overline{F}(\lambda_{+,X}+\id)}
  &
  {F(F^+ X+X)} 
  \\
  {F(F^+ TX+TX)}
             \kar[u]_{\dot{\lambda}_{\Fplus TX,X}}
  &
%
  \\
  {F^+ TX}
  \kar[u]_{\substack{\cong}}^{J(\initalg^F_{TX})^{-1}}  
  \kar@{-->}[r]^(.5){\lambda_{+,X}}
  \ar@{}[ruu]|{=}
  &
  {F^+ X}
       \kar[uu]_{\cong}^{J(\initalg_X^F)^{-1}}   
}}
\vcenter{  \xymatrix@R=.8em@C+.6em{
  {F(\Fplusinf TX+X)}
  \kar@{->}[r]^(.5){\overline{F}(\lambda_{\plusinf,X}+\id)}
  &
  {F(\Fplusinf X+X)} 
  \\
  {F(\Fplusinf TX+TX)}
             \kar[u]_{\dot{\lambda}_{\Fplusinf TX,X}}
  &
%
  \\
  {\Fplusinf TX}
  \kar[u]_{\substack{\cong}}^{J\fincoalg_{TX}^F}  
  \kar@{->}[r]^(.5){\lambda_{\plusinf,X}}
  \ar@{}[ruu]|{\color{blue}=_\nu}
  &
  {\Fplusinf X}
       \kar[uu]_{\cong}^{J\fincoalg_X^F}   
 \mathrlap{\hspace{.2cm}\normalsize\enspace\qed}
}}
\]
\end{myproposition}

\section{Decorated Trace Semantics for Other Branching Types}\label{sec:dtsobt}
\subsection{Decorated Trace Semantics for $T=\lift$}\label{sec:dtslift}
\begin{mydefinition}\label{def:ptae}
A \emph{parity tree automaton with an exception} is a quadruple 
$\mathcal{A}=(X,\Sigma,\delta,\Omega)$ consisting of a state space $X$,
a ranked alphabet $\Sigma$, a transition function $\delta:X\to\{\bot\}+\coprod_{n=0}^\infty\Sigma^n\times X^n$ and 
a priority function $\Omega:X\to[1,2n]$.

A \emph{run tree} over $\mathcal{A}$ is a $(\Sigma\times X)$-labeled tree $\rho$ such that
for each subtree $\rho_w =((a,x),(((a_0,x_0),t_{00},\ldots,t_{0n_0}),\ldots,((a_n,x_n),t_{n0},\ldots,t_{nn_n})))$ of $\rho$,
$\delta(x)$ is defined and $(a,x_0,\ldots,x_n)=\delta(x)$ holds.
\end{mydefinition}

\begin{mylemma}\label{lem:uniquelift}
For each $x\in X$, if a run tree of $\mathcal{A}$ whose root node is labeled with $x$ exists,
then it is unique.
\qed
\end{mylemma}

\begin{myproposition}\label{prop:asmsatisfiedlift}
For each $X,Y\in\Sets$, we define a partial order $\sqsubseteq$ over $\Kl(\lift)(X,Y)$ by 
$f\sqsubseteq g\defarrow \forall x \in X.\, g(x)=\bot\Rightarrow f(x)=\bot$.
Then the conditions in Asm.~\ref{asm:assumption} are satisfied by 
$(T,F)=(\lift,\FSigma)$ wrt.\ $\sqsubseteq$. 
\qed
\end{myproposition}

\begin{myproof}
Proved in a similar manner to Prop.~\ref{prop:asmsatisfied}.
\qed
\end{myproof}

\begin{myproposition}\label{prop:dtrlift}
Let $\mathcal{A}=(X,\Sigma,\delta,\Omega)$ be a parity tree automaton with an exception where $\Omega:X\to[1,2n]$.
We define a parity $(\lift,\FSigma)$-system $(c:X\to\lift\FSigma X,(X_1,\ldots,X_{2n}))$ 
by $c:=\delta$ and $X_i:=\{x\mid \Omega(x)=i\}$ for each $i\in[1,2n]$.
Then for each $i\in[1,2n]$ and $x\in X_i$, 
with respect to the isomorphism in Prop.~\ref{prop:Fijpow},
we have:
\begin{align*}
\dtr_i(c)(x)=\begin{cases}
\Omega(\rho) &  (\text{an accepting run tree $\rho$ whose root node is labeled with $x$ exists}) \\
\bot & (\text{otherwise})\,.
\end{cases}
\end{align*}
\end{myproposition}

\begin{myproof}
Proved in a similar manner to Prop.~\ref{prop:dtrpow}.
\qed 
\end{myproof}

\subsection{Decorated Trace Semantics for $T=\giry$}\label{sec:dtsgiry}
For parity $(T,F)$-systems, the weakened conditions in \S{}\ref{sec:forprob} becomes as follows.
\begin{myassumption}\label{asm:assumptionGiry}
When $n$ is odd, the following conditions are satisfied.
\begin{enumerate}
\item[\ref{item:asm:gfp-preserving}'-1.] $T$ and $\Fj{n}(\place+A)$ satisfy the gfp-preserving condition wrt.\ an algebra 
$\overline{\Fj{n}}((\Fj{i})^\plusinf B+A)\xkrightarrow{\overline{\Fj{n}}(\id+f)}\overline{\Fj{n}}((\Fj{n})^\plusinf B+B)\xkrightarrow{J(\fincoalg^{\Fj{n}}_B)^{-1}}(\Fj{n})^\plusinf B$ for each $f:A\kto B$;

\item[\ref{item:asm:gfp-preserving}'-2.] $T$ and $\Fplus(\place+A)$ satisfy the gfp-preserving condition wrt.\ an algebra
$\Fplus (F^{\plusinf\plusinf} A+A)\xkrightarrow{J\plustoinf}\Fplusinf (F^{\plusinf\plusinf} A+A)\xkrightarrow{J(\fincoalg^{\Fplusinf}_A)^{-1}}F^{\plusinf\plusinf}A$ where $\plustoinf$ is the unique homomorphism from $(\initalg^{F}_{F^{\plusinf\plusinf} A+A})^{-1}$ to $\fincoalg^{F}_{F^{\plusinf\plusinf} A+A}$; and

\item[\ref{item:asm:gfp-preserving}'-3.] $T$ and $F(\place+A)$ satisfy the gfp-preserving condition wrt.\ an algebra
$F(\Fplusinf A+\Fplusinf A+A)\xkrightarrow{JF([\id,\id]+\id)}F(\Fplusinf A+A)\xkrightarrow{J(\fincoalg^{F}_A)^{-1}}\Fplusinf A$. 
\end{enumerate}
\end{myassumption}

By carefully checking the proofs of Prop.~\ref{prop:plusinffunctor}, Lem.~\ref{lem:1710161317} and Lem.~\ref{lem:1710170852}
where the gfp-preserving condition is used,
we can prove the following proposition.

\begin{myproposition}\label{prop:modifyok}
Thm.~\ref{thm:mainthm} still holds if we replace Cond.~\ref{item:asm:gfp-preserving} of Asm.~\ref{asm:assumption}
with the conditions in Asm.~\ref{asm:assumptionGiry}.
\qed
\end{myproposition}

\begin{mydefinition}\label{def:parttree}
Let $\Sigma$ be a ranked alphabet.
A \emph{$\Sigma$-labeled partial tree} is a finite $\Sigma+\{*\}$-labeled tree. 
Hence the set of $\Sigma$-labeled partial tree is denoted by $\Treefin(\Sigma+\{*\})$.
We say that $t=(D,l)\in\Treefin(\Sigma+\{*\})$ is a \emph{prefix} of $t'=(D',l')\in\Treeinf(\Sigma)$ 
and write $t\preceq t'$
if $D\subseteq D'$ and  $l(w)\neq *$ implies $l(w)=l'(w)$.
\end{mydefinition}

\begin{mydefinition}\label{def:cyltree}
For each $t\in\Treefin(\Sigma+\{*\})$, we define $\Cyl(t)\subseteq \Treeinf(\Sigma)$ by:
$\Cyl(t):=\{t'\in\Treeinf(\Sigma)\mid t\preceq t'\bigr\}$.
We define a $\sigma$-algebra $\sigalg_{\Treeinf(\Sigma)}\subseteq\pow \Treeinf(\Sigma)$ over 
$\Treeinf(\Sigma)$ as the smallest $\sigma$-algebra that contains $\Cyl(t)$ for each $t\in\Treefin(\Sigma+\{*\})$.
\end{mydefinition}

\begin{mydefinition}\label{def:ppta}
A \emph{probabilistic parity tree automaton (PPTA)} is a quadruple 
$\mathcal{A}=(X,\Sigma,\delta,\Omega)$ consisting of a countable state space $X$,
a ranked alphabet $\Sigma$, a transition function $\delta:X\to[0,1]^{\coprod_{n=0}^\infty\Sigma^n\times X^n}$ such that 
$\sum_{\mathbf{x}\in \coprod_{n=0}^\infty\Sigma^n\times X^n}\delta(x)(\mathbf{x})\leq 1$ for each $x\in X$,  and 
a \emph{priority function} $\Omega:X\to[1,2n]$.
We identify the set $X$ with a measurable space $(X,\pow X)$.

We inductively define a function $f:X\times \Treefin(\Sigma\times X+\{*\})\to[0,1]$ as follows:
\begin{itemize}
\item 
A function $f(\place,\Treeinf(\Sigma\times X)):X\to[0,1]$ is defined as the greatest fixed-point of the following function:
\[
g\mapsto \int_{(a,x_0,\ldots,x_{m-1})\in \sigalg_{\coprod_{m}\Sigma_m\times X^m}}\prod_{t=0}^{m-1}g(x_t) 
\;d\delta(x)\,.
\]
Here $\sigalg_{\coprod_{m}\Sigma_m\times X^m}$ denotes the $\sigma$-algebra over $\coprod_{m=0}^\infty\Sigma_m\times X^m$.

\item If $t=((a,x),t_0,\ldots,t_{m-1})$, then we let 
$f(t):=\sum_{(x_0,\ldots,x_{m-1})\in X^m}\prod_{k=0}^{m-1}\delta(x)(a,x_0,\ldots,x_{m-1})\cdot f(x_k,t_k)$\,.
\end{itemize}
By the Carath\'eodory theorem (see e.g.~\cite{ashD2000probability}), 
for each $x\in X$, there exists a unique probability measure $\ProbRun_{\mathcal{A}}(x)$ over $(\Treeinf(\Sigma\times X),\sigalg_{\Treeinf(\Sigma\times X)})$ 
such that $\ProbRun_{\mathcal{A}}(x)(\Cyl(t))=f(x,t)$ for each $t\in \Treefin(\Sigma\times X+\{*\})$.
\end{mydefinition}

\begin{myproposition}\label{prop:asmsatisfiedgiry}
For each $(X,\sigalg_X),(Y,\sigalg_Y)\in\Meas$, we define a partial order $\sqsubseteq$ over $\Kl(\giry)((X,\sigalg_X),(Y,\sigalg_Y))$ by 
$f\sqsubseteq g\defarrow \forall x\in X.\, \forall A\in\sigalg_Y.\, f(x)(A)\leq g(x)(A)$.
Then Cond.~\ref{item:asm:assumption01}--\ref{item:asm:assumption2}, \ref{item:asm:det-greatest}--\ref{item:asm:assumption101}
in Asm.~\ref{asm:assumption} and Cond.~\ref{item:asm:gfp-preserving}'-1--3 in \S{}\ref{sec:forprob}
are satisfied by 
$(T,F)=(\giry,\FSigma)$ wrt.\ $\sqsubseteq$. 
\end{myproposition}

\begin{myproof}
Cond.~\ref{item:asm:assumption01}, \ref{item:asm:assumption1}, \ref{item:asm:assumption2},
\ref{item:asm:assumption5}, \ref{item:asm:assumption6} and \ref{item:asm:assumption101}
are proved in a similar manner to Prop.~\ref{prop:asmsatisfied}.

We next prove that Cond.~\ref{item:asm:gfp-preserving}'-1 is satisfied.
Let $c:X\kto\Fj{i}(X+A)$. Let $l:X\kto (\Fj{i})^{\plusinf}A$ be the greatest homomorphism from 
$c$ to $J\fincoalg^{\Fj{i}}_A$. 
Let $m:(\Fj{i})^{\plusinf}A\kto(\Fj{i})^{\plusinf}B$ be the greatest arrow such that 
$m=(J(\fincoalg^{\Fj{i}}_B)^{-1}\odot\overline{\Fj{i}}(\id+f))\odot\overline{\Fj{i}}(m+\id))\odot\fincoalg^{\Fj{i}}_A$.

For each $k\in\omega$, we inductively define $\pi_k:(\Fj{i})^\plusinf A\to(\Fj{i}(\place+A))^k1$ as follows:
$\pi_0:=!_{(\Fj{i})^\plusinf A}$ and $\pi_{k+1}:=(\Fj{i}(\place+\id_A))^k!\circ \Fj{i}(\pi_k+\id_A)\circ \fincoalg^{\Fj{i}}_A$.
We define $\pi'_k:(\Fj{i})^\plusinf B\to(\Fj{i}(\place+B))^k1$ in a similar manner.

By Thm.~\ref{thm:constinitfin}, $((\Fj{i})^\plusinf A,(\pi_k)_{k\in\omega})$ is a limit over a final sequence 
$1\xleftarrow{!}\Fj{i}(1+A)\xleftarrow{\Fj{i}(!+\id_A)}\Fj{i}(\Fj{i}(1+A)+A)\xleftarrow{\Fj{i}(\Fj{i}(!+\id_A)+\id_A)}\ldots$.
Similarly, $((\Fj{i})^\plusinf B,(\pi'_k)_{k\in\omega})$ is a limit over 
$1\xleftarrow{!}\Fj{i}(1+B)\xleftarrow{\Fj{i}(!+\id_B)}\Fj{i}(\Fj{i}(1+B)+B)\xleftarrow{\Fj{i}(\Fj{i}(!+\id_B)+\id_B)}\ldots$.

It is known that $\giry:\Meas\to\Meas$ preserves a limit of an $\omega^\op$-sequence that consists of standard Borel 
sets~\cite{schubert09terminalcoalgebras}.
This means that $((\Fj{i})^\plusinf A,(J\pi_k)_{k\in\omega})$ is a limit over a  sequence 
$1\xkleftarrow{J!}\Fj{i}(1+A)\xkleftarrow{J\Fj{i}(!+\id_A)}\Fj{i}(\Fj{i}(1+A)+A)\xkleftarrow{J\Fj{i}(\Fj{i}(!+\id_A)+\id_A)}\ldots$.
It is easy to see that it is also a $2$-limit, i.e.\ 
for two cones $(X,(\gamma^1_k)_{k\in\omega})$ and $(X,(\gamma^2_k)_{k\in\omega})$ over
$1\xkleftarrow{J!}\Fj{i}(1+A)\xkleftarrow{J\Fj{i}(!+\id_A)}\Fj{i}(\Fj{i}(1+A)+A)\xkleftarrow{J\Fj{i}(\Fj{i}(!+\id_A)+\id_A)}\ldots$,
if $\gamma^1_k\sqsubseteq \gamma^2_k$ for each $k\in\omega$, 
then we have $l^1\sqsubseteq l^2$ where $l^1$ (resp.\ $l^2$) is a mediating arrow from 
$(X,(\gamma^1_k)_{k\in\omega})$ (resp.\ $(X,(\gamma^2_k)_{k\in\omega})$) to
$((\Fj{i})^\plusinf A,(J\pi_k)_{k\in\omega})$.
Similarly, 
$((\Fj{i})^\plusinf B,(J\pi'_k)_{k\in\omega})$ is a $2$-limit over 
$1\xkleftarrow{J!}\Fj{i}(1+B)\xkleftarrow{J\Fj{i}(!+\id_B)}\Fj{i}(\Fj{i}(1+B)+B)\xkleftarrow{J\Fj{i}(\Fj{i}(!+\id_B)+\id_B)}\ldots$.

We inductively define a cone $(X,(\gamma_k:X\kto (\Fj{i}(\place+A))^k1)_{k\in\omega})$ over
$1\xkleftarrow{J!}\Fj{i}(1+A)\xkleftarrow{J\Fj{i}(!+\id_A)}\Fj{i}(\Fj{i}(1+A)+A)\xkleftarrow{J\Fj{i}(\Fj{i}(!+\id_A)+\id_A)}\ldots$ as 
follows:
\begin{itemize}
\item For each $t\in\omega$, we define an arrow $f_t:X\kto 1$ as follows:
i) $f_0:=J!_X$ and ii) $f_{t+1}:=J!\odot F(f_t+\id_A)\odot c$.
It is easy to see that $f_0\sqsupseteq f_1\sqsupseteq\dots$.
We define $f_\omega:X\kto 1$ by $f_\omega:=\bigsqcap_{t\in\omega}f_t$.
As the composition $\odot$ in $\Kl(T)$ is $\omega^\op$-continuous, by the Kleene fixed point theorem,
$f_\omega$ is the greatest fixed point of a function $f\mapsto J!\odot F(f+\id_A)\odot c$.
We let $\gamma_0:=f_\omega$.

\item $\gamma_{k+1}:=(\Fj{i}(\place+\id_A))^k!\odot \Fj{i}(\pi'_k+\id_A)\odot c$.
\end{itemize}
For an arbitrary cone $(X,(\gamma'_k:X\kto (\Fj{i}(\place+A))^k1)_{k\in\omega})$ over
$1\xkleftarrow{J!}\Fj{i}(1+A)\xkleftarrow{J\Fj{i}(!+\id_A)}\Fj{i}(\Fj{i}(1+A)+A)\xkleftarrow{J\Fj{i}(\Fj{i}(!+\id_A)+\id_A)}\ldots$,
we have $\gamma'_k\sqsubseteq\gamma_k$ for each $k$.
Hence the mediating arrow from $(X,(\gamma_k)_{k\in\omega})$ to $((\Fj{i})^\plusinf A,(J\pi_k)_{k\in\omega})$
is the greatest homomorphism $l$. 

We inductively define a cone $(X,(\delta_k:X\kto (\Fj{i}(\place+B))^k1)_{k\in\omega})$ over
$1\xkleftarrow{J!}\Fj{i}(1+B)\xkleftarrow{J\Fj{i}(!+\id_B)}\Fj{i}(\Fj{i}(1+B)+B)\xkleftarrow{J\Fj{i}(\Fj{i}(!+\id_B)+\id_B)}\ldots$
in a similar manner. 
The mediating arrow $l':X\kto (\Fj{i})^\plusinf B$ from 
$(X,(\gamma_k)_{k\in\omega})$ to $((\Fj{i})^\plusinf B,(J\pi'_k)_{k\in\omega})$
is the greatest fixed point of $\Phi_{c,J(\fincoalg^{\Fj{i}}_B)^{-1}\odot\overline{\Fj{i}}(\id+f)}$ on the one hand. 

On the other hand, we can easily show that $(\overline{\Fj{i}}(\id+\place))^kf\odot \gamma_{k}=\delta_k\odot l$.
This implies that $m\odot l$ is a mediating arrow from $(X,(\gamma_k)_{k\in\omega})$ to $((\Fj{i})^\plusinf B,(J\pi'_k)_{k\in\omega})$.
Hence we have $l'=m\odot l$.

Cond.~\ref{item:asm:gfp-preserving}'-2 and Cond.~\ref{item:asm:gfp-preserving}'-3 are similarly proved.

It is easy to see that if $Jf\sqsubseteq g:X\kto Y$, then $g=Jf$.
Hence Cond.~\ref{item:asm:det-greatest} holds.

This concludes the proof.
\qed
\end{myproof}

\begin{mylemma}\label{lem:Fijgiry}
For $i\in\mathbb{N}$ and $j\in[1,i]$,
$\FSigmaij{i}{j}A\cong(\AccTreeij{i}{j}(\Sigma,A),\sigalg_{\AccTreeij{i}{j}(\Sigma,A)})$
 where $\sigalg_{\AccTreeij{i}{j}(\Sigma,A)}:=\sigalg_{\Treeinf((\Sigma+A)\times [1,i])} \cap\AccTreeij{i}{j}(\Sigma,A)$.
Moreover, if $i$ is odd then
the function $\decompij{i}{j}$ in Prop.~\ref{prop:Fijpow} is given by
 $(\initalg^{\Fj{j-1}}_{\coprod_{k=j+1}^{i}\FSigmaij{i}{k}A+A})^{-1}$.
If $i$ is even, then it 
is given by  $\fincoalg^{\FSigmaj{j-1}}_{\coprod_{k=j+1}^{i}\FSigmaij{i}{k}A+A}$.
\qed
\end{mylemma}

\begin{myproposition}\label{prop:dtrgiry}
Let $\mathcal{A}=(X,\Sigma,\delta,\Omega)$ be a probabilistic parity tree automaton where $\Omega:X\to[1,2n]$.
We define a parity $(\giry,\FSigma)$-system $(c:X\to\lift\FSigma X,(X_1,\ldots,X_{2n}))$ 
by $X:=(X,\pow X)$, $c(x)(\{\mathbf{x}\}):=\delta(x)(\mathbf{x})$ and $X_i:=\{x\mid \Omega(x)=i\}$ for each $i\in[1,2n]$.
Then for each $i\in[1,2n]$, $x\in X_i$ and $A\in\sigalg_{\AccTreeij{i}{j}(\Sigma,A)}$, 
with respect to the isomorphism in Lem.~\ref{lem:Fijgiry},
\begin{equation*}
\dtr_i(c)(x)(A)=\ProbRun_{\mathcal{A}}(x)\bigl(\Omega^{-1}(A)\bigr)\,.
\end{equation*}
\end{myproposition}

\begin{myproof}
Proved in a similar manner to \cite[Thm.~A.13]{urabeH15extended}.
\qed
\end{myproof}

\end{document}